\documentclass[rmp,aps,twocolumn,nofootinbib]{revtex4}

\usepackage{amsmath}   
\usepackage{latexsym}
\usepackage{graphicx}
\usepackage[usenames]{color}

\newcommand{\MSun}{M_{\odot}}
\newcommand\degrees[1]{\ensuremath{#1^\circ}}
\newcommand{\Hz}{{\rm Hz}}
\newcommand{\etal}{\emph{et al.} \,}
\newcommand{\ahat}{(a/M)}
\newcommand{\etaB}{\eta_{\beta}}
\newcommand{\Jorb}{\vec{L}}
\newcommand{\rhat}{\hat{r}}
\newcommand{\thetahat}{\hat{\theta}}
\newcommand{\phihat}{\hat{\phi}}
\newcommand{\kms}{{\,\mbox{\rm km s}}^{-1}}

\begin{document}

\title{Black-hole binaries, gravitational waves, and numerical relativity}

\author{Joan Centrella}
\email{Joan.Centrella@nasa.gov}
\affiliation{Gravitational Astrophysics Laboratory, NASA/GSFC, 8800 Greenbelt Road, Greenbelt, Maryland 20771, USA}

\author{John G. Baker}
\email{John.G.Baker@nasa.gov}
\affiliation{Gravitational Astrophysics Laboratory, NASA/GSFC, 8800 Greenbelt Road, Greenbelt, Maryland 20771, USA}

\author{Bernard J. Kelly}
\email{Bernard.J.Kelly@nasa.gov}
\affiliation{CRESST and Gravitational Astrophysics Laboratory, NASA/GSFC, 8800 Greenbelt Road, Greenbelt, Maryland 20771, USA}
\affiliation{Department of Physics, University of Maryland, Baltimore County, 1000 Hilltop Circle, Baltimore, Maryland 21250, USA}

\author{James R. van Meter}
\email{James.R.vanMeter@nasa.gov}
\affiliation{CRESST and Gravitational Astrophysics Laboratory, NASA/GSFC, 8800 Greenbelt Road, Greenbelt, Maryland 20771, USA}
\affiliation{Department of Physics, University of Maryland, Baltimore County, 1000 Hilltop Circle, Baltimore, Maryland 21250, USA}

\date{\today}

\begin{abstract}
  Understanding the predictions of general relativity for the
  dynamical interactions of two black holes has been a long-standing
  unsolved problem in theoretical physics.  Black-hole mergers are
  monumental astrophysical events, releasing tremendous amounts of
  energy in the form of gravitational radiation, and are key sources
  for both ground- and space-based gravitational-wave detectors.  The
  black-hole merger dynamics and the resulting gravitational waveforms
  can only be calculated through numerical simulations of Einstein's
  equations of general relativity.  For many years, numerical
  relativists attempting to model these mergers encountered a host of
  problems, causing their codes to crash after just a fraction of a
  binary orbit could be simulated. Recently, however, a series of
  dramatic advances in numerical relativity has allowed stable, robust
  black-hole merger simulations. This remarkable progress in the rapidly
  maturing field of numerical relativity, and the new understanding of
  black-hole binary dynamics that is emerging is chronicled. Important
  applications of these fundamental physics results to astrophysics, to
  gravitational-wave astronomy, and in other areas are also discussed.
\end{abstract}

\maketitle

\tableofcontents

\section{Prelude}
\label{sec:intro}

The final merger of two black holes in a binary system releases more
power than the combined light from all the stars in the visible
Universe.  This vast energy comes in the form of gravitational waves,
which travel across the Universe at the speed of light, bearing the
waveform signature of the merger. Today, ground-based gravitational-wave
detectors stand poised to detect the mergers of stellar
black-hole binaries, the corpses of massive stars. In addition,
planning is underway for a space-based detector that will observe the
mergers of massive black holes, awesome behemoths at the centers of
galaxies, with masses of $\sim (10^4 - 10^9)\MSun$, where $\MSun$ is
the mass of the Sun.  Since template matching forms the basis of most
gravitational-wave data analysis, knowledge of the merger waveforms is
crucial.

Calculating these waveforms requires solving the full Einstein
equations of general relativity on a computer in three spatial
dimensions plus time. As you might imagine, this is a
formidable task. In fact, numerical relativists have attempted to
solve this problem for many years, only to encounter a host of
puzzling instabilities causing the computer codes to crash before they
could compute any sizable portion of a binary orbit.  Remarkably, in
the past few years a series of dramatic breakthroughs has occurred in
numerical relativity (NR), yielding robust and accurate simulations of
black-hole mergers for the first time.

In this article, we review these breakthroughs and the wealth of new
knowledge about black-hole mergers that is emerging, highlighting key
applications to astrophysics and gravitational-wave data analysis.  We
focus on comparable-mass black-hole binaries, with component mass
ratios $1 \le q \le 10$, where $q = M_1/M_2$ and $M_1$, $M_2$ are the
individual black-hole masses. We will frequently also refer to the
\emph{symmetric mass ratio}
\begin{equation}
\label{eq:eta_def}
\eta \equiv \frac{M_1 M_2}{(M_1 + M_2)^2} = \frac{q}{(1+q)^2}.
\end{equation}
For simplicity, we choose to set $c=1$ and $G=1$; with this, we can
scale the dynamics and waveforms for black-hole binaries with the
total system mass $M$.  In particular, we can express both length and
time scales in terms of the mass, giving $M \sim 5 \times 10^{-6}
M/\MSun {\rm s} \sim 1.5 M/\MSun {\rm km}$.

We begin by setting both the scientific and historical contexts.  In
Sec.~\ref{sec:astro_grav} we provide a brief overview of astrophysical
black-hole binaries as sources for gravitational-wave detectors.  We
next turn to a historical overview in Sec.~\ref{sec:history},
surveying efforts to evolve black-hole mergers on computers, spanning
more than four decades and culminating with the recent
triumphs. Having thus set the stage, we focus on more in-depth
discussions of the key components underlying successful black-hole
merger simulations, discussing computational methodologies in
Sec.~\ref{sec:comp}, including numerical-relativity techniques and
black-hole binary initial data.  Section~\ref{sec:GWs} is the heart of
this review.  Here we discuss the key results from numerical
relativity simulations of black-hole mergers, following a historical
development and concentrating on the merger dynamics and the resulting
gravitational waveforms.  These results have opened up a variety of
exciting applications in general relativity, gravitational waves, and
astrophysics.  We discuss synergistic interactions between numerical
relativity and analytic approaches to modeling gravitational dynamics
and waveforms in Sec.~\ref{sec:nrpn}, and applications of the results
to gravitational-wave data analysis in Sec.~\ref{sec:gwda}. The impact
of merger simulations on astrophysics is presented in
Sec.~\ref{sec:astro}, which includes discussions of recoiling black
holes and potential electromagnetic signatures of the final merger.
We conclude with a look at the frontiers and future directions of this
field in Sec.~\ref{sec:frontiers}.

Before we begin, we mention several other resources that may
interest our readers.  The review articles by \citet{Lehner:2001wq}
and \citet{Baumgarte:2002jm} provide interesting surveys of numerical
relativity several years before the breakthroughs in black-hole merger
simulations.  The article by \citet{Pretorius:2007nq} is an early
review of the recent successes, covering some of the same topics that
we discuss here. \citet{Hannam:2009rd} reviewed the status of black-hole
simulations producing long waveforms (including at least ten 
cycles of the dominant gravitational-wave mode) and their application
to gravitational-wave data analysis.  Finally, the text books by
\citet{Bona05} and \citet{Alcubierre08} provide many more details on
the mathematical and computational aspects of numerical relativity
than we can include here, and serve as useful supplements to our
discussions.

\section{Black-Hole Binaries and Gravitational Waves}
\label{sec:astro_grav}

Black holes and gravitational waves are surely among the most exotic
and amazing predictions in all of physics.  These two offspring of
Einstein's general relativity are brought together in black-hole
binaries, expected to be among the strongest emitters of gravitational
radiation.

\subsection{Basic Properties}
\label{sec:astro_grav:basic}

We begin by presenting some basic properties of black holes and
gravitational waves.  For fuller discussions and more details, see
\citet{Misner73} and \citet{Schutz09}.

\subsubsection{Black-Hole Basics}
\label{sssec:astro_grav:basic:bh}

A black hole forms when matter collapses to infinite density,
producing a singularity of infinite curvature in the fabric of
spacetime.  Each black hole is surrounded by an event horizon, at
which the escape velocity is the speed of light. The event horizon is
a global property of the spacetime, since it is defined by the paths
of ``outgoing" photons that are the boundary between photon
trajectories that must fall inward, and those that can escape to
infinity.  The photons defining the event horizon hover at finite
radius at the surface of the black hole. Since, in principle, mass
(energy) can fall into the event horizon at late times -- which will
move the location at which photon paths can hover -- we must know the
entire future development of the system to locate the event horizon.

When black holes merge, a single event horizon forms whose area is at
least as large as the sum of the individual horizons.  Since numerical
relativists want to know when this occurs during the course of a
calculation, they rely on a related concept known as an \emph{apparent
horizon}, whose location depends only on the properties of the
spacetime at any given time \cite{Poisson04}. For quiescent black
holes, the apparent and event horizons coincide; for more general
holes, the apparent horizon is always inside the event horizon (with
restrictions on the behavior of the matter involved).  So, in terms of
causality in a numerically-generated spacetime, any physical
phenomenon found inside an apparent horizon should not leak out and
affect the spacetime outside.

The simplest black hole is nonrotating and is described by the
spherically symmetric Schwarzschild solution to the Einstein equations
of general relativity in vacuum (i.e., with no ``matter'' sources in
the spacetime). A Schwarzschild black hole is fully specified by one
quantity, its mass $M$. The horizon is located at coordinate $r = 2M$
(in Schwarzschild coordinates); its area is $4\pi(2M)^2$.

More general black holes can have both charge and spin. Since a
charged black hole in astrophysics will generally be neutralized
rapidly by any surrounding plasma, we can consider only rotating,
uncharged black holes.  Stationary (i.e., time independent) black
holes are described by the axisymmetric Kerr solution. A Kerr black
hole is fully specified by two quantities, its mass $M$ and its
angular momentum per unit mass $a$. The event horizon is located at
the Boyer-Lindquist \cite{Misner73} radius $r_+$, where
\begin{equation}
r_+ = M + (M^2 - a^2)^{1/2}.
\label{eq:horizon}
\end{equation}
The area of the event horizon is $8\pi M r_+$.

Equation~\eqref{eq:horizon} requires $a \le M$; when $a = M$ the black
hole is said to be maximally rotating or ``extremal''.  Notice that
$r_+ = 2M$ when $a=0$, and that $r_+$ decreases as $a$ increases, thus
bringing the location of the horizon deeper into the potential well as
the black-hole spin increases.

Photons and test particles in the vicinity of a single black hole can
experience either stable or unstable orbits.  For a Schwarzschild
black hole of mass $M$, the \emph{innermost stable circular orbit} (ISCO)
occurs at $r = 6M$ for massive test particles.  In the case of a
Kerr black hole, the ISCO is closer in for co-rotating test particles,
and farther out for counter-rotating particles.
 
While the concept of an ISCO is strictly defined only for massive test
particles, it has proven useful for studies of the spacetime around
two black holes spiralling together on quasicircular orbits.  Imagine
that you put the two black holes on an {\em instantaneously} circular
orbit around each other; at that moment they have neither nonzero
radial velocity nor nonzero radial acceleration.  At any given
separation, the black holes have some angular momentum.  The ISCO is
the separation where that angular momentum is a minimum, in analogy to
the test particle definition.  Black holes at closer separations would
be expected to fall inward, toward the center, even without
radiating angular momentum via gravitational radiation.

\subsubsection{Gravitational Wave Primer}
\label{sssec:astro_grav:basic:gw}

Gravitational waves are ripples in the curvature of spacetime itself.
They carry energy and momentum and travel at the speed of light,
bearing the message of disturbances in the gravitational field.

As with electromagnetic waves, gravitational waves can be decomposed
into multipolar contributions that reflect the nature of the source
that generates them.  Recall that electromagnetic radiation has no
monopole contribution due to the conservation of total charge.  By
analogy, conservation of total mass-energy guarantees that there can
be no monopole gravitational radiation.  Since dipolar variations of
charge and currents are possible, electromagnetic waves can have a
dipole character.  However, conservation of linear and angular
momenta removes any possibility of dipolar gravitational waves, so the
leading-order contribution to gravitational radiation is quadrupolar.
 
Gravitational waves are thus generated by systems with time-varying
mass quadrupole moments \cite{Misner73,Flanagan:2005yc}.  In the wave
zone, a gravitational wave is described as a perturbation $h$ to a
smooth underlying spacetime.  The wave amplitude is
\begin{equation}
\label{h-amp}
 h \sim \frac{G}{c^4} \frac{\ddot{Q}}{r}
 \sim \frac{GM_{\rm quad}}{rc^2} \frac{v^2}{c^2},
\end{equation}
where $Q$ is the quadrupole moment of the source, $r$ is the distance
from the source, and $M_{\rm quad}$ is the mass in the source that is
undergoing quadrupolar changes.  This shows that the strongest
gravitational waves will be produced by large masses moving at high
velocities, such as binaries of compact stars and black holes.

\begin{figure}
\includegraphics*[width=3.5in,angle=0]{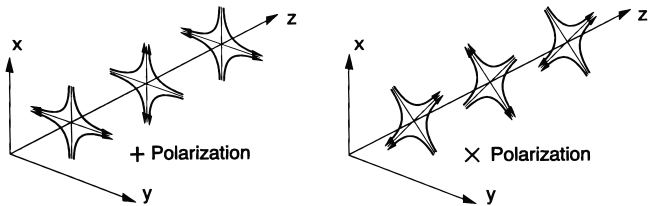}
\caption{Lines of force for plane gravitational waves propagating
  along the $z$ axis.  The wave on the left is purely in the $+$
  polarization state, and the one on the right is purely in the
  $\times$ polarization state.  The gravitational waves produce tidal
  forces in planes transverse to the propagation direction. From
  \cite{Abramovici:1992ah}. Reprinted with permission from AAAS.}
\label{fig:forcelines}
\end{figure}
A gravitational wave is purely transverse, acting tidally in
directions perpendicular to its propagation direction.  When a
gravitational wave impinges on a detector of length scale $L$, it
produces a length change in that detector $\delta L/L \sim h/2$.  By
substituting in typical values for compact objects in the Universe
into Eq.~\eqref{h-amp}, one can see that astrophysical sources
typically yield wave amplitudes of $h \lesssim 10^{-21}$ at the Earth.
Consequently, precision measurements are needed to make detections.

Gravitational waves have two polarization components, known as $h_+$
and $h_{\times}$ for a linearly polarized
wave. Figure~\ref{fig:forcelines} shows the corresponding lines of
force for sinusoidal gravitational waves propagating along the $z$
axis.  In the purely $+$ polarization on the left, the wave stretches
along one axis and squeezes along the other, alternating sinusoidally
as the wave passes.  The $\times$ polarization wave on the right acts
similarly, stretching and squeezing along axes rotated by
$\degrees{45}$.  In general, a gravitational wave is a superposition
of these two states, conveniently written as a complex waveform strain
$h$, where
\begin{equation}
h = h_+ + ih_{\times}.
\label{eq:h}
\end{equation}
 
\subsection{Astrophysical Black Holes}
\label{sec:astro_grav:astro_BHS}

Astronomers have found evidence for black holes throughout the
Universe on a remarkable range of scales.  The smallest of these,
stellar black holes, have masses in the range $\sim (3 - 30) \MSun$
and form as the end-products of massive star evolution.  There is good
observational evidence for the existence of stellar black holes, based
on dynamical measurements of the masses of compact objects in
transient systems that undergo X-ray outbursts.  Since neutron stars
cannot have masses $\gtrsim 3 \MSun$, compact objects more massive
than this must be black holes \cite{Remillard:2006fc}.

Intermediate-mass black holes (IMBHs) have masses in the range $\sim
(10^2 - 10^4) \MSun$.  IMBHs may form as the result of multiple
mergers of smaller objects in the centers of dense stellar clusters in
the present Universe \cite{PortegiesZwart:2002jg,Miller:2003sc},
assuming mass loss from stellar winds is not significant
\cite{Glebbeek:2009jr}.  They may also arise from the evolution of
very massive stars early in the history of the Universe, forming black-hole
``seeds'' in the centers of massive halos (the precursors of the
galaxies we see today) early in the history of the Universe, to
redshifts $z \gtrsim 10$ \cite{Madau:2001sc}.  Currently the best
observational evidence for IMBHs comes from models of ultraluminous
X-ray sources \cite{Colbert:2004sz}.
 
Finally, massive black holes (MBHs) have masses in the range $\sim
(10^4 - 10^9) \MSun$ and are found at the centers of galaxies,
including our own Milky Way galaxy.  The observational case for the
existence of MBHs is quite strong, based on dynamical models of stars
and gas believed to be moving in the potential well of the central MBH
\cite{Kormendy:1995er,Richstone:1998ky,Ferrarese:2004qr,Desroches:2009vd}.

Black-hole binaries are binary systems in which each component is a
black hole.  As mentioned above, we focus here on comparable-mass
binaries, which are expected to produce the strongest gravitational-wave
signals. Stellar black-hole binaries may form as the result of
binaries composed of two massive stars; see \citet{Bulik:2009ew} and
references therein.  Stellar black-hole binaries may also arise from
dynamical processes in which a black hole is captured into an orbit
around another black hole in dense stellar environments
\cite{O'Leary:2007qa, Miller:2008yw}.  IMBH binaries can also form
through dynamical processes in stellar clusters \cite{Gurkan:2005xz}
and from mergers of massive halos at high redshifts.  Since both
stellar and IMBH binaries are ``dark'' -- that is, they are generally
not surrounded by gas which might produce electromagnetic radiation --
we have few observational constraints on these types of black-hole
binaries.  This situation will change dramatically, however, with the
detection of gravitational radiation from these systems
\cite{Bulik:2009ew,Miller:2008fi}, as gravitational waves bring direct
information about the dynamical behavior of the orbiting masses and do
not rely on electromagnetic emissions from nearby matter.

Since essentially all galaxies are believed to contain an MBH at the
center and to undergo a merger with another galaxy at least once
during the history of the Universe, MBH binaries can arise when their
host galaxies merge \cite{Begelman:1980vb}; see also
\citet{Djorgovski:2008hc} and references therein.  However, due to the
vast cosmic distances involved, and the small angular separations on
the sky expected for MBH binaries, only a few candidates are currently
known through electromagnetic observations \cite{Komossa:2003wz,
  Komossa:2002tn, Rodriguez:2006th}.  When they form, MBH binaries
typically have relatively wide separations, and the gravitational
radiation they emit is very weak and insufficient to cause the binary
to coalesce within the age of the Universe.  However, various
processes such as gaseous dissipation and $N$-body interactions with
stars can remove orbital energy from the binary and cause the black
holes to spiral together \cite{Gould:1999ia,Armitage:2002uu}; see also
\citet{Berentzen:2008yw} and \citet{Colpi:2009ny} and references
therein.  Eventually, the black holes reach separations at which
gravitational radiation reaction becomes the dominant energy-loss
mechanism, leading to the final coalescence of the black holes and the
emission of strong gravitational waves \cite{Sesana:2008ur}.

\subsection{Gravitational Waves from Black-Hole Binaries}
\label{sec:astro_grav:gws}

Mergers of comparable-mass black-hole binaries are expected to be
among the strongest sources of gravitational waves.  This final death
spiral of a black-hole binary encompasses three stages: inspiral,
merger, and ringdown \cite{Flanagan:1997sx,Hughes:2009iq}.

In the early stages of the inspiral, the orbits of most astrophysical
black-hole binaries will circularize due to the emission of
gravitational radiation \cite{Peters:1963ux,Peters:1964zz}.  During
the inspiral, the orbital time scale is much shorter than the time
scale on which the orbital parameters change; consequently, the black
holes spiral together on quasicircular orbits.  Since the black holes
have wide separations, they can be treated as point particles.  The
inspiral dynamics and waveforms can be calculated using post-Newtonian
(PN) equations, which result from a systematic expansion of the full
Einstein equations in powers of $\epsilon \sim v^2/c^2 \sim
GM/Rc^2$, where $R$ is the binary separation \cite{Blanchet:2006LR}.
The inspiral phase produces gravitational waves in the characteristic
form of a {\em chirp}, which is a sinusoid with both frequency and
amplitude increasing with time.

As the black holes spiral inward, they eventually reach the
strong-field, dynamical regime of general relativity.  In this merger
stage, the orbital evolution is no longer quasi-adiabatic; rather, the
black holes plunge together and coalesce into a single, highly
distorted remnant black hole, surrounded by a common horizon.  Since
the point-particle and PN approximations break down, numerical
relativity simulations of the Einstein equations in three dimensions are needed to
calculate the merger. Due to the difficulty of these simulations, the
resulting gravitational waveforms were completely unknown until
recently.

Finally, the highly distorted remnant black hole settles down into a
quiescent rotating Kerr black hole by shedding its nonaxisymmetric
modes through gravitational-wave emission.  We call this process the
``ringdown,'' in analogy to how a bell that has been struck sheds its
distortions as sound waves.  Various analytic techniques of black-hole
perturbation theory
\cite{Regge:1957td,Zerilli:1970se,Teukolsky:1973ha} form the basis of
ringdown calculations, producing gravitational waveforms in the shape
of exponentially damped sinusoids \cite{Leaver:1985ax,Berti:2009kk}.

The characteristic gravitational-wave frequency of a quasicircular
black-hole binary, produced by the dominant (highest-order) quadrupole
component, is
\begin{equation}
\label{eq:freq}
f_{\rm GW} \sim 2f_{\rm orb} \sim (M/R^3)^{1/2},
\end{equation} 
where $f_{\rm orb}$ is the orbital frequency. Astrophysical black-hole
binaries produce gravitational waves that span three frequency
regimes, depending on the black-hole masses \cite{Flanagan:2005yc}.
Stellar black-hole binaries and the lower mass end of the IMBH
binaries radiate in the high frequency band, $f_{\rm GW} \sim (10 -
10^4) \Hz$, which is already being observed by ground-based laser
interferometer detectors such as LIGO \cite{Abbott:2009qj}, and will
be observed by the advanced detectors by $\sim 2016$
\cite{Smith:2009bx}.  Low frequency gravitational waves cover the band
$f_{\rm GW} \sim (10^{-5} - 1) \Hz$ and will be observed by the
space-based laser interferometer LISA, currently under development
\cite{Jennrich:2009ti}.  MBH binaries with masses $M \sim (10^{4.5} -
10^7) \MSun$ will be very strong sources for LISA, with the lower mass
systems visible out to redshifts $z \gtrsim 10$ \cite{Arun:2008zn};
the inspirals of IMBH binaries will also be detectable
\cite{Miller:2008fi}.  Finally, the very low frequency band $f_{\rm
  GW} \sim (10^{-9} - 10^{-7}) \Hz$ will be observed by pulsar timing
arrays \cite{Verbiest:2009kb}.  This band is expected to dominated by
gravitational waves from a very large population of unresolved MBH
binaries \cite{Sesana:2008mz} with possibly a few discrete sources
\cite{Sesana:2008xk}.

\section{Historical Overview}
\label{sec:history}

The quest to calculate the gravitational-wave signals from the merger
of two black holes spans more than four decades and encompasses key
developments in theoretical and experimental general relativity,
astrophysics, and computational science.  In this section, we begin by
delineating these threads in general terms, and then turn to a more
detailed account of select milestones along the path toward successful
simulations of black-hole mergers.

\subsection{Setting the Stage}
\label{sec:history:stage}

At the end of the 18th century \citet{Michell:1784} and
\citet{Laplace:1796} first speculated, using Newtonian gravity, that a
star could become so compact that the escape velocity from its surface
would exceed the speed of light.  In the 20th century,
scientists realized that such black holes could form as the final
state of total gravitational collapse in general relativity
\cite{Oppenheimer:1939ue,HarrisonEtAl_1965}; John Wheeler would later
popularize the term ``black hole'' to describe such an object
\cite{MisnerEtAl_2009,RuffiniWheeler_1971}.  Beginning in the 1960s,
many highly energetic astrophysical phenomena were discovered with
physical properties pointing to extremely strong gravitational fields
as underlying mechanisms; among these are quasars and X-ray binaries
such as Cygnus X-1 \cite{Overbeck:1967,Overbeck:1968}, the first
credible black-hole candidate.  As discussed in Sec.~\ref{sec:astro_grav:astro_BHS},
today astrophysical black holes are believed to exist on a vast range
of scales throughout the Universe, and black-hole binaries are
considered to be strong sources of gravitational waves.

Einstein's equations of general relativity form a coupled set of
nonlinear partial differential equations, in which dynamic curved
spacetime takes the role of Newton's gravitational field and interacts
nonlinearly with massive bodies.  Gravitational waves were first
recognized as solutions to the linearized, weak-field Einstein
equations early in the past century.  By mid-century, gravitational
waves were recognized as real physical phenomena, carrying energy and
being capable of producing a response when impinging on a detector.
This development spawned a major branch of experimental general
relativity, with concepts for the first gravitational-wave detectors
appearing in the 1960s; see \citet{Camp:2004gg} for a review.  The
discovery of two neutron stars in a binary system by
\citet{Hulse:1974eb} provided an astrophysical laboratory for the
first {\em indirect} detection of gravitational radiation. Decades of
observation revealed the binary orbit to be shrinking by precisely the
amount expected if the system were emitting gravitational waves
according to general relativity \cite{Weisberg:2004hi}; Hulse and
Taylor were awarded the Nobel Prize in 1993. Today, the prognosis for
{\em direct} detection of gravitational waves is excellent, with the
first events expected from the advanced ground-based interferometers
around the middle of this coming decade.

As you might expect, Einstein's equations pose formidable obstacles to
anyone who would dare to probe the physics within.  Throughout most of
the 20th century, relativists uncovered a fairly small number
of exact solutions by exploiting symmetries, and made progress toward
more general problems using various perturbative expansions.  By the
1960s, computers had become powerful enough to encourage attempts at solving
Einstein's equations numerically, to uncover physics beyond the realm
of perturbation theory.  The subsequent development of numerical
relativity has been made possible in part by continued increases in
computer power and advances in algorithms and computational methods.

Most numerical-relativity simulations start with the idea of decomposing
four-dimensional spacetime into a stack of curved three-dimensional spacelike slices
threaded by a congruence of timelike curves \cite{York79}.  Arnowit,
Deser, and Misner (ADM) pioneered this ``3+1'' approach as the basis
for a canonical formulation of the dynamics of general relativity
\cite{Arnowitt:1962hi}.  In this Hamiltonian formulation, the three-metric
$\gamma_{ij}$ on the spatial slices takes the role of the
``configuration variables.'' Quantities based on the extrinsic
curvature $K_{ij}$, which is roughly the time derivative of
$\gamma_{ij}$, play the role of ``conjugate momenta.''  Variation of
an action with respect to $\gamma_{ij}$ produces a set of six
first-order evolution equations for the conjugate momenta; varying the
momenta gives six first-order evolution equations for $\gamma_{ij}$.
ADM also introduced four Lagrange multipliers as freely specifiable
gauge or coordinate conditions, representing the four coordinate
degrees of freedom in general relativity. Variation of these Lagrange
multipliers yields four equations that must hold on each slice: a
Hamiltonian constraint, and three momentum constraints.

Originally intended as a tool for quantizing gravity, the ADM
formalism later became the basis for most work in numerical
relativity.  As we discuss in Sec.~\ref{sec:comp} below, key elements
in this approach are solving the Cauchy problem, beginning with the
initial data on a spacelike slice, and then evolving that data forward
in time.  The constraint equations form the basis for this initial
value problem. Appropriate choices for the gauge conditions are
crucial ingredients for today's successful black-hole merger
simulations.

\subsection{Numerical Relativity Milestones}

\citet{Hahn64} made the first known attempt to simulate the head-on
collision of two equal-mass black holes on a computer in 1964, using a
two-dimensional axisymmetric approach. Their simulation ran for 50 timesteps to a
duration of $\sim 1.8 M$; at this point, they decided that the
simulation was no longer accurate enough to warrant continued
evolution, and stopped the code.  Smarr and Eppley
\cite{Eppley75,Smarr75,Smarr:1976qy,Smarr77} returned to this problem in
the mid-1970s, again employing two-dimensional axisymmetry but now using the ADM
formalism, specialized coordinates, and improved coordinate
conditions.  Although they encountered problems with instabilities and
large numerical errors, they managed to evolve the collision and
extract information about the emitted gravitational waves. Smarr and
Eppley had used the most powerful computers of their day. Going to the
next step, orbiting black holes in three dimensions, was deemed to be not
feasible at the time, due to unresolved questions about the
instabilities and insufficient computer power.  Consequently, the
black-hole merger problem lay dormant for over a decade.

In the 1990s, attention returned to black-hole mergers as the LIGO
project began to move forward, and black-hole mergers were recognized
as the strongest sources for this detector. Since the signal-to-noise ratio
for such ground-based detectors is fairly modest, having a template
for the merger waveform is a key part of the data analysis strategy.
Numerical relativists revisited the problem of colliding two black
holes head-on with modern techniques and more powerful computers
\cite{Bernstein:1994wt, Anninos:1993zj}. In the mid-1990s, the
National Science Foundation funded a Computational Grand Challenge
grant for a large multi-institution collaboration aimed at evolving
black-hole mergers in three dimensions and calculating the resulting gravitational
waveforms. Around the same time, a large and very active numerical
relativity group arose at the newly-formed Albert-Einstein Institut
(AEI) in Potsdam, Germany.  During the late 1990s and into the early
2000s, two developments on the experimental side further increased the
desire for black-hole merger simulations: the ground-based
gravitational-wave detectors started taking data, and interest grew in
LISA and its potential for observing gravitational waves from massive
black-hole binary mergers.

While no one expected the task at hand -- developing computer codes to
solve the full Einstein equations in three dimensions for the final few orbits and
merger of two black holes -- to be simple, numerical relativists found
that the problem was far more difficult than anticipated.  Producing a
waveform useful for gravitational-wave detection purposes typically
would require running a simulation for a duration of several hundred
$M$.  However, a variety of instabilities plagued the codes, causing
them to crash well before any significant portion of an orbit could be
achieved.

Nevertheless, during a period encompassing a little over a decade,
much important work was accomplished that laid the foundations for
later success.  Key milestones include these developments:
\begin{itemize}
\item{initial data for binary black holes near the ISCO
    \cite{Cook:1994va,Cook:2004kt,Cook:2001wi,Baumgarte:2000mk};}
\item{new methods for representing black holes on computational grids
    such as punctures \cite{Brandt:1997tf} and excision
    \cite{Seidel:1992vd,Anninos:1994dj,
      Alcubierre:2000yz,Shoemaker:2003td};}
\item{recognition of the importance of hyperbolicity in formulating
    the Einstein equations for numerical solution
    \cite{Bona92,Abrahams:1996hh, Anderson:1997jn,Friedrich:2000qv};}
\item{improved formulations of the Einstein equations
    \cite{Nakamura:1987zz,Shibata:1995we,Baumgarte:1998te};}
\item{fully three-dimensional evolution codes and their use in evolving distorted
    black holes \cite{Brandt:2002wa,Camarda:1998wf}, boosted black holes
    \cite{Cook:1997na}, head-on collisions \cite{Sperhake:2005uf}, and
    grazing collisions
    \cite{Bruegmann:1997uc,Brandt:2000yp,Alcubierre:2000ke};}
\item{coordinate conditions that keep the slices from crashing into
    singularities and the spatial coordinates from falling into the
    black holes as the evolution proceeds
    \cite{Bona:1994dr,Alcubierre:2002kk};}
\item{the Cactus Computational Toolkit\footnote{\tt
      http://www.cactuscode.org/}, which provided a framework for
    developing numerical-relativity codes and analysis tools used by
    many groups;}
\item{modern adaptive mesh refinement finite-difference (including
    Carpet\footnote{{\tt http://www.carpetcode.org/}} with Cactus, BAM
    \cite{Bruegmann:1997uc}, and Paramesh \cite{MacNeice00}), and
    multi-domain spectral \cite{Pfeiffer:2002wt} infrastructures for
    numerical relativity.}
\end{itemize}

Throughout this period, the length of black-hole evolutions gradually
increased. Improvements in the formalisms allowed simulations of
single black holes and, later, two black holes to increase in duration
to $\gtrsim 10M$.  The addition of new slicing and shift conditions
again increased the evolution times to $\gtrsim 30M$.  The Lazarus
project took a novel approach to combine these relatively short-duration
binary simulations with perturbation techniques for the
late-time behavior to produce a hybrid model for a black-hole merger
\cite{Baker:2000zh,Baker:2001sf,Baker:2001nu,Baker:2002qf}.  By late
2003, \citet{Bruegmann:2003aw} carried out the first complete orbit of
two equal-mass, nonspinning black holes.  While this simulation lasted
$\sim 100M$, the code crashed shortly after the orbit was completed
and the gravitational waves were not extracted.  Overall, progress was
slow, difficult, and incremental.  However, the situation was about to
change dramatically.

\subsection{Breakthroughs and the Gold Rush}
\label{sec:history:break}

In early 2005, Frans Pretorius electrified the relativity community
when he achieved the first evolution of an equal-mass black-hole
binary through its final orbit, merger, and ringdown using techniques
very different from those employed by the rest of the community
\cite{Pretorius:2005gq}.  Later in 2005 the groups at the University
of Texas at Brownsville (UTB) and NASA's Goddard Space Flight Center
independently and simultaneously discovered a new method, called
``moving punctures,'' that also produced successful black-hole mergers
\cite{Campanelli:2005dd,Baker:2005vv}.  Their presentations were given
back-to-back at a workshop on numerical relativity, to the amazement
of each other and the assembled participants.

Since the moving-puncture approach was based on underlying techniques
used by several other groups, it was rapidly and readily adopted by most
in the community, producing a growing avalanche of results.  2006 was
a year of many firsts: the first simulations of unequal-mass black
holes and the accompanying recoil of the remnant hole
\cite{Baker:2006vn}, the first mergers of spinning black holes
\cite{Campanelli:2006uy}, the first long waveforms ($\sim 7$ orbits)
\cite{Baker:2006kr}, the first comparisons with PN results
\cite{Buonanno:2006ui,Baker:2006ha}, and the first systematic
parameter study in numerical relativity \cite{Gonzalez:2006md}.  The
year 2007 opened with the discovery of ``superkicks'' -- recoil
velocities exceeding $1000 \kms$
\cite{Gonzalez:2007hi,Campanelli:2007ew}, and in 2008 a black-hole
binary merger with mass ratio $q = 10$ was accomplished
\cite{Gonzalez:2008bi}, while \citet{Campanelli:2008nk} carried out
the first long-term evolution of generic spinning and precessing
black-hole binaries.  As this article was being written in 2009 the
state of the art continues to advance, with the first simulations of
black-hole mergers using spectral numerical techniques
\cite{Scheel:2008rj,Chu:2009md,Szilagyi:2009qz}, and the first steps
towards modeling the flows of gas around the merging black holes
\cite{vanMeter:2009gu}.  Applications of the merger results in areas
such as comparisons with PN expressions for the waveforms,
astrophysical computations of black-hole merger rates, and the
development of templates for gravitational-wave data analysis have
accompanied these technical developments in the simulations.  The
study of black-hole mergers using numerical relativity is thriving
indeed.

\section{Numerical Development}
\label{sec:comp}

In this section, we discuss the mathematical and numerical foundations
underlying current black-hole merger simulations, highlighting the key
issues involved in achieving successful evolutions.  For more detailed
treatments, we direct the interested reader to \citet{Alcubierre08} or
\cite{Gourgoulhon:2007ue}.

\subsection{Einstein's Equations}
\label{sec:Einstein}

The central task of numerical relativity is solving Einstein's field
equations
\begin{equation}
\label{Einsteinsequations}
G_{\mu\nu}=8\pi T_{\mu\nu},
\end{equation}
where Einstein's tensor $G_{\mu\nu}$ represents the curvature of
spacetime, the energy-momentum tensor $T_{\mu\nu}$ contains the matter
sources, and $\mu, \nu = 0,1,2,3$. By convention, an index $\mu = 0$
selects a `` time'' component, and $\mu = 1,2,3$ selects a ``space''
component.  $G_{\mu\nu}$ depends on the first and second derivatives
of the metric tensor $g_{\mu\nu}$. For vacuum black-hole spacetimes,
$T_{\mu\nu} = 0$.  Note that all of the tensor fields discussed here
are symmetric in the indices, e.g. $g_{\mu\nu} = g_{\nu\mu}$.

The metric characterizes the geometry of spacetime by giving the
infinitesimal spacetime interval $ds$ through the following
definition:
\begin{equation}
ds^2=g_{\mu\nu}dx^{\mu}dx^{\nu},
\label{eq:interval}
\end{equation}
where we use the Einstein summation convention which implies summation
over all values of a given index that appears twice in an expression.
Many physical implications of the metric are immediately apparent from
Eq.~(\ref{eq:interval}).  For example, when $ds^2 = 0$, the resulting
metric-determined relationship between the time and space coordinates
yields the paths that light rays must follow in this spacetime.

The dependence of Einstein's tensor on the metric can be simply
illustrated in coordinates known as ``harmonic"; in vacuum
Eq.~\eqref{Einsteinsequations} takes the form:
\begin{equation}
\label{eq:harmonic}
\Box g_{\mu\nu}-t_{\mu\nu}=0,
\end{equation}
where $\Box$ is the flat-space wave-operator and $t_{\mu\nu}$
represents all terms nonlinear in the metric.  If $t_{\mu\nu}$ is
interpreted as an effective source term, this is a simple wave
equation.  The familiar form of this equation suggests that its Cauchy
problem can be solved by specifying the metric and its first
derivative on an initial, spatial surface and then integrating in
time, as for an ordinary wave equation.

Einstein's equations admit various formulations and coordinate
conditions, which should be tailored to the problem at hand -- in this
case, numerical simulation of black-hole spacetimes.  Regardless of
these choices, current numerical practices universally involve an
initial three-dimensional slice of spacetime that is evolved forward
in time.  Here, we review the history and current most common
choices of initial data, black-hole representations, formulation,
coordinate conditions, and some details of the numerics.  The impetus,
and most important outcome, of all these developments is the ability
to generate gravitational waveforms from black-hole binary sources
over many cycles.

\subsection{The Cauchy Problem}
\label{sec:Cauchy} 

The Cauchy problem concerns solution of the field equations given
initial data specified on an initial (typically spatial) hypersurface.
In practice, the Cauchy problem is more conveniently investigated in a
``3+1'' formulation, explicitly based on a foliation of the spacetime
into three-dimensional spatial slices parametrized by a time
coordinate.  A common 3+1 formulation inspired by ADM
\cite{Arnowitt:1962hi} divides up the components of the metric
according to their relationships with space and time, such that the
line element takes the form:
\begin{equation}
ds^2 = (-\alpha^2 +\beta^i\beta_i)dt^2 + 2\beta_i dt dx^i +\gamma_{ij}dx^idx^j,
\end{equation}
where $\alpha$ is called the lapse function, $\beta^i$ the shift
vector, and $\gamma_{ij}=g_{ij}$ is the spatial three-metric.  We write
the time coordinate $x^0 = t$ and the spatial coordinate indices $i,j
= 1,2,3$.  Note that contraction with $g_{\mu\nu}$ or its inverse
$g^{\mu\nu}$ is used to lower or raise indices of four-dimensional
tensors, respectively, while $\gamma_{ij}$ or its inverse
$\gamma^{ij}$ is used to lower or raise indices of three-dimensional
tensors.  The lapse and shift represent coordinate freedom in the
metric; we can choose these quantities arbitrarily.  However, since
the three-metric $\gamma_{ij}$ (and its first and second spatial
derivatives) determines the intrinsic curvature of the slice, it
carries the information about the gravitational field and thus is
constrained by the physics.
 
\begin{figure}
\includegraphics*[width=3.5in,angle=0]{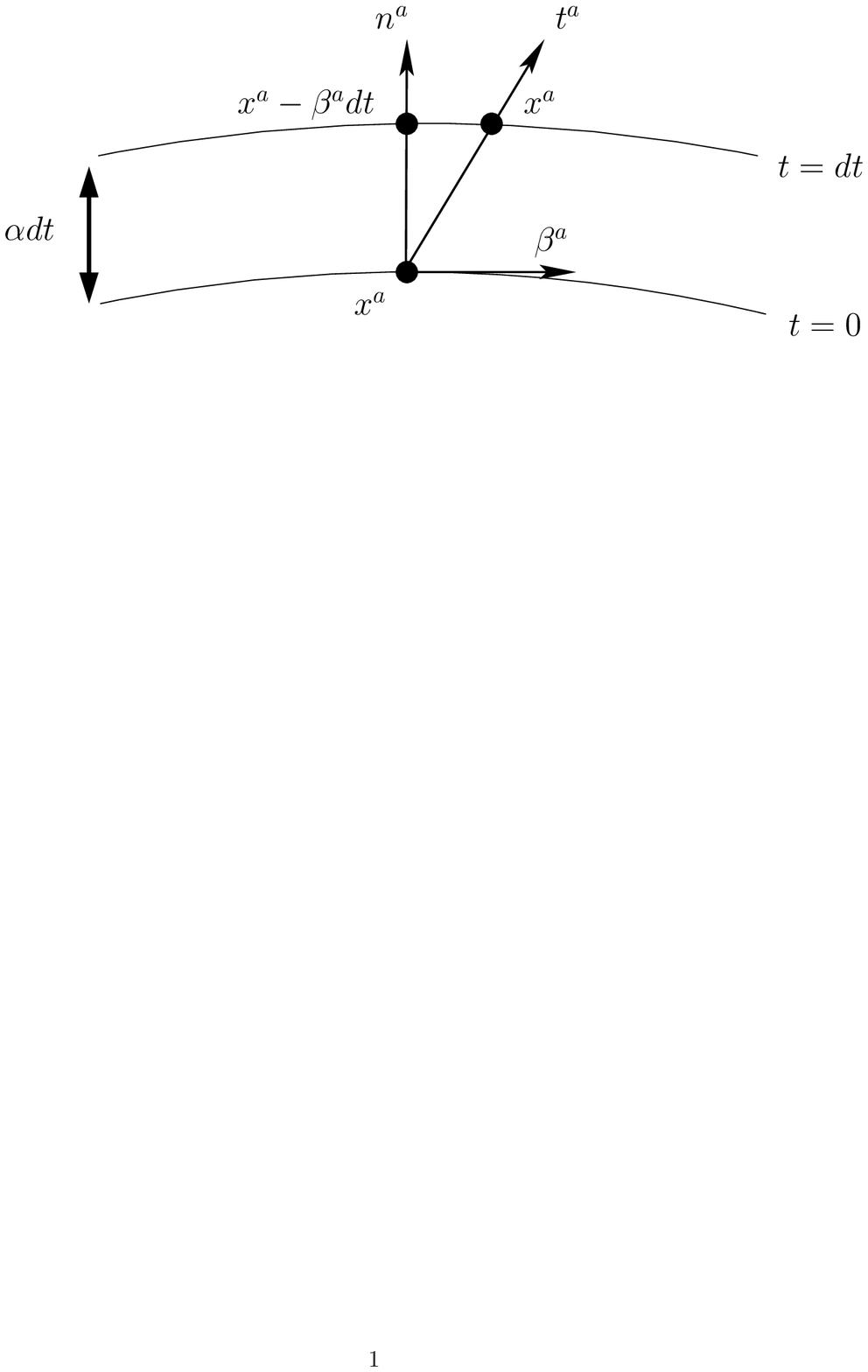}
\caption{The 3+1 split into space and time. Two spatial slices at
  $t=0$ and $t=dt$ are depicted.  $\alpha$ is the lapse, and $\alpha
  dt$ represents the proper time lapse between slices.  $\beta^a$ is
  the shift, and $\beta^a dt$ represents the amount by which the
  spatial coordinates shift between slices. $n^a$ is normal to the
  slice at $t=0$.  If a ray parallel to $n^a$ intersects the $t=0$
  slice at a point $x^a$, then it will intersect the $t=dt$ slice at
  $x^a-\beta^adt$. $t^a\equiv\alpha n^a+\beta^a$ is a coordinate time
  vector. If a ray parallel to $t^a$ intersects the $t=0$ slice at
  point $x^a$, then it will also intersect the $t=dt$ slice at
  $x^a$. }
\label{fig:smarr_3plus1}
\end{figure}

The meaning of the lapse and shift can be understood by considering
two successive spatial slices separated by an infinitesimal time
interval $dt$ (Fig.~\ref{fig:smarr_3plus1}).  An observer along a
vector normal to the first slice will measure an elapsed proper time
of $d\tau=\alpha dt$ in evolving to the second slice, and a change in
spatial coordinate of $dx^i=-\beta^i dt$.

Since Einstein's equations are second order in time, we must also
specify the initial time derivative of the three-metric.  Rather than
specifying this derivative directly, we define a new quantity:
\begin{equation}
\label{eq:Kij}
K_{ij}=-\frac{1}{2\alpha}\left(\partial_t\gamma_{ij}-D_i\beta_j-D_j\beta_i\right),
\end{equation}
where $\partial_t = \partial/\partial t$ is an ordinary partial
derivative and $D_i$ is the spatial covariant derivative.  Note that
the spacetime covariant derivative $\nabla_{\mu}$ is a partial
derivative with a correction such that it transforms as a vector and
satisfies $\nabla_{\lambda}g_{\mu\nu}=0$.  $D_i$ is the projection of
$\nabla_{\mu}$ onto the spatial slice and is equivalent to a
three-dimensional covariant derivative formed from $\gamma_{ij}$.  For the
case of Euclidean normal coordinates, $\alpha = 1$ and $\beta^i = 0$,
Eq.~(\ref{eq:Kij}) reduces to the simple expression $K_{ij} =
-(1/2)\partial_t\gamma_{ij}$.

If we define a unit vector $n_{\mu}$ normal to the spatial slice, we
can show that Eq.~(\ref{eq:Kij}) is equivalent to $K_{ij}=-D_i n_j$.
As suggested by this expression, $K_{ij}$ is a measure of the change
of the normal vector as it is transported along the slice.  In this
way $K_{ij}$ gives an extrinsic measure of the curvature of a
three-dimensional spatial slice with respect to its embedding in
four-dimensional spacetime.  It is therefore called the extrinsic
curvature.  Depending on the formulation, the extrinsic curvature
might or might not come into the evolution equations; however,
$K_{ij}$ is almost universally utilized when calculating initial data.

Of the ten component equations of Eq.~\eqref{Einsteinsequations}, six
determine the time-evolution of the metric, while four must be
satisfied on a spatial slice at any given time, and are thus
constraint equations.  With an appropriate choice of time coordinate,
and assuming vacuum spacetime, these four constraint equations are
equivalent to the condition $G_{0\nu}=0$.  The time-time component
$G_{00}=0$ is called the \emph{Hamiltonian constraint}, and the
time-space components $G_{0i}=0$ are called the \emph{momentum
  constraint}.  These take the form of conditions on the extrinsic
curvature:
\begin{eqnarray}
^{(3)}R+K^2-K_{ij}K^{ij}&=&0 
\label{eq:ham} \\
D_j(K^{ij}-\gamma^{ij}K)&=&0
\label{eq:mom}
\end{eqnarray}
where $^{(3)}R$ is the three-dimensional Ricci scalar associated with
the three-metric $\gamma_{ij}$, and $K \equiv {\rm trace}K_{ij} =
\gamma^{ij} K_{ij}$, sometimes called the \emph{mean curvature}.
These constraint equations must be solved in order to obtain an
initial spatial slice consistent with Einstein's equations.

\subsection{Representing Black Holes in Numerical Spacetimes} 
\label{sec:BHonGrid}

How does one represent an exotic object such as a black hole in a
numerical simulation?  In particular, how can one use finite
computational methods repeatedly to model an object which,
analytically, contains physical and/or coordinate singularities?
Fortunately, two successful strategies have emerged to meet this
challenge.

\begin{figure}
\includegraphics*[scale=0.50]{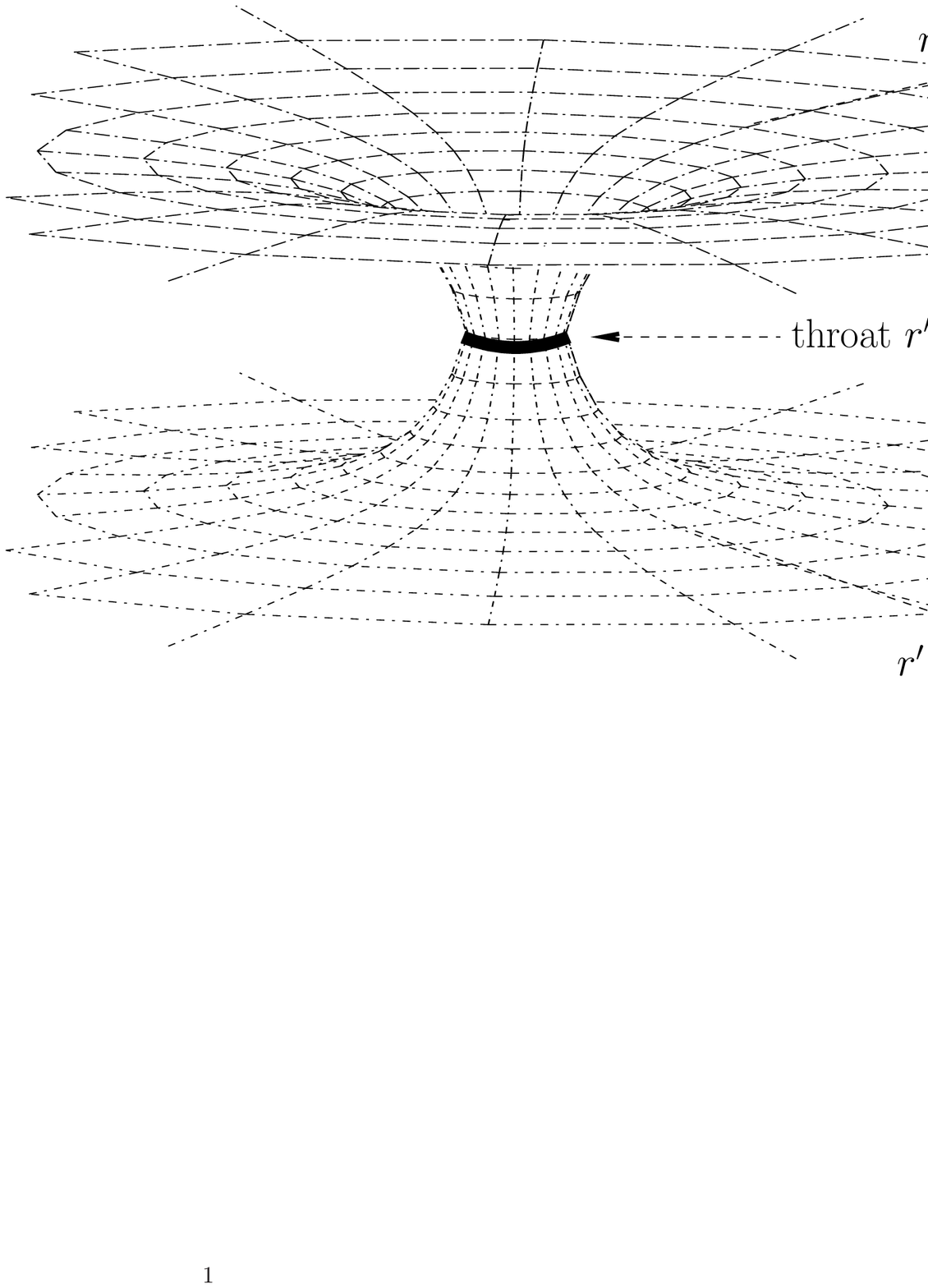}
\caption{In the wormhole representation of a black hole, the initial
  slice typically just touches the horizon. The upper ``sheet''
  represents the exterior space, while the lower sheet is a duplicate,
  joined to the upper sheet by a ``throat''.}
\label{fig:wormhole_flatspace}
\end{figure}

The unusual topology of black holes offers one way out.  As
\citet{Einstein:1935tc} originally showed, a black hole can be considered a
``bridge" or ``wormhole" connecting one Universe or ``worldsheet", to
a second worldsheet (see Fig.~\ref{fig:wormhole_flatspace}).
Exploiting this topology, a continuous spatial slice which avoids the
physical singularity contained within the event horizon of each black
hole can be constructed as follows.  Starting with a spatial slice of
Schwarzschild spacetime, remove the interior of the event horizon.
Identify the resulting spherical boundary with the spherical boundary
of an identical copy of this space.  The two-dimensional analog would
be to take two sheets of paper, cut out a disk from each, and then
glue the resulting circular edges together.  Each sheet, or copy of
Schwarzschild spacetime in this example, is called a worldsheet.  The
identified spherical boundaries connecting the worldsheets form what
is referred to as the ``throat" of the wormhole.

To complete this construction, we require an appropriate coordinate
system to continuously cover both worldsheets.  Brill and Lindquist
discovered coordinates that will prove convenient, in which the
three-metric of a particular spatial slice of Schwarzschild spacetime is
given by \cite{Brill:1963yv}
\begin{equation}
\label{eq:BLS}
\gamma_{ij}=\left(1+\frac{m}{2r}\right)^4\delta_{ij}, 
\end{equation}
where $r$ is a radial coordinate.  In these coordinates, the event
horizon is at $r=m/2$.  We can consider each of the worldsheets
described above as being separately labeled by such coordinates.
Designate one worldsheet as $A$ and the other as $B$, and call their
radial coordinates $r_A$ and $r_B$ respectively. Since the interior of
the event horizon has been removed from each worldsheet, assume that
$r_A\geq m/2$ and $r_B\geq m/2$. The metric on each worldsheet has the form of
Eq.~(\ref{eq:BLS}). As noted by Brill and Lindquist, this form of the
metric is unchanged by the transformation $r'=m^2/4r$, and $r'=r$
when $r=m/2$.  So if we define a new coordinate $r'$ by
\begin{equation}
        r_A = r' \mbox{ for } r' \geq \frac{m}{2} , \;\; r_B = \frac{m^2}{2r'} \mbox{ for } r' \leq \frac{m}{2},
\end{equation}
mapping spatial infinity on worldsheet $B$ to $r'=0$, we obtain a
single continuous coordinate system that applies to worldsheet $A$ for
$r'\geq m/2$ and worldsheet $B$ for $r'\leq m/2$.

This avoidance of the physical singularity comes at the expense of a
coordinate singularity in the metric at $r'=0$, called a ``puncture''.
However, it turns out this coordinate singularity can be confined to a
single scalar variable, as suggested in Eq.~(\ref{eq:BLS}).  With a
suitable change of variables and other means, numerical simulations
haven proven capable of handling this irregular scalar field.  Thus
the ``puncture" method is one way to represent a black hole that is
amenable to computation.

Another strategy is excision, first proposed by Unruh
\cite{Thornburg93}.  Given that no physical information can escape an
event horizon to influence the exterior, the interior of a black hole
can in principle be excised from the computational domain\footnote{As
  unphysical coordinate ``gauge modes'' may couple to physical modes,
  we also assume no superluminal coordinate effects are present.}.
This relies on the fact that all physical information propagates
inwards from the event horizon towards the physical singularity,
i.e. lightcones tilt inwards, and nothing physical propagates outward
from the horizon.  Extrapolation can be used for the boundary
condition of the excised region, and any nonphysical numerical error
that escapes the horizon should be negligible.  This approach is not
constrained to the particular coordinates required by the puncture
method, but it can be more difficult due to the need for precise
positioning of the excision boundary and accurate extrapolation.
Excision was used successfully for orbiting binary simulations by
\citet{Pretorius:2005gq} and continues to be used by the
Caltech-Cornell dual-coordinate spectral code
\cite{Scheel:2006gg,Scheel:2008rj}.

Either of these methods can be useful in representing black holes on
the initial data slice.  Surprisingly, we will also see that either of
these representations can be made robust enough to persist as the
black holes evolve.

\subsection{Initial Data}
\label{sec:comp:initialdata}

The starting point for successful simulations of black-hole mergers is
finding initial data for astrophysically realistic inspiralling black
holes.  If we were simulating the orbits of stars in Newtonian
gravity, this would be a simple procedure.  For example, we could
simply specify the masses and spins of the stars, along with their
positions and velocities on orbits derived from the dynamics of point
particles, and then evolve the system numerically to allow it to
``relax'' into orbits appropriate for bodies of finite size.  In
general relativity, however, the initial data must satisfy the
constraint equations~(\ref{eq:ham}) -~(\ref{eq:mom}), which are cast
in terms of the $3$-metric $\gamma_{ij}$ and the extrinsic curvature
$K_{ij}$.  Since there is no obvious, or natural, connection between
these field variables and the astrophysical properties of inspiralling
black holes, obtaining suitable initial data is a major challenge.

Building on the earlier work of \citet{Lichnerowicz44}, York developed
a general procedure for solving the constraint equations to produce
initial data for the Cauchy problem in the 1970s (see \citet{York79}
for a review).  This approach generally requires solving a coupled
elliptic system of four nonlinear field equations. We can break this
problem down into more manageable pieces with some simplifying
assumptions. While these choices do come at the cost of some loss of
generality and astrophysical realism in the initial data for two black
holes, it has been seen that for sufficiently long evolutions the
final orbits and waveform signatures from the black-hole evolution are
largely insensitive to this level of detail in the initial data, at
least in the case of equal-mass nonspinning black holes.

The first simplification is to choose traceless extrinsic curvature,
$K=0$.  With this, the Hamiltonian constraint is decoupled from the
momentum constraints, and can be solved separately.  To find solutions
corresponding to multiple black holes, we generally further assume
that the initial slice is conformally flat. That is, the three-metric is
the product of a scalar conformal factor with a flat metric,
\begin{equation}
\gamma_{ij}=\psi^4\delta_{ij}.
\end{equation}
With this, the problem reduces to solving a (typically nonlinear)
equation for the scalar field $\psi$.

\citet{Brill:1963yv} found a simple solution, representing $N$ black
holes momentarily at rest, which gives
\begin{equation}
\psi=1+\sum_i^N \frac{m_i}{2r_i}
\end{equation}
where $m_i$ is the mass associated with the $i$th black hole and $r_i$
represents its coordinate center.  Each $r_i$ corresponds to the
location of a puncture, as described in the last section.  This is
a valuable solution, but not generally useful, because these black
holes lack momentum and spin.

An explicit solution by Bowen for the momentum constraint was the last
crucial step in defining a procedure for calculating initial data for
multiple black holes with specified linear and angular momenta
\cite{Bowen:1980yu}.  The Bowen-York prescription employed a
two-sheeted topology found by Misner for the black-hole interiors
\cite{Misner63}. Later Brandt and Br\"{u}gmann generalized the
procedure for the Brill-Lindquist topology \cite{Brandt:1997tf}.  The
Brandt-Br\"{u}gmann puncture data for arbitrary momenta and spin is
now widely used because of its ease of implementation.

More recently York developed another modeling ansatz, known as the
Conformal-Thin-Sandwich (CTS) approach
\cite{York:1998hy,Pfeiffer:2005jf}, which has certain additional
advantages.  For example, it turns out the conventional
Brandt-Br\"{u}gmann puncture cannot yield a spin parameter $\ahat$
greater than $\sim 0.93$ \cite{Dain:2002ee}, while CTS data can go
higher \cite{Lovelace:2008tw}.  As discussed above, the spatial metric is chosen
to be conformally flat, then instead of providing an ansatz for the
extrinsic curvature $K_{ij}$, the initial time-derivative of the
conformal metric is specified (generally to vanish). In addition, a
condition can be imposed on the slicing (see below).  The result is a
coupled system of elliptic equations which is solved to enforce the
constraints and optionally the choice of slicing condition.  Boundary
conditions may additionally be supplied to enforce rotational motion.

Either approach provides an ansatz for constructing three-dimensional
binary black-hole initial {\em field} data for a specified choice of
particle-like parameters, masses, positions, momenta, and spins.
Inevitably there are differences from the nearly quiescent evolving
systems that we seek to represent.  Generally there is some level of
spurious radiation generated from a period of initial transient
dynamics through which the system relaxes to become quiescent on
sub-orbital time scales.  In particular, extraneous radiation content
is an unavoidable consequence of conformal flatness, which
post-Newtonian analysis has shown must deviate from the physically
relevant inspiralling binary solution \cite{Damour:2000we}.  This
spurious radiation is seen in plots of the gravitational waveforms
produced by mergers; see Fig.~\ref{fig:universal_wf} and other
waveform plots in Sec.~\ref{sec:GWs:universal}.  Often the simulations
will also undergo a period of initial gauge-evolution which, though
physically inconsequential, may affect the quality of the simulation
numerically.  For puncture initial data \citet{Hannam:2009ib} analyzed
some of these gauge dynamics, and developed a promising approach
(``trumpet" data) to mitigate it (see Sec.~\ref{sec:MovePunc} for
related gauge issues).  However, for simulations lasting for several
orbits these modest transient effects are generally negligible.

Most black-hole binary simulation studies are designed to represent
the astrophysical population of systems, which have circularized
before the gravitational radiation becomes observationally
significant.  These simulations begin with circularly inspiralling
initial data configurations.  Even before reliable numerical
relativity simulations were possible, considerable attention had been
given to prescribing initial data for these near-circular
configurations (see \citet{Cook:2000LR} for a review).  Within the CTS
approach, it is particularly natural to impose initially circular
motion upon the system, resulting in an initial data prescription for
quasicircular orbits \cite{Cook:2001wi,Cook:2004kt,Caudill:2006hw}.
For either the CTS or Brandt-Br\"ugmann data, quasicircular parameters
may be chosen by constraints on either an effective gravitational
potential or total energy of the system
\cite{Cook:1994va,Gourgoulhon:2001ec,Caudill:2006hw}.  For more
realistic inspiralling trajectories, the PN approximation may be used
\cite{Husa:2007rh}, or, for higher accuracy, an iterative procedure
involving short numerical evolutions \cite{Pfeiffer:2007yz}.

\subsection{Numerically Friendly Formulations of the Evolution Equations}
\label{sec:comp:evolution}

The Einstein evolution equations, which determine the time-development
of the initial data, form a set of at least six coupled, nonlinear
propagation equations.  The exact formulation of these equations
depends on the choice of evolved variables as well as how constraint
identities are incorporated.  Although many formulations are possible,
not all are equivalent from a numerical point of view.  The choice of
formulation can affect how numerical errors grow in time, and whether
these errors can converge to zero as the resolution of the
computational grid is increased.

\subsubsection{Hyperbolicity and Well-Posedness}

An ideal formulation is well-posed, meaning that: (i) for a given set
of initial data, a unique solution exists, and (ii) solutions depend
continuously on perturbations of the initial data.  To determine
whether a given formulation has the desirable property of
well-posedness, we first consider it as a matrix equation:
\begin{equation}
\label{eq:evolution}
\partial_t \mathbf{u} = \mathbf{A u} + f(\mathbf{u},\partial_i \mathbf{u})
\end{equation}
where $\mathbf{u}$ is a vector, the components of which represent the
evolved variables such as components of the metric; the linear
operator $\mathbf{A}$ is the principal part, containing the highest-order
spatial derivatives; and $ f(\mathbf{u},\mathbf{u}_{,i})$
contains additional terms which may be nonlinear. Certain conditions
on the eigenvalues and eigenvectors of the operator $\mathbf{A}$ in
Eq.~\eqref{eq:evolution} are used to characterize the equations,
typically as either ``strongly hyperbolic'' or ``weakly hyperbolic''
\cite{Reula:2004xd,Nagy:2004td,Calabrese:2002ei}.

If the eigenvalues are real and the eigenvectors form a complete set
(so any solution can be written as a linear combination of them), the
equations are strongly hyperbolic.  Then, assuming adequate boundary
conditions, the equations are well-posed and all solutions are bounded
by a function that grows (at most) exponentially at a rate independent
of the initial data.  Strong hyperbolicity is a key ingredient for
successful black-hole merger simulations.

On the other hand, if the eigenvalues are real but the eigenvectors
are not complete, the equations are weakly hyperbolic. In this case
the equations are not well-posed, and they permit solutions which grow
at rates that depend on the initial data.  Weak hyperbolicity implies
that small numerical errors in the initial data may grow at a rate
which depends on the resolution.  It then becomes difficult to show
that the numerical solution converges to a well-defined continuum
solution, and at high resolutions the simulation may become unstable
\cite{Calabrese:2002ej}.

\subsubsection{Harmonic Formulations}
\label{sec:harm}

The quest for workable formulations of Einstein's equations has
proceeded along two parallel lines of development.  One originated
with consideration of ``harmonic coordinates", so called because the
coordinates satisfy the wave equation $\Box_g x^{\mu}=0$, where
$\Box_g$ is the curved-space wave operator.  In these coordinates,
Einstein's equations can be written such that the principal part
resembles a wave equation in terms of the metric as in
Eq.~\eqref{eq:harmonic}.  In this form, Einstein's equations are
manifestly hyperbolic \cite{Choquet62}.  However, the harmonic
coordinate condition is too restrictive for numerical purposes, so
generalized harmonic coordinates were eventually developed by
introducing a source term into the coordinate condition, i.e. $\Box_g
x^{\mu}=H^{\mu}$ \cite{Friedrich85,Garfinkle:2001ni}, a suitable
choice for which preserves strong hyperbolicity.  The subsequent
introduction of {\it constraint-damping} terms, which tend to drive
the constraints towards zero, further ensured stability
\cite{Gundlach:2005eh}.  This formulation is manifestly second-order
in both time and space, and has been implemented numerically as such
\cite{Pretorius:2006tp}, but for more efficient numerical integration
a first-order-in-time formulation was also developed
\cite{Lindblom:2005qh}, and is currently being used by some groups.

\subsubsection{ADM-based Formulations}

A second line of development originated with the invention of the ADM
formulation of Einstein's equations \cite{Arnowitt:1962hi}.  This
formulation was refined by \citet{York79}, who suggested evolving the
specific quantity of the extrinsic curvature (Eq.~\eqref{eq:Kij}).
The associated evolution equations specify the first-order
time-derivatives for the three-metric $\gamma_{ij}$ and the extrinsic
curvature $K_{ij}$, as well as for the gauge variables $\alpha$ and
$\beta^i$.  However, the ADM formulation is weakly hyperbolic (see,
e.g., Chapter 5 of \citet{Alcubierre08}), and attempts at stable
numerical evolutions with it were not successful.

There are essentially two ways to modify the ADM equations that may
affect their hyperbolicity (while keeping them first-order in time and
maintaining their 3+1 character).  One way is to add the constraints,
which may change the principal part of the equations without changing
the associated physics.  The other is to define new independent
variables.  Simple addition of constraints failed to give a strongly
hyperbolic formulation however.  Eventually it was found that strongly
hyperbolic versions could be constructed from the ADM equations by the
promotion of certain derived quantities to the status of independently
evolved variables \cite{Bona:1994dr,Kidder:2001tz,Nagy:2004td}.

Although strong hyperbolicity is important for a stable formulation,
it is not the only property critical for accurate evolution.  Of the
various strongly hyperbolic 3+1 formulations, one eventually emerged
as more successful than the rest.  Its development began with the
observation that the numerical accuracy of $\gamma$, the determinant of the
three-metric, and $K$, the trace of the extrinsic curvature,
could best be preserved by evolving these two quantities independently
\cite{Nakamura:1987zz}.  Subsequently, to remove the redundancy in evolving
the full three-metric and extrinsic curvature, the conformal three-metric
\begin{equation}
\tilde{\gamma}_{ij}\equiv \gamma^{-1/3}\gamma_{ij}
\end{equation}
and conformal traceless extrinsic curvature 
\begin{equation}
\tilde{A}_{ij}\equiv\gamma^{-1/3}(K_{ij}-\frac{1}{3}K\gamma_{ij})
\end{equation}
were substituted \cite{Shibata:1995we}.  To eliminate certain terms
with second derivatives (which contribute to the principal part of
this system), a new independently evolved variable
\begin{equation}
\tilde{\Gamma}^i\equiv-\partial_j\tilde{\gamma}^{ij}
\end{equation}
was also introduced \cite{Shibata:1995we}.  Further improvements were
made shortly afterwards, such as the addition of constraints to
eliminate more second-derivative terms \cite{Baumgarte:1998te}.  The
resulting formulation, now commonly known as BSSN, proved to be
robustly stable and accurate for numerical purposes.  Note that the
above choices which comprise its form were empirically motivated, by
identifying and eliminating terms which tended to compromise numerical
accuracy.  Some analytic justification for its success was later found
by \citet{Alcubierre:1999rt}, who showed that the BSSN formulation
avoids exponentially growing modes and most zero-speed modes, which
accumulate numerical error.  Finally \citet{Sarbach:2002bt} and
\citet{Gundlach:2006tw} proved that BSSN is also strongly hyperbolic,
given an appropriate choice of gauge.

A few noteworthy refinements of the BSSN formulation followed.
\cite{Yo:2002bm} and \citet{Duez:2004uh} suggested adding specific
terms proportional to the ``Gamma constraint" ($\tilde{\Gamma}^i$
minus its definition) and the Hamiltonian constraint into the evolution
equations.  These terms have the effect of damping the constraints and
thereby improving stability.

Later, for evolution of punctures specifically,
\citet{Campanelli:2005dd} evolved the conformal factor in the form
$\chi=\gamma^{-1/3}$, which vanishes at the puncture (an improvement
in accuracy over the previous standard of $\phi=\ln(\gamma)/12$, which
is singular at the puncture).  This change of variables, however,
introduces the potentially singular factor $1/\chi^2$ into the
evolution equations, necessitating an arbitrary restriction on the
minimum value of $\chi$.  An alternative variable
$\chi'=\gamma^{-1/6}$ was subsequently suggested because resulting
appearances of $1/\chi'^2$ mostly cancel with factors of
$\gamma^{1/3}$ \cite{vanMeter:2006g2n,Tichy:2007hk,Marronetti:2007wz}.

\subsection{Gauge Conditions}
\label{sec:comp:gauge}

The choice of gauge or coordinate conditions, like the choice of
formulation, has important consequences on the numerics, especially
the stability of the simulation.  Important considerations include how
to deal with the extreme conditions of black holes such as the
physical singularities, the possible coordinate singularities, the
strong-field gradients, and the dynamical, surrounding spacetime. The
coordinates must accommodate these features in a way that is
numerically tractable.

\subsubsection{Choosing the Slicing and Shift}

An early consideration was how to avoid the physical singularity of a
black hole through an appropriate slicing condition.  One such choice,
\emph{maximal slicing}, required solution of a numerically expensive
elliptic equation \cite{Lichnerowicz44,Estabrook:1973ue,Smarr:1977uf}.
Later, an interest in constructing a hyperbolic evolution system led
to a generalization of harmonic slicing known as the Bona-Mass\'{o}
family \cite{Bona:1994dr}.  One particular member of this family,
called 1+log slicing, was found to also be singularity-avoiding, like
maximal slicing, but at less numerical cost
\cite{Bernstein93a,Anninos:1995am}.

Meanwhile, it was recognized that a shift vector was required to
counter the large field gradients, or ``slice-stretching'', incurred
in the presence of a black hole \cite{Alcubierre:2000yz}.  A
hyperbolic shift condition called the ``Gamma driver'' was introduced
that fulfilled this requirement
\cite{Alcubierre:2000yz,Alcubierre:2002kk,Alcubierre:2001vm}:
\begin{eqnarray}
\partial_t B^i &=& \partial_t\tilde{\Gamma}^i-\etaB B^i, \\
\partial_t\beta^i &=& F B^i,
\end{eqnarray}
where $\tilde{\Gamma}^i=-\partial_j\tilde{\gamma}^{ij}$ depends on a
conformal three-metric $\tilde{\gamma}_{ij}$ of the evolving spatial
slice, $B^i$ is an auxiliary variable, $\beta^i$ is the shift, and $F$
is some scalar field.  $\etaB$ is a damping parameter that fine-tunes
the growth of the shift, which affects the coordinate size of the
black-hole horizons, which in turn has bearing on the required
numerical resolution \cite{Gonzalez:2008bi,Bruegmann:2006at}.  This
shift condition also has the desirable tendency to drive the
coordinates to quiescence in synchrony with the physics, for example
after a binary merges into a stationary black hole.  Clearly if
$\tilde{\Gamma}^i$ vanishes, $B^i$ is damped to zero and $\beta^i$
approaches a (typically small) constant.

\subsubsection{Moving Punctures}
\label{sec:MovePunc}

Initially these conditions were intended for use with punctures that
remained at fixed coordinate positions (i.e. comoving coordinates), by
choosing $F$ to vanish at the punctures.  However, this led to large
gradients between the merging black holes since there the metric had
to vanish as the physical distance contracted. Pathologies also arose
from the twisting of the coordinates as the black holes orbited each
other.

These issues, which resulted in large numerical errors and
instabilities, were eventually resolved with the breakthrough
discovery that slight modifications of the 1+log and Gamma-driver
conditions allowed arbitrary motion of the punctures with robust
stability \cite{Campanelli:2005dd,Baker:2005vv}.  In particular, in
the shift condition, a critical alteration was ``unpinning" the
puncture by no longer requiring the factor $F$ to vanish at the
puncture but rather to remain constant.  These modifications were
subsequently refined to eliminate zero-speed modes, thus preventing
the possibility of error build-up \cite{vanMeter:2006vi}.  This was
accomplished in part by the addition of advection terms:
\begin{eqnarray}
\partial_t B^i&=&\partial_t\tilde{\Gamma}^i-\beta^j\partial_j\tilde{\Gamma}^i-\etaB B^i+\beta^j\partial_jB^i, \label{eq:movingshift_B}\\
\partial_t\beta^i&=&\frac{3}{4}B^i+\beta^j\partial_j\beta^i.\label{eq:movingshift_beta}
\end{eqnarray}
It is noteworthy that these particular gauge choices, coupled with the
BSSN equations, also result in strong hyperbolicity
\cite{Gundlach:2006tw,Sarbach:2002bt}.  A further refinement resulted
from the observation that, assuming that
$\beta^i=B^i=\tilde{\Gamma}^i=0$ initially,
Eqs.~\eqref{eq:movingshift_B} and \eqref{eq:movingshift_beta} can be
integrated to yield $B^i=\tilde{\Gamma}^i-\frac{4}{3} \etaB \beta^i$;
this substitution allows the removal of the equation for $B^i$,
leaving only a single shift equation \cite{vanMeter:2006vi}:
\begin{equation}
\partial_t\beta^i=\frac{3}{4}\tilde{\Gamma}^i+\beta^j\partial_j\beta^i-\etaB\beta^i.
\end{equation}
Use of this or similar gauge conditions became known as the ``moving
puncture" method, and proved to be very successful as it became
increasingly widespread among the numerical-relativity community.
Analysis has verified that moving punctures are valid black-hole
solutions, although the initial character of the punctures changes
significantly during the course of evolution.  The second worldsheet
shown in Fig.~\ref{fig:wormhole_flatspace}, for example, invariably
becomes disconnected in numerical simulations, at a point within the
horizon near the throat, due to the action of the shift vector
effectively shifting computational grid points onto the first
worldsheet.  Meanwhile the spatial coordinates evolve such that the
$r^{-1}$ singularities in the conformal factor $\psi$ become
$r^{-1/2}$ singularities.  For a single nonspinning black hole, the
numerical result rapidly asymptotes to an exact form of the
Schwarzschild solution called a ``trumpet", recently investigated by
\citet{Hannam:2006vv,Hannam:2006xw,Hannam:2008sg,Hannam:2009ib,Brown:2007tb}
and \citet{Baumgarte:2007ht}, which turns out to be a type of solution
first considered by \citet{Estabrook:1973ue}.

\subsubsection{Generalized Harmonic Coordinates}

Development of generalized harmonic coordinates initially proceeded
independently of the above 3+1-formulated conditions.  As mentioned,
in harmonic coordinates the D'Alembertian of each coordinate vanishes.
In generalized harmonic coordinates, the wave equation for each
coordinate is allowed a source term, i.e.
\begin{equation}
\Box x^{\mu}=H^{\mu}.
\end{equation}
These ``gauge driving" source terms $H^{\mu}$ can be either
algebraically specified or evolved such that hyperbolicity is
preserved
\cite{Friedrich85,Garfinkle:2001ni,Lindblom:2005qh,Pretorius:2006tp}.

The first successful numerical orbit of black holes involved a source
term for the time coordinate that effectively kept the lapse close to
its Minkowski value of unity, while the spatial coordinates remained
harmonic \cite{Pretorius:2006tp}.  This was accomplished by evolving
the source term itself, according to
\begin{equation}
\Box H_0 = \left[-\xi_1(\alpha-1)+\xi_2(\partial_t-\beta^i\partial_i)H_0\right]\alpha^{-1}
\end{equation}
where $\xi_1$ and $\xi_2$ are constants.  More recently, to dampen
extraneous gauge dynamics during the inspiral and merger of generic
binaries, \citet{Szilagyi:2009qz} found the following gauge driver to
be successful:
\begin{eqnarray}
H_0 &=& \mu_0\left[\log\left(\frac{\sqrt{g}}{\alpha}\right)\right]^3 \\
H_i &=& -\mu_0\left[\log\left(\frac{\sqrt{g}}{\alpha}\right)\right]^2\frac{\beta_i}{\alpha}
\end{eqnarray}
where $\mu_0$ is a specified function of time that starts at zero and
eventually increases monotonically to unity.

\subsubsection{Other Coordinate Techniques}

Additional noteworthy advances in coordinate conditions were developed
for the generalized harmonic formulation to facilitate spectral
methods (see Sec.~\ref{sec:comp:computation} below) but are in
principle generalizable to other frameworks.  One of these is the use
of multiple coordinate patches, where the coordinates are chosen
according to the local physical geometry for optimal accuracy.  In the
generalized harmonic formulation, characteristic speeds are readily
available for use in constructing physical boundary conditions between
patches \cite{Lindblom:2005qh,Scheel:2006gg,Pazos:2009vb}.

The other advance is that of ``dual coordinates".  Because moving
punctures involve irregular fields, they are not easily made
compatible with spectral methods, which are sensitive to
irregularities.  An alternative, that of a moving, excised region, can
leave a large trail of interpolation error in its wake.  The remaining
option is to use comoving coordinates \cite{Bruegmann:2003aw}, but
these can result in instabilities due to the steep field gradients and
other pathologies mentioned in Sec.~\ref{sec:comp:gauge}. Dual
coordinates were invented to exploit the advantages of both comoving
and non-comoving coordinates while avoiding their disadvantages.  This
is accomplished by computing all tensor field components in the
non-comoving coordinate system (thus avoiding steep gradients), yet
evolving them as functions of the comoving coordinates (thus allowing
stationary excision boundaries) \cite{Scheel:2006gg}.

\subsection{Numerical Approximation Methods}
\label{sec:comp:computation}

The initial data, formulation, and gauge are all, in principle,
analytic choices applicable to an infinite, continuous manifold.  One
must finally make choices pertaining explicitly to the finite,
discrete mechanics of the computations that will numerically
approximate the above analytic specifications.  An immediate
consideration is the fact that the simulated domain must have finite
extent.  Various conformal compactification schemes which map spatial
or null infinity to a finite boundary have been tested
\cite{Rinne:2009qx,Pretorius:2006tp,vanMeter:2006mv}, but currently it
is more common to impose artificial boundaries at finite spatial
coordinates and apply some form of either radiative boundary condition
\cite{Alcubierre:2002kk} or constraint-preserving boundary condition
\cite{Lindblom:2005qh}.

To mitigate the effect of inward-propagating errors from the outer
boundaries, some form of mesh refinement is typically employed to push
the outer boundaries as far away as possible from regions of interest
(e.g. wave sources and extraction regions).  With a limited number of
grid points available, their density is judiciously chosen to be
highest near the strong field gradients of the black holes, moderate
throughout regions where wave propagation is studied, and coarser
beyond that.  If the simulated black holes are very dynamic then some
algorithm for automatically adapting the mesh refinement is necessary.
Meanwhile the interfaces between refinement regions require
interpolation.

On this grid, spatial derivatives are computed in one of two ways.
They may be computed with finite differencing stencils across
uniformly spaced grid points, derived from Taylor expansions,
currently up to eighth-order accurate
\cite{Lousto:2007rj,Pollney:2009yz}.  Alternatively, the spectral
approach may be used, in which coefficients of an expansion in basis
functions are computed to some order, on a number of collocation
points comparable to the number of basis modes, from which the
derivatives are then obtained analytically \cite{Boyle:2006ne}.  In
the former case, dissipative terms are often added to the evolution
equations to reduce noise \cite{Kreiss73}.  In the latter case,
spectral methods often include a smoothing step for a similar end.

Last, time must be advanced in discrete steps.  Various explicit
time-integration algorithms have been tested, for example the iterated
Crank-Nicholson scheme, but the fourth-order Runge-Kutta algorithm has
become most widely used, due to its superior accuracy and efficiency.
In some codes the timestep size is made to vary with spatial
resolution, for even greater efficiency, albeit at the expense of the
complication of time-interpolation.  Implicit timestepping schemes are
also being investigated as a means to greatly increase the timestep
size \cite{Lau:2008fb}.

\subsection{Extracting the Physics}
\label{sec:extract}

One of the most important end-results of a simulation of merging black
holes is a computation of the emitted gravitational radiation.  For
this purpose, it is useful to calculate the ``Weyl tensor",
$C_{abcd}$.  The Weyl tensor, like the Einstein tensor, is constructed
from derivatives of the metric.  Unlike the Einstein tensor, the Weyl
tensor has degrees of freedom that do not necessarily depend on a
massive source; it can be nonzero while the Einstein tensor vanishes.

Certain components of the Weyl tensor form a complex quantity called
$\Psi_4$, one of five ``Weyl scalars" used to classify spacetimes
\cite{Newman:1961qr}.  In a special, ``transverse-traceless",
spherical coordinate system, $\Psi_4$ can be expressed as follows:
\begin{equation}
\Psi_4=C_{\rhat \thetahat \rhat \thetahat}-C_{\rhat \phihat \rhat \phihat}-2iC_{\rhat \thetahat \rhat \phihat},
\end{equation}
where the subscripts $\{\rhat, \thetahat, \phihat\}$ denote
orthonormal tetrad components.  In a spacetime with gravitational
radiation this quantity typically falls off as $\sim 1/r$
\cite{Newman:1961qr}, and can be associated with outgoing radiation at
spatial infinity ($r\to\infty$) in asymptotically flat spacetimes
\cite{Szekeres65}.  In terms of the strain introduced in the last
section, this can be written
\begin{equation}
\lim_{r\to\infty}(r\Psi_4)=\lim_{r\to\infty}  \left[-r(\ddot{h}_+-i\ddot{h}_{\times})\right] ,
\end{equation}
where, in terms of a metric perturbation
$h_{\mu\nu}=g_{\mu\nu}-\eta_{\mu\nu}$,
\begin{eqnarray}
h_+ & = & \frac{1}{2}(h_{\thetahat \thetahat}-h_{\phihat \phihat}), \\
h_{\times} & = & h_{\thetahat \phihat}.
\end{eqnarray}
What makes $\Psi_4$ a particularly useful measure of the radiation is
that to first order in $h_{\mu\nu}$ it is coordinate-invariant.

Although the linearized radiation interpretation for $\Psi_4$ is only
strictly valid in the limit as $r\to\infty$, in numerical simulations
it is often extracted on a sphere of large but finite radius.  This
will approximate the expected radiative behavior if $|h_{\mu\nu}|<<1$;
if spheres of multiple radii are used then it is also possible to
extrapolate the results to infinity.  Typically $\Psi_4$ is computed
on the computational grid points from the metric variables and then
interpolated onto the spherical extraction surface.  Recently,
Cauchy-characteristic extraction methods have been applied to allow
direct evaluation of radiation at future null infinity
\cite{Reisswig:2009us}.

The product of $\Psi_4$ with a spherical harmonic function is then
integrated over the sphere, as it is useful to extract specific modes
of the field.  The structure of the gravitational field is such that
it is convenient to expand in spin-weight -2 spherical harmonics
$_{-2}Y_{\ell m}(\theta,\phi)$ \cite{Newman:1966ub,Teukolsky:1972my}
(just as, analogously, it proves convenient to use the spin-weighted
-1 harmonics in expanding the electromagnetic field).  Thus:
\begin{equation}
\Psi_4=\sum_{\ell=2}^{\infty}\sum_{m=-\ell}^{m=\ell}\Psi_{4,\ell m} \,{}_{-2}Y_{\ell m}
\end{equation}
Unlike electrodynamics, there is no dipole moment of the gravitational
field in conventional general relativity, so the dominant contribution
is the $\ell=2$ quadrupole moment.

Other information that is important in analyzing spacetimes relates to
the properties of the black holes themselves. This is commonly taken
from apparent horizons, surfaces from which all light rays must go
inward.  It is useful to compute this during a simulation, for example
by finding surfaces on which the expansion of light rays are minimized
\cite{Thornburg:2003sf}.  When a black hole is sufficiently isolated,
it is possible to define a mass and spin magnitude from data evaluated
on the horizon
\cite{Dreyer:2002mx,Campanelli:2006fg,Campanelli:2008nk}, as well as
linear momentum \cite{Krishnan:2007pu}.  These quantities can be used
to characterize the physical properties of the black holes, and to
label the spacetimes when comparing with PN theory.

\section{Black-Hole Merger Dynamics and Waveforms}
\label{sec:GWs} 

The final merger of a black-hole binary takes place in the dynamical,
strong-field regime of general relativity.  For many years, numerical
relativists wondered what they would find when they probed this arena
using black-hole merger simulations.  What would the merger waveforms
look like?  How strongly might artifacts in the initial data affect
the mergers?  How might the effects of unequal masses and spins change
the picture obtained by studying the simplest case of nonspinning,
equal-mass mergers?  Would they uncover any unexpected phenomena? 
Recent breakthroughs in black-hole merger simulations are providing
important tools for addressing such questions.

In this section we explore the dynamics of black-hole mergers and the
resulting gravitational waveforms.  We start with a look at the
Lazarus approach, which provides hybrid waveforms by combining
analytic methods with brief numerical simulations.  We then begin
a more comprehensive discussion of black-hole merger physics, starting
with mergers of nonspinning, equal-mass black holes. Next we consider
mergers of unequal-mass holes, followed by mergers with spin.
Throughout this discussion we aim to provide a historical context
while highlighting the key physical results.

\subsection{First Glimpses of the Merger: The Lazarus Approach}
\label{sec:GWs:laz}

In the late 1990s, numerical relativity had advanced sufficiently to
allow brief simulations of two black holes in three dimensions. By ``brief'' we
mean here durations of $\sim 10M$, which is a small fraction of the
estimated binary orbital period of $\gtrsim 100M$ near the ISCO. While
most of the community focused on technical developments aimed at
extending the duration of these simulations, a small band of
collaborators took a different approach \cite{Baker:2000zh}.

Their novel idea was to use full numerical-relativity simulations to
calculate the strong-field approach to merger during $\sim 10M$ of
simulation time, and then to calculate the remaining evolution using
perturbative techniques \cite{Baker:2001sf}.  They began with
traditional puncture black holes (which remain fixed in the grid) on
quasicircular orbits near the ISCO, and evolved them towards
merger. Then, just before the simulation became unstable, they stopped
the calculation and extracted the physical data for the merging black
holes and the emerging gravitational waves.  This data was then
interpreted as initial data for a highly distorted single black hole,
and evolved using techniques of black-hole perturbation theory.  They
called this method of reviving a (nearly) dead calculation the {\em
 Lazarus} approach.

\begin{figure}
\includegraphics*[scale=0.3,angle=0]{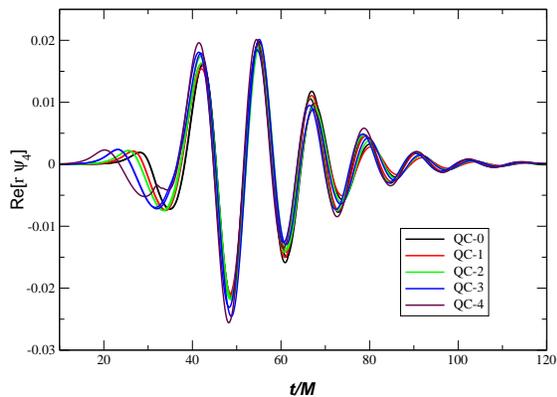}
\caption{Lazarus waveforms for equal-mass, nonspinning black-hole
  mergers, from \citet{Baker:2002qf}. The real, $\ell=2, m=2$ part of
  $r \Psi_4$, corresponding to the $+$ polarization (measured on the
  system's axis), is shown for ten simulations: five different initial
  black-hole separations (designated QC-0, etc.), each with two
  transition times to perturbative evolution.  }
\label{fig:Laz_wf}
\end{figure}

The hybrid waveforms produced by the Lazarus collaboration gave the
first indications of what might be expected from the final merger of
black-hole binaries \cite{Baker:2001nu,Baker:2002qf}.  They ran a
suite of simulations, varying the initial black-hole separations and
the time at which they stopped the calculation and made the transition
from the fully numerical evolution to the perturbed black hole
evolution.  Figure~\ref{fig:Laz_wf} shows the dominant quadrupole
$\ell=2, m=2$ component of Re($\Psi_4$), which corresponds to the $+$
polarization, for the case of equal-mass, nonspinning black holes.
Note the simple shape of the waveform smoothly tying together
what might be the end of an (inspiral) chirp with a damped (ringdown)
sinusoid.

By tuning the mass and spin of the background hole for the
perturbative evolution, \citet{Baker:2002qf} also determined that the
merger remnant was a Kerr (spinning) hole of mass $M_{\rm Kerr}
\approx 0.97 M_{\rm initial}$, and spin $a_{\rm Kerr} \approx 0.7
M_{\rm Kerr}$.

The Lazarus method was also applied to mergers of black holes with
spins either aligned or anti-aligned with the orbital angular momentum
\cite{Baker:2003ds}; in all cases, a similar waveform shape was seen.
How generic was this simple shape and, in particular, would it also
arise in situations where the black holes complete one or more orbits
before merging?

\subsection{Mergers of Equal-Mass, Nonspinning Black Holes}
\label{sec:GWs:eq}

\subsubsection{The First Merger Waveforms}
\label{sec:GWs:first_wfs}

\begin{figure}
\includegraphics*[scale=0.475,angle=0]{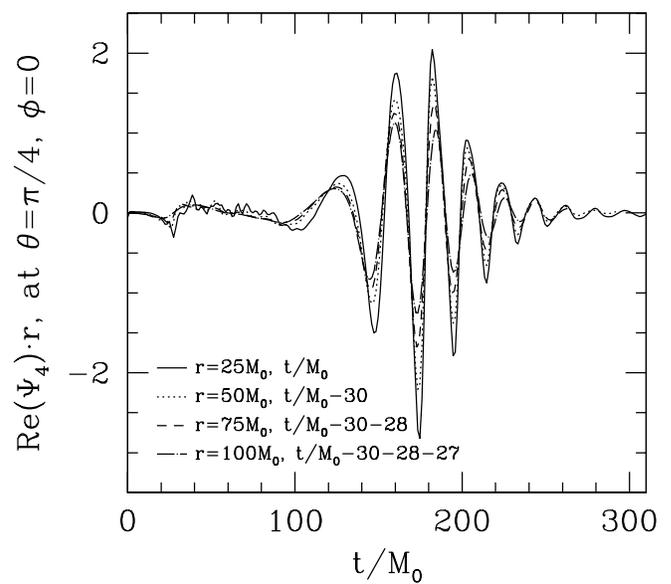}
\caption{The first gravitational waveform for black holes evolving
  through a single plunge orbit, merger, and ringdown, achieved by
  \citet{Pretorius:2005gq}.  The waves were extracted at four radii,
  and shifted in time to overlap for comparison.  Note the amplitudes
  decrease for larger $r$, due to lower resolution in the outer
  regions}
\label{fig:frans_wf}
\end{figure}

Early in 2005, Frans Pretorius stunned the community by achieving the
first robust simulation of two equal-mass black holes through a single
plunge orbit, merger, and ringdown \cite{Pretorius:2005gq}.  The
resulting gravitational waveforms are shown in
Fig.~\ref{fig:frans_wf}, where $r$Re($\Psi_4$) is plotted versus
$t/M_0$.  Here $r$ is the coordinate distance from the center of the
grid to the sphere on which the waveform is extracted (see
Sec.~\ref{sec:extract}), and $M_0 \sim 0.5M$ is the mass of a single
black hole.  (Note that the time axes for all other waveform plots in
this paper are scaled by either the total system mass $M$ or the mass
of the final, remnant black hole $M_f$).
Here curves are plotted for waveforms extracted on four spheres with
successively larger radii.  The curves have been shifted in time so
that the waveforms overlap.  Note that the overall waveform shape is
simple. The merger yields a single black hole that is spinning, with
spin parameter $a_f \sim 0.70 M_f$, where $M_f$ is the mass of the
final black hole that forms.

To carry out these mergers, Pretorius \cite{Pretorius:2004jg,Pretorius:2006tp}
used techniques that are very different from
the more traditional approach based on BSSN and punctures, used by
nearly all other numerical relativists.  As discussed in
Sec. \ref{sec:harm}, he used a generalized harmonic formulation in
which the Einstein equations are written with second-order time and
space derivatives.  The spatial domain was compactified, so that the
outer boundary of each slice was mapped to spatial infinity. Moreover, he
excised the black holes (each formed from the collapse of scalar field
``blobs'') and evolved their motion across the grid using adaptive
mesh refinement (see Secs.~\ref{sec:BHonGrid} and~\ref{sec:comp:computation}).

Pretorius had clearly developed a robust method for evolving
black-hole binary mergers.  In the wake of his remarkable success,
many numerical relativists began studying his methods. However, another
surprise was soon to emerge.

In the autumn of 2005 two research groups, one at UTB and the other at
Goddard, independently developed a powerful technique for evolving
mergers of puncture black holes within the traditional BSSN approach
\cite{Campanelli:2005dd,Baker:2005vv}. As discussed in
Sec.~\ref{sec:MovePunc}, the standard puncture method requires the
black holes to remain fixed on the numerical grid.  In this new
``moving puncture'' approach, simple but novel gauge conditions allow
the punctures to move across the grid, producing accurate and stable
merger evolutions \cite{vanMeter:2006vi}.

Although the UTB and Goddard codes were both based on the BSSN
approach, they were independently written and featured somewhat
different implementations of moving punctures.  In addition, the
Goddard code used second-order finite differencing (standard in the
community at the time).  They employed a box-in-box fixed mesh
refinement to produce the high resolution in the region around the
black holes needed to compute the dynamics accurately, while
maintaining adequate resolution in more distant regions and a large
enough computational domain to allow accurate extraction of the
gravitational waves.  The UTB code used fourth-order finite
differencing, and the ``fish-eye'' coordinate transformation to
produce higher resolution around the black-hole orbits and lower
resolution in the wave extraction regions.

\begin{figure}
\includegraphics*[scale=0.5,angle=0]{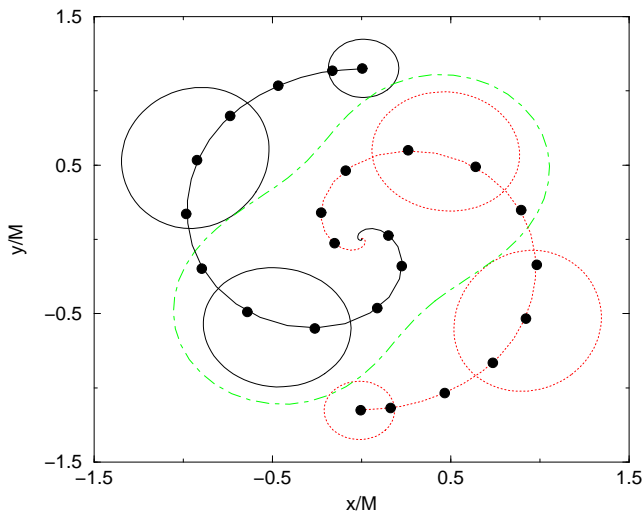}
\caption{The trajectories of the black-hole punctures from a run by
  \citet{Campanelli:2005dd}. The individual apparent horizons are
  shown at times $t = 0, 10M$ and $18.8M$.  The solid circles denote
  the centroids of the apparent horizons every $2.5M$ during the
  evolution.  The first common horizon, marking the time of merger,
  forms at $18.8M$, just before the punctures complete $1/2$ orbit.  }
\label{fig:UTB_tracks}
\end{figure}

\begin{figure}
\includegraphics*[scale=0.315,angle=0]{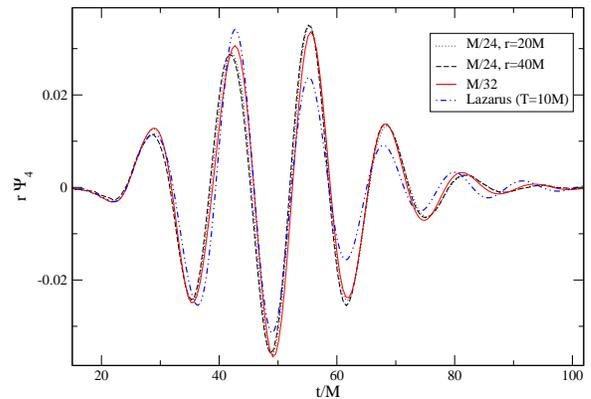}
\caption{Gravitational waveforms from the black-hole merger
  simulations by \citet{Baker:2005vv}.  The real part of the $\ell=2$,
  $m=2$ mode of $r\Psi_4$ was extracted from the numerical simulation
  on spheres of radii $r = 20M$, and $40M$ for the medium- and
  high-resolution runs. The waveforms extracted at different radii
  have been rescaled by $r$, and shifted in time to account for the
  wave propagation time between the two extraction spheres. Curves are
  shown for medium ($M/24$ in the finest grid) and high ($M/32$)
  resolutions. A waveform from a Lazarus calculation starting with the
  same initial data is shown for comparison \cite{Baker:2002qf}. The
  Lazarus waveform shown in this figure is scaled differently from
  those in Fig.~\ref{fig:Laz_wf}.}
\label{fig:GSFC_wf}
\end{figure}

Both groups successfully evolved an equal-mass, nonspinning binary
through the final plunge orbit, merger, and ringdown, and extracted
the gravitational waveforms.  In these simulations, the black holes
completed $\sim 1/2$ orbit before merger, defined as the time at
which a single apparent horizon was
formed. Figure~\ref{fig:UTB_tracks} shows the tracks traced by the
black-hole punctures, with the apparent horizons superimposed, for the
UTB run \cite{Campanelli:2005dd}.  These simulations produced plunge
waveforms with a simple shape similar to that found by Pretorius;
compare Fig.~\ref{fig:frans_wf} from Pretorius's simulations with
Fig.~\ref{fig:GSFC_wf}, which shows waveforms from the Goddard
simulation plus a corresponding Lazarus waveform.

\subsubsection{Universal Waveform}
\label{sec:GWs:universal}

In most cases, astrophysical black-hole binaries spiral together over
many orbits, radiating away any orbital eccentricity in the form of
gravitational waves.  By the time the black holes reach the final
inspiral, their orbits will be quasicircular.  All mergers of such
equal-mass nonspinning binaries should thus produce the same waveform
signature, subject only to rescaling with the total system mass $M$.

Over the years, concerns had been raised within the relativity
community about the effects of deviations from astrophysical initial
conditions on the waveforms.  For example, spurious eccentricity in
the orbits could arise from starting conditions that did not
approximate quasicircular inspiral accurately. Moreover, the commonly-used
conformal flatness prescription (see Sec.~\ref{sec:comp:initialdata})
for the initial data is different from the conditions experienced by
an astrophysical binary; these differences would result in spurious
gravitational radiation being present in the initial conditions for
the binary simulations.  How would factors such as these influence the
merger waveforms?

Having developed robust techniques for evolving black-hole mergers,
numerical relativists eagerly pursued longer simulations with the
black holes starting at wider separations, completing more orbits, and
producing longer wavetrains.  The UTB group ran a simulation in which
the black holes completed nearly 1.5 orbits before the formation of a
single apparent horizon \cite{Campanelli:2006gf}.  They observed that
the resulting waveform was quite similar to those produced by their
earlier simulation that started near the plunge
\cite{Campanelli:2005dd}, with a brief oscillatory signal at the
beginning. \cite{Campanelli:2006gf} anticipated
``that the plunge waveform, when starting from quasicircular orbits,
has a generic shape that is essentially independent of the initial
separation of the binary.''

The Goddard group, simultaneously pursuing this same goal, produced
the first ``universal'' waveform for the merger of equal-mass,
nonspinning black holes \cite{Baker:2006yw}.  Using approximately
quasicircular initial conditions, they ran a series of four
simulations, starting the black holes at increasingly wider
separations.  In all, the binaries completed $\sim 1.8$, 2.5, 3.6, and
4.2 orbits before forming a common apparent horizon; the systems
emitted just under 4\% of their energy as gravitational radiation, and
the final black holes were spinning with $\ahat_f \sim 0.69$.  To
compare the results of these models, they chose the fiducial time
$t=0$ to be the moment of peak amplitude in the gravitational
radiation; this typically occurs within a few $M$ of merger.  It
is worth noting how close these results are to those obtained from the
Lazarus approach to the same system (see Sec.~\ref{sec:GWs:laz}).
\begin{figure}
\includegraphics*[scale=0.45,angle=0]{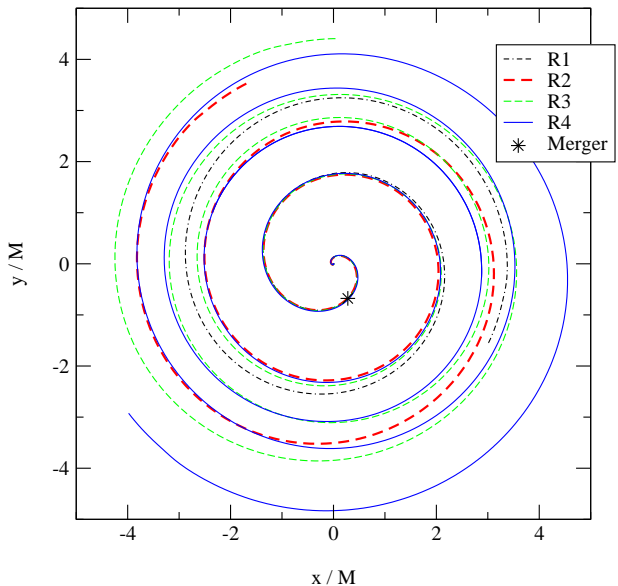}
\caption{Puncture tracks from equal-mass, nonspinning black-hole
  merger simulations starting at increasingly larger separations by
  \citet{Baker:2006yw}.  For clarity, the trajectory of only one black
  hole in each run is shown.  The tracks lock on to a universal
  trajectory $\sim 1$ orbit before merger (denoted by *).}
\label{fig:tracks}
\end{figure}

The binary orbital dynamics of these runs is clearly seen in
Fig.~\ref{fig:tracks}, which shows the tracks of the black-hole
punctures.  Here only one puncture track is shown for each binary,
with the trajectories oriented to superpose at the fiducial $t=0$.
The tracks show the effects of eccentricity in the early
stages of each run; plots of the separation $r$ versus time show that
the initial eccentricity decreases as the initial separation
increases.  As the holes spiral together into the strong-field regime,
the eccentricity diminishes.  The puncture tracks lock on to a single
universal trajectory, independent of the initial conditions, through
the final orbit, plunge, and merger.

\begin{figure}
\includegraphics*[scale=0.315,angle=0]{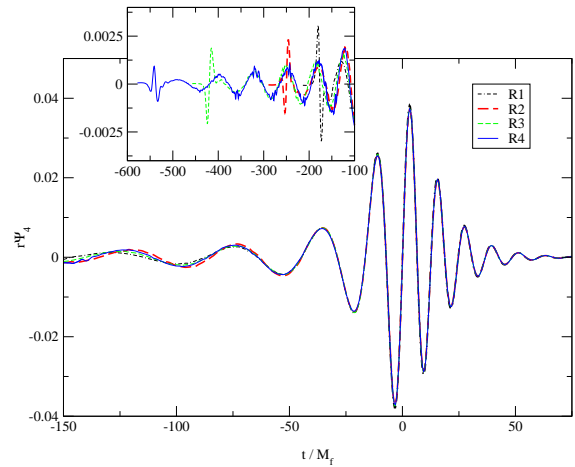}
\caption{The universal waveform ($\ell = 2$, $m=2$ mode) produced by
  the four simulations whose puncture tracks are shown in
  Fig.~\ref{fig:tracks} \cite{Baker:2006yw}. The signals have been
  shifted in time so that the peak radiation amplitude occurs at $t =
  0$.  The agreement among the waveforms is excellent, with
  differences $\sim 1\%$ for the final burst of radiation starting at
  $t\sim -50M_f$.}
\label{fig:universal_wf}
\end{figure}

Figure~\ref{fig:universal_wf} shows the corresponding universal
gravitational waveform for equal-mass, nonspinning black
holes. Specifically, it shows the real part of the $\ell=2$, $m=2$
mode of $r\Psi_4$ versus time; for this case, the quadrupole $(2,\pm
2)$ modes strongly dominate all other modes.  Here the signals
produced by each run have been shifted in time so that $t=0$ marks the
moment of peak radiation amplitude.  Starting from $t \sim -50 M_f$,
the waveforms for the final burst of radiation show nearly perfect
agreement, with differences at the level of $\sim 1\%$. For the
preceding few orbits, the waveforms agree in amplitude and phase to
$\sim 10\%$, except for the brief initial bursts of spurious radiation
(see Sec.~\ref{sec:comp:initialdata}).  Overall, the merger stage
lasts $\sim 100M$, converting $\sim 4\%$ of the initial total mass $M$
into gravitational-wave energy.  The gravitational wave released
during this burst has a luminosity $L \sim 10^{23} L_{\odot}$, which
is greater than the total luminosity of all the stars in the visible
Universe.  For stellar black-hole binary mergers, this luminosity will
last for a few milliseconds, while for MBH binaries it will last for
several minutes.

Note that this universal waveform has a simple shape. The signal
starts with a short inspiral chirp, increases smoothly in amplitude
during the plunge and merger, and then transitions to the damped
ringdown sinusoid.  Overall, the frequency increases monotonically,
reaching a maximum value that stays constant during the ringdown
\cite{Baker:2006yw}.

\begin{figure}
\includegraphics*[scale=0.315,angle=0]{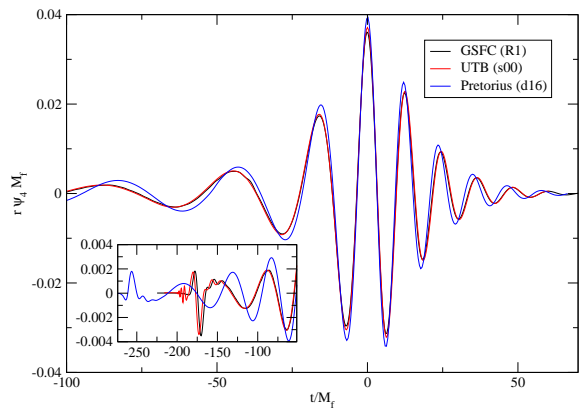}
\caption{Comparison of three waveforms from simulations of equal-mass
  black-hole mergers computed by Pretorius and the UTB and Goddard
  groups \cite{Baker:2007fb}.  The UTB and Goddard simulations used
  nonspinning black holes, whereas Pretorius' initial conditions
  produced co-rotating black holes, each with a small spin
  $\ahat_{1,2} = 0.08$. The differences between Pretorius' waveform
  and the others during the ringdown, $t/M_f \gtrsim 25$, are
  consistent with his simulation producing a slightly faster rotating
  final black hole due to these initial spins. Reproduced by permission of the Institute of Physics.}
\label{fig:compare_wf}
\end{figure}

Since black-hole-merger waveforms will be used to help find signals in
data from gravitational-wave detectors, it was important to verify
that various groups consistently achieve the same results.
Figure~\ref{fig:compare_wf} provides a comparison of waveforms
computed by Pretorius and the groups at UTB and Goddard for
equal-mass, nonspinning black-hole mergers \cite{Baker:2007fb}; here
the real part of $r\Psi_4$'s dominant $\ell = 2$, $m=2$ mode is
shown, with all three waveforms aligned in time so that the moment of
peak gravitational radiation amplitude occurs at $t = 0$.  All three
simulations evolved through the last $\sim 1.8 - 2.5$ orbits before
merger; the Goddard run is R1 from \citet{Baker:2006yw} and shown in
Figs.~\ref{fig:tracks} and~\ref{fig:universal_wf}.  Note that all
evolutions produce the same overall waveform shape.  The Goddard and
UTB groups evolved black holes with zero spin starting on similar
quasicircular initial orbits, and produced nearly identical waveforms.
Pretorius used co-rotating initial conditions which impart a small
spin $\ahat_{1,2} = 0.08$ to each black hole.  This difference is
consistent with the slightly higher frequencies seen in Pretorius'
results during the ringdown, $t \gtrsim 25 M_f$.

\begin{figure}
\includegraphics*[scale=1.0,angle=0]{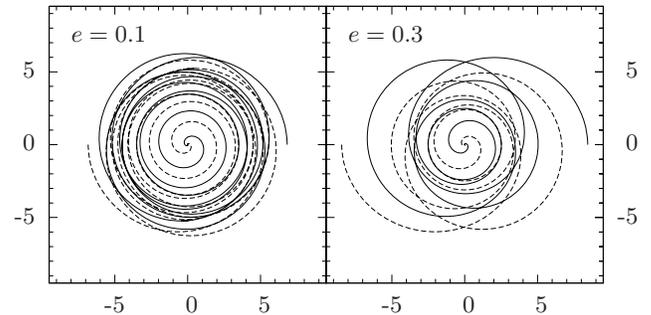}
\caption{Puncture tracks for equal-mass black-hole mergers with
  initial orbital eccentricity $e = 0.1$ (left) and $e=0.3$ (right)
  from \citet{Hinder:2007qu}. In these cases, the initial eccentricity
  is radiated away and the binaries circularize before merger.}
\label{fig:PSU-ecc-tracks}
\end{figure}

\begin{figure}
\includegraphics*[scale=1.0,angle=0]{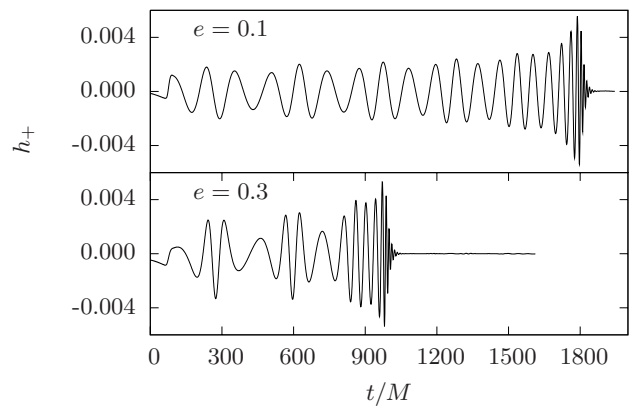}
\caption{Real part of $(2,2)$ strain mode (labeled $h_+$ here) for
  black-hole mergers shown in Fig.~\ref{fig:PSU-ecc-tracks} from
  \citet{Hinder:2007qu}.  These runs have initial orbital eccentricity
  $e=0.1$ (top) and $e=0.3$ (bottom) and enter a universal plunge by
  $t \sim 50 M_f$ before the gravitational radiation reaches its peak
  value.}
\label{fig:PSU-ecc-wfs}
\end{figure}

How robust is this universal waveform to larger amounts of orbital
eccentricity and gravitational radiation in the initial data?
\citet{Hinder:2007qu} studied a series of equal-mass, nonspinning
binaries starting from roughly the same initial orbital period and
having varying amounts of initial orbital eccentricity $e$.  For $e
\lesssim 0.4$, the initial eccentricity is radiated away, and the
binaries circularize and begin a universal plunge at $t \sim 50 M_f$
before the time of peak radiation amplitude, producing a final black
hole with $a \sim 0.69 M_f$.  For $e \gtrsim 0.5$, the black holes do
not complete any orbits, but rather plunge together and merge; the
final black hole spin reaches a maximum value $a \sim 0.72 M_f$ around
$e \sim 0.5$. Figure~\ref{fig:PSU-ecc-tracks} shows the puncture tracks,
and Fig.~\ref{fig:PSU-ecc-wfs} shows the real part of the strain's
$\ell = 2$, $m=2$ mode, for two nonspinning, equal-mass black-hole mergers with nonzero
eccentricity from this study; in both cases, \citet{Hinder:2007qu} show
quantitatively that the waveform for the final burst of radiation is
the same as the universal waveform.

In a complementary study, \citet{Sperhake:2007gu} found that the final
spin of the remnant black hole is essentially insensitive to the
eccentricity for binaries that do not plunge immediately.
\citet{Bode:2007dv} examined the effects of spurious gravitational
waves by superposing a tunable packet of gravitational radiation on an
equal-mass, nonspinning binary.  They found the binary evolution and
the spin of the remnant black hole to be robust to modest amounts of
added gravitational radiation.

\subsubsection{Longer Waveforms}
\label{sec:GWs:longer_wfs}

The black-hole simulations discussed so far cover the last $\lesssim$
four orbits before merger.  These achievements marked the triumph of
solving a long-standing problem of fundamental importance to general
relativity: the two-body problem for the final merger of equal-mass
Schwarzschild black holes driven by gravitational radiation reaction.
These simulations provided the first look at the dynamics and
waveforms of black holes evolving and merging in the nonlinear,
strong-field regime.  They also pointed the way towards applications
in detecting the final gravitational-wave burst from black-hole
mergers \cite{Baumgarte:2006en}.  However, for more advanced
applications to gravitational-wave data analysis and comparisons with
PN analytical treatments, longer waveforms that start many cycles
before merger are essential.

The Goddard group \cite{Baker:2006ha,Baker:2006kr} produced the first
such long waveforms for equal-mass, nonspinning black holes starting
$\sim 7$ orbits or $\sim 14$ gravitational-wave cycles before merger.
They used improved initial conditions with low eccentricity $e <
0.01$, and focused on improving accuracy while the black holes
traversed the relatively long inspiral.  They investigated the
observability of black-hole mergers with ground- and space-based
gravitational-wave detectors \cite{Baker:2006kr} (see
Sec.\ref{sec:gwda:direct}) and successfully applied their long
waveforms to comparisons with PN results, focusing on the waveform
phases \cite{Baker:2006ha}.  \footnote{Recall that the PN
  approximation is an expansion in powers $\epsilon=v^2/c^2$ and
  applies when the black holes are far enough apart that the black
  hole speeds remain well below the speed of light.  We refer to PN
  results by the order at which the series is truncated.  For example,
  ``2~PN'' means that terms of order $\epsilon^2=v^4/c^4$ are
  retained.  See Sec.~\ref{sec:nrpn} for a deeper discussion of the PN
  approximation in the context of numerical relativity.}

Shortly thereafter, the Jena group simulated a binary inspiralling for
nine orbits ($18$ gravitational-wave cycles) before merger
\cite{Hannam:2007ik}.  With this, they made the first quantitative
comparisons with both the PN phase \emph{and} amplitude, and
quantified the level of error in the quadrupole approximation. They
used higher-order finite differencing \cite{Husa:2007hp} and initial
binary parameters calculated using the PN approximation to reduce the
initial eccentricity significantly \cite{Husa:2007rh}, enabling a
precise measurement of the waveform phase.

\begin{figure}
\includegraphics*[scale=0.45,angle=0]{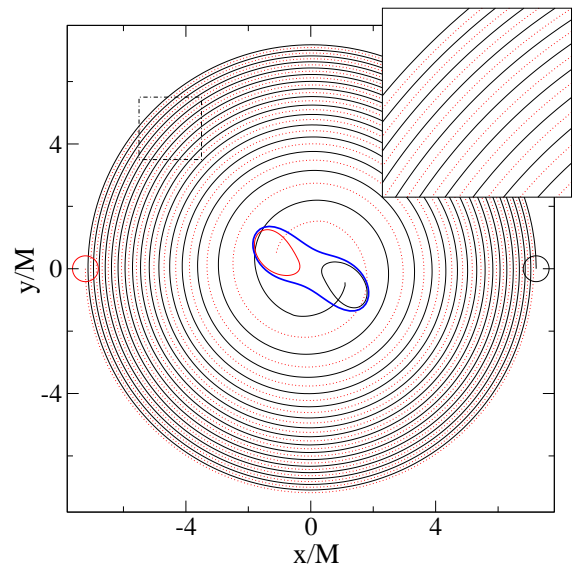}
\caption{Trajectories for the merger of equal-mass, nonspinning black
  holes computed by the Caltech-Cornell group using their spectral
  evolution code \cite{Scheel:2008rj}.  The red and black
  circles/ellipses are the initial/final coordinate locations of the
  apparent-horizon surfaces, and the blue thick contour is the
  common apparent horizon just after it appeared. The black holes
  complete 16 orbits before merging.  Figure kindly provided by
  H. Pfeiffer.  }
\label{fig:cc_traj}
\end{figure}

\begin{figure}
\includegraphics*[scale=0.50,angle=0]{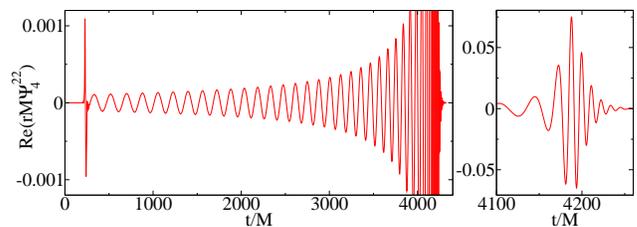}
\caption{Gravitational waveforms from the Caltech-Cornell merger
  simulation seen in Fig.~\ref{fig:cc_traj} showing the $\ell=2$,
  $m=2$ component of Re($r\Psi_4$) \cite{Scheel:2008rj}.  The left
  panel shows a zoom of the inspiral waveform, and the right panel a
  zoom of the merger and ringdown.}
\label{fig:cc_wf}
\end{figure}

The Caltech-Cornell group currently holds the record for the longest
and most accurate black-hole binary evolution, starting 16 orbits and
32 gravitational-wave cycles before merger \cite{Scheel:2008rj}.
Using their spectral code (see Sec.~\ref{sec:comp:computation}),
they begin with a very small initial orbital eccentricity $e \sim 5
\times 10^{-5}$ and evolve with very high accuracy through a
relatively long inspiral, then merger and ringdown.  The
impressive trajectories of their black holes are shown in
Fig.~\ref{fig:cc_traj} and the accompanying gravitational waveforms in
Fig.~\ref{fig:cc_wf}.


\begin{figure}
\includegraphics*[scale=0.515,angle=0]{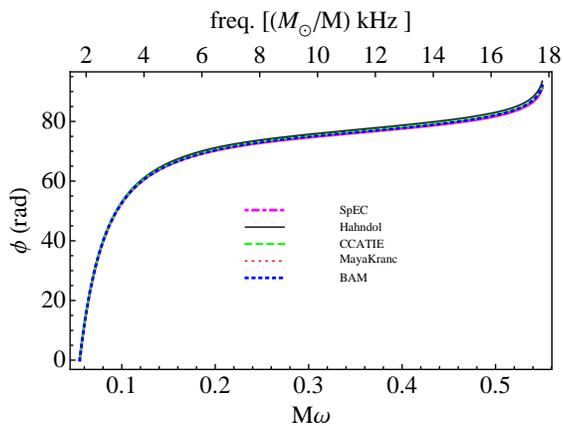}
\caption{The gravitational-wave phase $\phi$ as a function of the
  dimensionless frequency $M\omega$ for the five codes from the
  Samurai comparison \cite{Hannam:2009hh}.  The scale along the top of
  the panel labels the frequency in kHz scaled with respect to the
  total binary mass in solar units.}
\label{fig:samurai-phase}
\end{figure}

\begin{figure}
\includegraphics*[scale=0.55,angle=0]{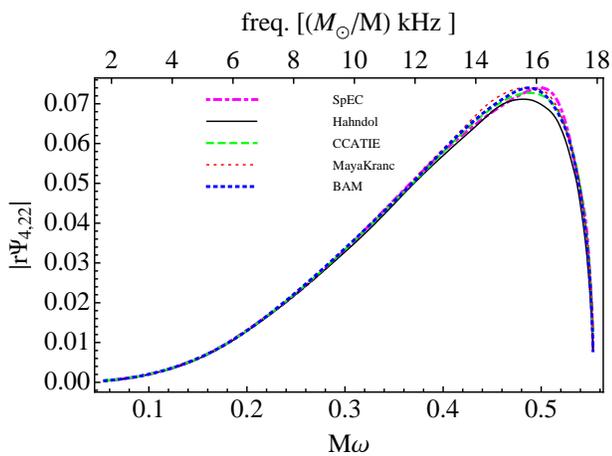}
\caption{Amplitudes of the $\ell=2$, $m=2$ component of the
  gravitational waves produced by the five codes in the Samurai
  comparison \cite{Hannam:2009hh} are shown as a function of
  frequency.}
\label{fig:samurai-amps}
\end{figure}

Comparison of waveforms from different groups remains important in the
push for higher accuracy and use in gravitational-wave data analysis.
The Samurai project \cite{Hannam:2009hh} sets the current state of the
art for studying the consistency of black-hole binary waveforms.  This
effort focuses on comparing waveforms from equal-mass, nonspinning
binaries, starting with at least six orbits (or 12 gravitational-wave
cycles) before merger and continuing through the ringdown.  They
focus on comparing the amplitude $A(t)$ and phase $\phi(t)$ for the
$\ell=2$, $m=2$ mode of $r\Psi_4$, defined as
\begin{equation}
\label{eq:def-psi4}
r\Psi_{4,22}(t) = A(t) e^{-i\phi(t)};
\end{equation}
the gravitational-wave frequency of this mode is then $\omega(t) =
\dot{\phi}(t)$.  They compare the results from five independent
numerical codes: the moving-puncture codes from the AEI, Goddard,
Jena, and Penn State groups, and the Caltech-Cornell spectral code.
Figure~\ref{fig:samurai-phase} compares the gravitational-wave
amplitudes and Fig.~\ref{fig:samurai-amps} the gravitational-wave
phases as a function of frequency for the five
waveforms. Qualitatively, the results appear to be quite consistent.
Quantitatively, they concluded that these waveforms have sufficient
accuracy to be used for detection with all current and planned
ground-based detectors.

\subsection{Mergers of Unequal-Mass, Nonspinning Black Holes}
\label{sec:GWs:uneq}

\begin{figure}
\includegraphics*[scale=0.45,angle=-0]{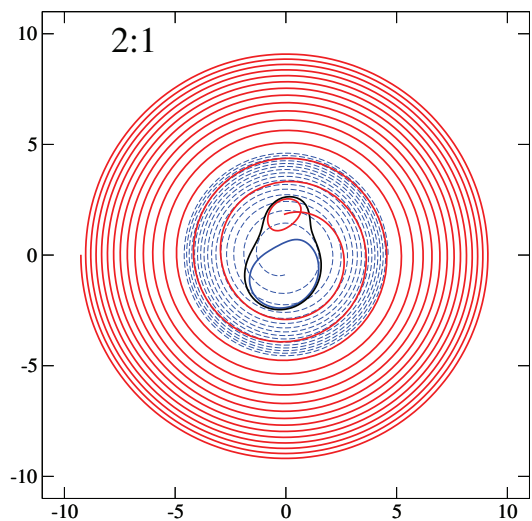}
\caption{Puncture tracks for an unequal-mass nonspinning binary with
  $q = 2$ produced by the Caltech-Cornell spectral code; figure kindly
  provided by H. Pfeiffer.  The solid (red) curve gives the trajectory
  of the smaller black hole, and the dotted (blue) curve the path of
  the larger one. The red and blue contours are the final coordinate
  locations of the apparent-horizon surfaces, and the black
  contour is the common apparent horizon just after it appeared.}
\label{fig:unequal_tracks}
\end{figure}

Early in 2006, numerical relativists took the next step in opening up
the parameter space of binary black-hole mergers by simulating
nonspinning binaries with unequal masses. This added a new parameter,
the mass ratio $q$, to the problem and brought an additional need for
adaptive mesh refinement to achieve adequate resolution around the
smaller black hole. As shown in Fig.~\ref{fig:unequal_tracks}, the
smaller black hole also moves faster, completing an orbit around the
center of mass in the same time as the larger hole and thus requiring
smaller timesteps for its evolution.  These factors combine to make
simulations of unequal-mass binaries more technically challenging;
currently, numerical relativists are able to simulate mass ratios up
to $q = 10$ \cite{Gonzalez:2008bi}.

\subsubsection{Mode Analysis and Gravitational Waveforms}
\label{sec:GWs:uneq-wfs}

Decomposition into spin-weighted spherical harmonic modes (see
Sec.~\ref{sec:extract}) provides the basis for an in-depth study
of black-hole mergers.  For the equal-mass case $q=1$, the $\ell = 2$,
$m=\pm2$ quadrupole mode is dominant and the odd-$m$ modes are
suppressed by symmetry.  As $q$ increases, the sub-dominant modes
become more important, affecting both the evolution of the source and
the emitted radiation.

\begin{figure}
\includegraphics*[scale=0.35,angle=-90]{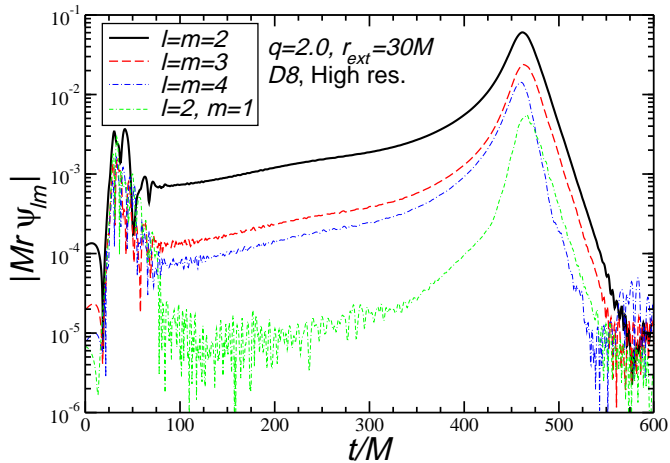}
\caption{The amplitudes of several $(\ell,m)$ modes of $Mr\Psi_4$ for
  mass ratio $q = 2$ from simulations by \citet{Berti:2007fi}.  The
  initial burst of radiation is an artifact of the initial data, and
  the wiggles at late times result from numerical noise.}
\label{fig:Berti-amps}
\end{figure}

\citet{Berti:2007fi} carried out a set of unequal-mass mergers with
mass ratios in the range $1 \le q \le 4$.  They found that, to leading
order, the total energy emitted during merger scales $\sim \eta^2$
while the spin of the final black hole scales $\sim \eta$, where
$\eta$ is the symmetric mass ratio \eqref{eq:eta_def}.  They also
studied the multipolar structure of the gravitational waves.
Figure~\ref{fig:Berti-amps} shows several $(\ell,m)$ modes of the
radiation produced in their simulations for the case $q=2$.  They
demonstrated that the higher modes carry larger fractions of the total
energy as $q$ increases; in particular, the $\ell=3$ mode generally
carries $\sim 10\%$ of the emitted energy for $q > 2$.

\begin{figure}
\includegraphics*[scale=0.33,angle=0]{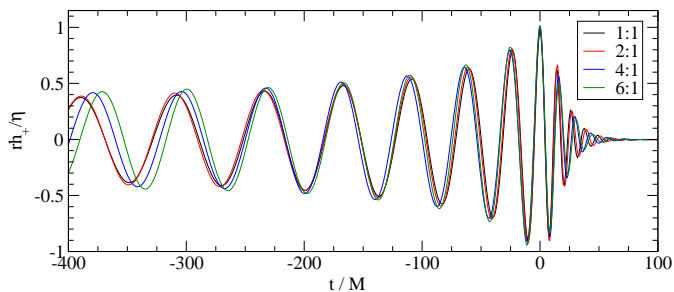}
\caption{Strain waveforms scaled by $\eta$ for different mass ratios,
  from \citet{Baker:2008mj}. $h_+$ is calculated for different mass
  ratios using the $\ell = 2$ and $\ell = 3$ modes. The observer is
  located at distance $r$ along the axis of the system.}
\label{fig:wfs_eta_scaled}
\end{figure}

The Goddard group \cite{Baker:2008mj} performed a complementary study
of nonspinning unequal-mass mergers for mass ratios $1 \le q \le 6$.
They found that the overall simple waveform shape first discovered for
equal-mass mergers extends to the cases with $q > 1$; this is easily
seen when the gravitational waves are scaled by $\eta$.
Figure~\ref{fig:wfs_eta_scaled} shows the strain $rh_+/ \eta$,
including the $\ell = 2$ and $\ell = 3$ modes, for an observer located
at distance $r$ on the system's orbital axis.  The waveforms are
aligned so that the maximum amplitude of the $\ell=2$, $m=2$ mode
occurs at $t=0$.  The differences in phase during the final ringdown
portion of the waveforms are consistent with the differing spins of
the final black holes; for example, with nonspinning black holes, a
larger mass ratio $q$ results in a smaller final spin for the remnant
black hole.  When viewed off-axis, the waveforms show modest amplitude
and phase modulations, while still preserving the overall simplicity
in shape.

\citet{Baker:2008mj} also found that each of the individual spherical
harmonic components is circularly polarized during the inspiral,
merger, and ringdown, as seen by distant observers on the system's
rotational axis.  During the inspiral, the phase and frequency of the
different $(\ell,m)$ components are the same for each mass ratio; this
is expected since the waveform phase is directly related to the
orbital phase.  More interestingly, for the $\ell = m$ modes, this
relationship continues to hold during merger and ringdown.

Based on these observations, they developed a simple conceptual
picture in which each $(\ell,m)$ mode of the gravitational radiation
is produced separately by the $(\ell,m)$ mode of some \emph{implicit
  rotating source}.  The fixed relationship between the phase and
frequency of the $\ell = m$ modes shows that these components of the
implicit source rotate rigidly during the entire coalescence from
inspiral through ringdown.  In comparison, the $\ell \ne m$ source
components peel away from the main trend of the $\ell = m$ parts
during merger, indicating less rigid rotation.

\begin{figure}
\includegraphics*[scale=0.33,angle=-90]{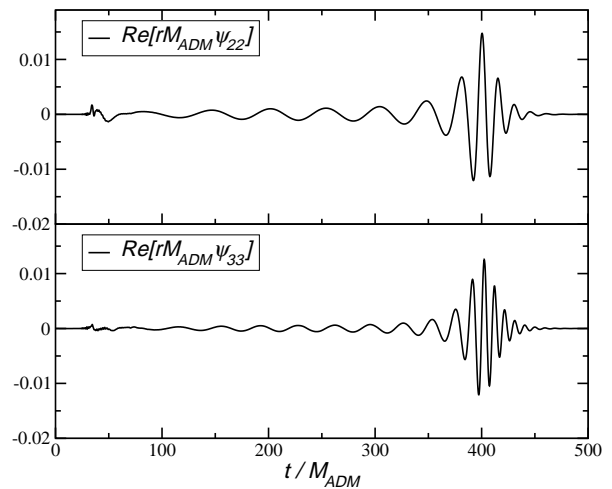}
\caption{The $\ell=m=2$ and $\ell=m=3$ component of Re($rM\Psi_4$)
  from the 10:1 mass ratio simulation by \citet{Gonzalez:2008bi}}
\label{fig:10to1_wfs1}
\end{figure}

\citet{Gonzalez:2008bi} have carried out the highest mass ratio merger
simulation to date, with $q = 10$.  This binary radiates $\sim 0.42\%$
of its mass as gravitational radiation as it undergoes $\sim 3$ orbits
before merging to form a black hole with $\ahat_f \sim 0.26$; these
values fit the scaling relations found by \citet{Berti:2007fi} and
\citet{Pan:2007nw}.  Figure~\ref{fig:10to1_wfs1}, from
\citet{Gonzalez:2008bi} shows that the $\ell = m =2$ and $\ell = m =
3$ modes of Re($rM\Psi_4$) exhibit the simple waveform shape
individually; this trend continues through $\ell = m = 5$, the highest
mode they studied.

\subsubsection{A Qualitatively New Feature: Kicks}

Unequal-mass black-hole binaries can form as a result of galaxy
mergers, dynamical processes in star clusters, and the final stages of
binary stellar evolution.  Mergers of unequal-mass binaries bring a
qualitatively new phenomenon of great importance to astrophysical
scenarios of black-hole growth and retention: recoils or kicks.

As discussed in Sec.~\ref{sec:astro_grav:basic}, the gravitational
waves emitted by a merging black-hole binary carry away linear
momentum.  If the pattern of gravitational-wave emission is not
symmetrical ({\em i.e.,} if there is more radiation emitted in some
direction than in others), then global conservation of momentum
requires the center of mass, and thus the remnant black hole that
forms, to recoil in the opposite direction.

\begin{figure}
\includegraphics*[scale=0.6,angle=0]{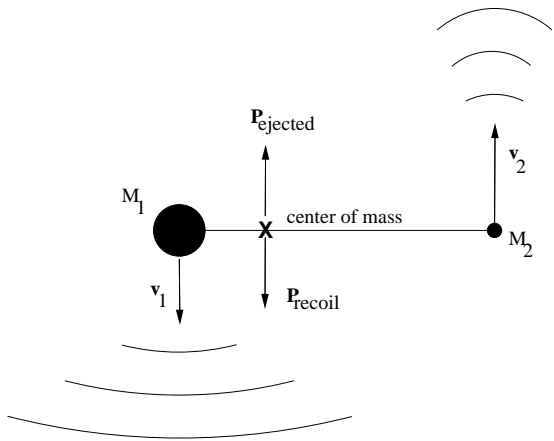}
\caption{Schematic drawing of the physical basis of the recoil or kick
  produced by a merging black-hole binary with unequal masses. A net
  flux of momentum $P_{\rm ejected}$ is emitted parallel to the smaller
  hole's velocity. Momentum conservation requires that the system center
  of mass recoil in the opposite direction, $P_{\rm recoil}$. Reproduced
  with kind permission from Springer Science+Business Media: Fig.~1 of
  \citet{Hughes:2004ck}.}
\label{fig:schematic_kick}
\end{figure}

The situation for an unequal-mass, nonspinning binary is shown
schematically in Fig.~\ref{fig:schematic_kick}, from
\citet{Hughes:2004ck}, who attributed this argument to Alan Wiseman; we
summarize their discussion here.  The two holes are orbiting in a
plane about their center of mass.  Since the smaller black hole moves
faster, its wave emission undergoes more ``forward beaming'' than that
of the larger hole.  Instantaneously this gives a net flux of momentum
parallel to the smaller hole's velocity, and an opposing recoil or
kick at the center of mass.  The direction of this kick changes
continually as the black holes orbit. If the orbit was circular, the
center of mass would move in a circle and suffer no net recoil.
However, since the black holes are spiralling together (due to the
energy and angular momentum carried away by the gravitational waves),
the instantaneous kicks do not cancel exactly but rather accumulate.
This causes the final merged black hole to have a non-zero net recoil
in the orbital plane.

This recoil has been studied by several authors, including
\citet{Peres:1962zz,Bekenstein1973,Fitchett_1983,Fitchett:1984qn,RR89}.
More recent analytic treatments using PN approximations were carried
out by \citet{Wiseman:1992dv,Favata:2004wz,Blanchet:2005rj,
  Damour:2006tr,Tiec:2009yg}, while some comparable-mass estimates
were also made with the Lazarus approach \cite{Campanelli:2004zw}.
However, since the dominant part of the kick depends sensitively on
the strong-field regime close to merger, where the orbits are less
circular, an accurate calculation of the kick requires full numerical
relativity simulations of the final stages of binary inspiral, merger
and ringdown \cite{Schnittman:2007ij}.

\begin{figure}
\includegraphics*[scale=0.33,angle=-90]{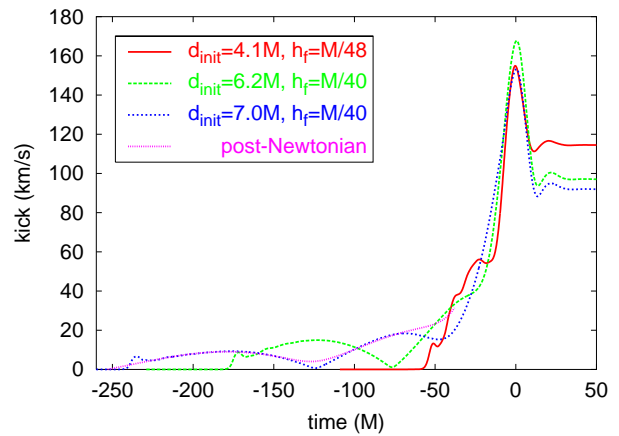}
\caption{The magnitude of the kick velocity for a nonspinning, unequal
  mass merger with $q = 1.5$ from \citet{Baker:2006vn}.  Results are
  shown from three simulations, with increasingly wider initial black-hole
  coordinate separations $d_{\rm init}$.  The small-dotted
  (purple) curve shows a 2~PN calculation starting with initial
  conditions commensurate with the $d_{\rm init} = 7.0M$ case (blue
  dotted line). Reproduced by permission of the AAS.}
\label{fig:GSFC-kick.eps}
\end{figure}

The Goddard group \cite{Baker:2006vn} carried out the first accurate
calculation of black-hole recoil for the merger of a nonspinning
binary with $q = 1.5$.  Figure~\ref{fig:GSFC-kick.eps} shows the kick
velocity as a function of time for three simulations in which the
black holes start out on orbits with increasingly larger initial
separations $d_{\rm init}$.  The kick velocity grows rapidly during
merger, reaching a peak value $\sim 10M$ after the time of peak
radiation amplitude in $\Psi_4$, and then dropping to a lower, final
value.  It is clear from this figure that the black holes must start
far enough apart in order to get a consistent and reliable value for
the final recoil velocity.  For this $q = 1.5$ case, their results
spanned the range $v_{\rm kick}=101 \pm 15 \kms$, with a best estimate
of $v_{\rm kick}=92 \pm 6 \kms$.

Following on from this, and other unequal-mass simulations by
\citet{Herrmann:2006ks}, \citet{Gonzalez:2006md} performed the first
systematic parameter study of nonspinning black-hole binary mergers
with mass ratios in the range $0.253 \le 1/q \le 1$.  They found that
the maximum recoil velocity $v_{\rm kick} = 175.2 \pm 11 \kms$ occurs
for the mass ratio $1/q = 0.36 \pm 0.03$.  For the case $q = 10$,
\citet{Gonzalez:2008bi} found a recoil velocity $v_{\rm kick} = 66.7
\pm 3.3 \kms$.

\subsection{Mergers of Spinning Black Holes}
\label{sec:GWs:spinning}

Astrophysical black holes are generally expected to be spinning.
Including the spin vector of both black holes in the binary adds six
more dimensions to the parameter space, giving seven in all when the
mass ratio $q$ is included.  Exploration of this large parameter space
began in 2006, with the simplest cases of equal masses and spins, and
is gradually growing to incorporate more generic binaries.

\subsubsection{Gravitational Waveforms}

The UTB group carried out the first black-hole binary mergers with
spin \cite{Campanelli:2006uy}.  They simulated three binaries, each
having equal masses and starting on quasicircular orbits with initial
orbital frequency $M \Omega = 0.5$, giving an initial orbital period
$\sim 125M$.  In the aligned case, both holes have spin parameters
$\ahat_{1,2} = 0.757$ and vectors parallel to the orbital angular
momentum $\Jorb$; in the anti-aligned case, the black holes have the
same spin parameters but the vectors are anti-parallel to $\Jorb$.  A
nonspinning equal-mass binary was run for comparison.

\begin{figure}
\includegraphics*[scale=0.5,angle=0]{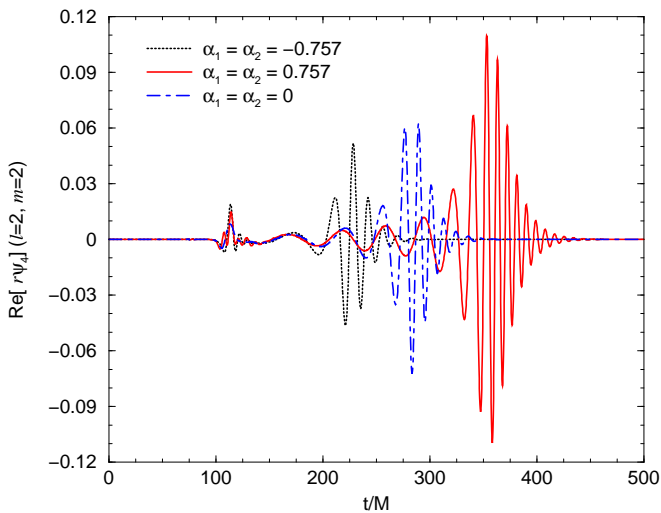}
\caption{Gravitational waveforms for mergers of equal-mass spinning
  binaries as in \citet{Campanelli:2006uy}. The dominant $\ell=2,m=2$
  mode is shown for three cases: aligned spins (solid/red),
  anti-aligned spins (dotted/black), and no spins (dashed/blue).  All
  three simulations start with the same initial orbital period at
  $t=0$.  Figure kindly provided by M. Campanelli.}
\label{fig:UTB-spin}
\end{figure}

The $\ell = 2$, $m=2$ components of the waveforms are plotted in
Fig.~\ref{fig:UTB-spin}, which shows all three cases starting out at
the same time $t=0$ at the same orbital period and gravitational-wave
frequency $f_{\rm GW} \sim 2 f_{\rm orb}$.  Notice that the aligned
system takes longer to merge, completing more orbits and producing a
longer wavetrain.  This behavior is caused by a spin-orbit interaction
that produces an effective repulsive force between the black holes.
In the anti-aligned case, this interaction yields an effective
attraction, causing the binary to merge more quickly and with fewer
orbits.  The nonspinning case is intermediate between the other two.
All three cases produce similar waveforms with a clean, simple shape.

It is also interesting to compare the total energy radiated as
gravitational waves $E_{\rm rad}$ and the spin parameter of the
final black hole $\ahat_f$ for these three cases. The aligned case
yields $E_{\rm rad} \sim 0.07M$ and $\ahat_f \sim 0.89$ while the
anti-aligned case gives $E_{\rm rad} \sim 0.02M$ and $\ahat_f \sim
0.44$.  The nonspinning case is again between these two cases, with
$E_{\rm rad} \sim 0.04M$ and $\ahat_f \sim 0.69$
\cite{Campanelli:2006uy}.

The black-hole mergers discussed so far have resulted from simple
binary dynamics with no precession and have produced simple waveforms
with a similar shape, at least when examined mode-by-mode in a
spin-weighted spherical harmonic decomposition.  Might more
``generic'' precessing binaries generate waveforms with more complex
patterns?  This important question pertains not only to basic orbital
dynamics in general relativity, but also to strategies for developing
templates to search for signals in data from gravitational-wave
detectors (see Sec.~\ref{sec:gwda}).

The parameter space for such binaries is very large, and explorations
of this space have only recently started.  The UTB group, now at the
Rochester Institute of Technology (RIT), pioneered the study of
precessing black-hole binary mergers.  They evolved a generic
black-hole binary with unequal masses and unequal, non-aligned
precessing spins that undergoes $\sim 9$ orbits before merger and
produces a relatively long wavetrain \cite{Campanelli:2008nk}.  This
system has a mass ratio $q = 1.25$ and arbitrarily oriented spins with
magnitudes $\ahat_1 \sim 0.6$ and $\ahat_2 \sim 0.4$.  The initial
conditions were determined from point-mass evolutions using 2.5~PN and
3.5~PN parameters.

\begin{figure}
\includegraphics*[scale=0.8,angle=0]{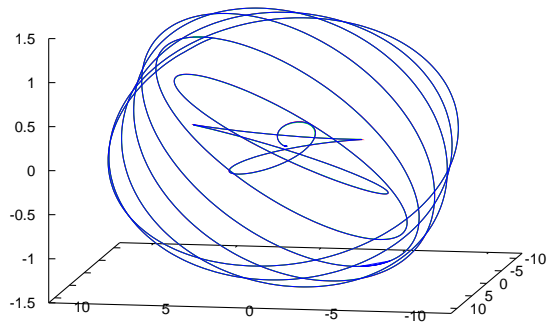}
\caption{The difference in the black-hole trajectories $\vec x_1 -
  \vec x_2$ for the generic binary evolution from
  \citet{Campanelli:2008nk}.  The precession of the system spin
  induces precession of the orbital plane.  }
\label{fig:RIT-3d-tracks}
\end{figure}

\begin{figure}
\includegraphics*[scale=0.45,angle=0]{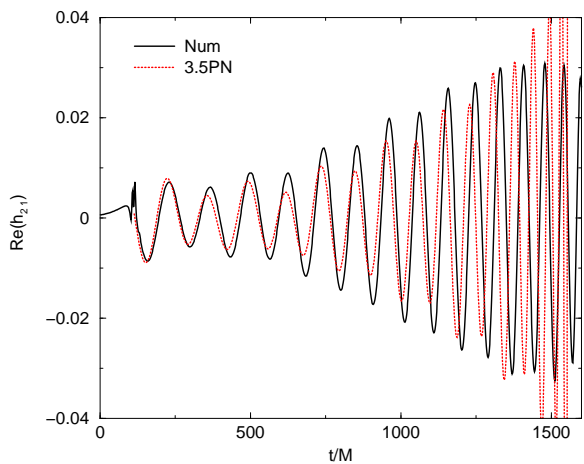}
\caption{Precession-induced amplitude modulations in the waveforms
  from the generic black-hole binary merger from
  \citet{Campanelli:2008nk}.  The $\ell= 2$, $m = 1$ mode of the
  strain ($h_+$) is shown for the numerical-relativity simulation
  (solid) and a 3.5~PN calculation (dotted). The amplitude
  oscillations are induced by precession.}
\label{fig:RIT-precess}
\end{figure}

Figure~\ref{fig:RIT-3d-tracks} shows the difference in the black-hole
trajectories $\vec x_1 - \vec x_2$.  For a non-precessing binary,
$\vec x_1 - \vec x_2$ would have no component in the $z$ direction.
In this generic case, the precession of the total system spin drives a
precession of the orbital plane, producing evolution of the trajectory
in the $z$ direction.
The resulting waveforms demonstrate the effects of precession
on the amplitudes of the sub-dominant modes.
Figure~\ref{fig:RIT-precess} shows the $\ell = 2$,$m = 1$ mode of the
strain $h_+$.  The numerical-relativity evolution is shown by the
solid curve; note the amplitude modulations induced by the precession.
The dotted curve shows a solution of the PN equations of motion at
3.5~PN order.

\subsubsection{Spinning Binary Mergers and Spin Flips}

The merger of a binary consisting of nonspinning black holes produces
a single spinning black hole with spin direction aligned with the
orbital angular momentum $\Jorb$ of the binary.  In this case, the
spin of the final hole arises purely from $\Jorb$.  When the
individual black holes have spins that are not aligned with $\Jorb$,
the spin of the final hole will, in general, not be aligned with the
initial spin but rather will have a different direction.  This
phenomenon, in which the spin direction of the final black hole
differs significantly from the spin directions of the individual holes
prior to merger, is known as a ``spin flip.''  The simplest situation
in which a spin flip can occur is a binary with both spins
anti-aligned with $\Jorb$.

\begin{figure}
\includegraphics*[scale=0.8,angle=0]{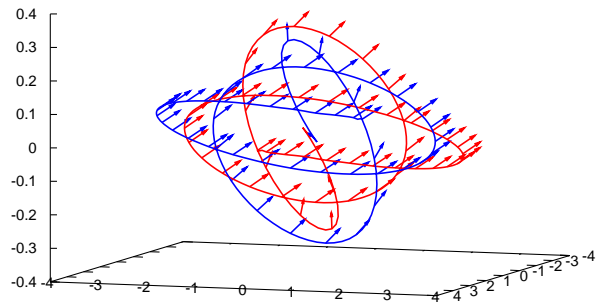}
\caption{Black hole trajectories for the spin flip case studied by
  \citet{Campanelli:2006fy}. The black holes have equal masses, and
  equal spins initially pointing $\degrees{45}$ above the initial
  orbital plane.  The spin directions are shown as arrows along each
  trajectory every $4M$ until merger.  }
\label{fig:RIT-3d-flip-tracks}
\end{figure}

\begin{figure}
\includegraphics*[scale=0.8,angle=0]{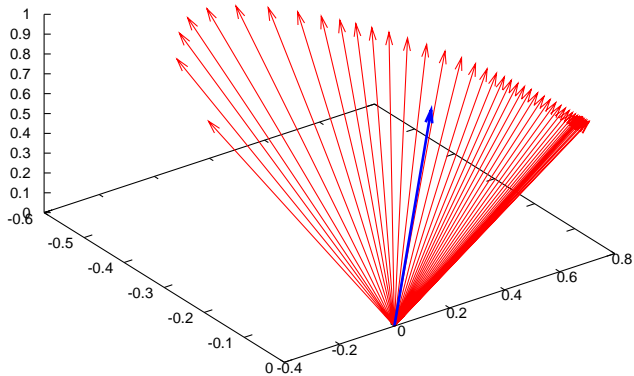}
\caption{The spin flip produced by the evolution in
  Fig.~\ref{fig:RIT-3d-flip-tracks}, from \citet{Campanelli:2006fy}.
  The spin direction of the component holes, shown by the red arrows
  plotted every $4M$ until merger, varies continuously due to
  precession.  The blue arrow shows the final black-hole spin, with
  direction flipped discontinuously from that of the component holes
  just prior to merger.}
\label{fig:RIT-spin-flip}
\end{figure}

The case of generic binaries, with unequal masses and
arbitrarily-oriented spins has a much larger parameter space.  The RIT
group \cite{Campanelli:2006fy,Campanelli:2007ew} has examined spin
flips for several such cases.  Figure~\ref{fig:RIT-3d-flip-tracks}
shows the puncture trajectories for an equal-mass binary with equal
spins having magnitude $\ahat_{1,2} = 0.5007$ and initially pointing
$\degrees{45}$ above the initial orbital plane.  The black holes
execute $\sim 2.25$ orbits before merger, during which time the spins,
shown as arrows along each trajectory, precess by $\sim
\degrees{151}$.  The evolution of the spin direction for this case is
shown in Fig.~\ref{fig:RIT-spin-flip}. The final black hole has
$\ahat_f \sim 0.8$ with a spin direction flipped by $\sim
\degrees{35}$ with respect to the component spins just before merger.

In a related phenomenon, the direction of the \emph{total} angular
momentum ($\Jorb + \vec{S}_1 + \vec{S}_1)$) may change.  This case was
studied using general principles and extrapolations from
point-particle results by \citet{Buonanno:2007sv}, who predicted that
binaries with $q > 6.78$ will experience a flip in total angular
momentum direction provided that the initial spins are equal and
anti-aligned with $\Jorb$, and that the individual black hole spins
obey $\ahat_{1,2} \ge 0.5$.  In particular, they predicted that, for $q
= 4$, a Schwarzschild black hole should form if $\ahat_1 = \ahat_2 =
0.8$.

Motivated by this work, \citet{Berti:2007nw} simulated a series of
binaries with $q = 4$ and having equal anti-aligned spins in the range
$\ahat_{1,2} \in [-0.75,-0.87]$.  They found that a Schwarzschild hole
is formed when the initial spin is $\ahat_{1,2} \simeq -0.842 \pm
0.003$.


\subsubsection{Kicks from Mergers of Spinning Black Holes}
\label{sec:GWs:kicks}

The mergers of asymmetrical spinning binaries will produce recoiling
remnant black holes.  In the simplest case, the binary components have
equal masses and spin magnitudes, and the asymmetry arises from the
spin vectors pointing in opposite directions.  In early 2007, several
new results appeared in quick succession, generating considerable
excitement among numerical relativists as well as astrophysicists
eager to apply these results.  In this section we discuss the basic
results and defer to Sec.~\ref{sec:astro} their impact on
astrophysics.

\citet{Herrmann:2007ac} simulated the merger of equal-mass black holes
with equal spin magnitudes $\ahat_{1,2} \in (0.2,0.4,0.6,0.8)$ and
having one spin vector aligned and the other anti-aligned with the
orbital angular momentum.  These mergers produced kicks directed
purely in the orbital plane with magnitudes $\sim 475 \ahat_{1,2}
\kms$. Shortly thereafter \citet{Koppitz:2007ev} showed that,
for the specific case $\ahat_{1,2} = 0.584$, $v_{\rm kick}= 257 \pm 15
\kms$; linear extrapolation of results to the maximally spinning case
$\ahat_{1,2} = 1$ yields $v_{\rm kick} \simeq 440 \kms$.
\citet{Pollney:2007ss} carried out a systematic study of equal-mass
mergers with spin magnitude $\ahat_2 = 0.584$ and direction aligned
with the orbital angular momentum.  The other black hole has spin
magnitude $\ahat_1 \in {(0, 0.25, 0.50, 0.75, 1.0) \ahat_2}$, and
direction both aligned and anti-aligned.  They found a maximum kick
velocity $v_{\rm kick} = 448 \pm 5 \kms$.

These results all demonstrated that mergers of spinning black
holes can produce significantly larger kick velocities than
nonspinning mergers.  However, new results would soon show that kicks
from the mergers of spinning black holes could get much larger indeed.

The idea of ``superkicks'' was first discussed by the RIT
group \cite{Campanelli:2007ew}.  They observed that a PN treatment
\cite{Kidder:1995zr} predicts that the recoil due to spin is maximized
when $M_1 = M_2$, $\ahat_1 = \ahat_2$, and the spin directions are
anti-aligned with each other and lying in the orbital plane.

\begin{figure}
\includegraphics*[scale=0.5,angle=0]{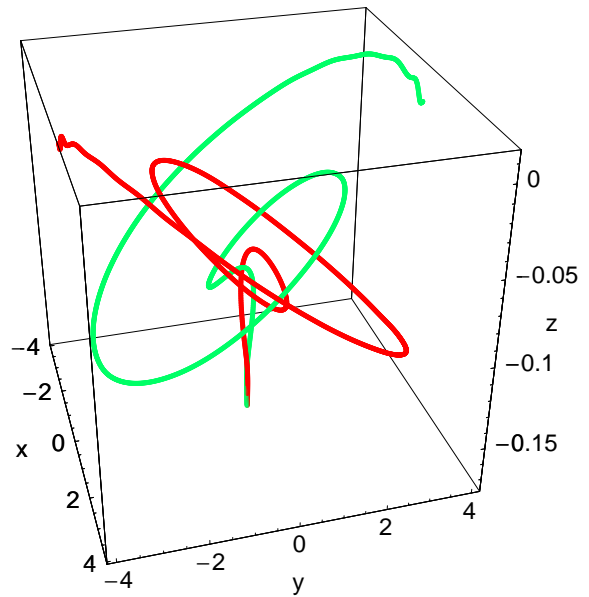}
\caption{Puncture trajectories from a superkick simulation carried out
  by \citet{Gonzalez:2007hi}.  The black holes have equal masses and
  spins $\ahat_{1,2} \sim 0.8$, initially oppositely directed in the
  orbital plane.  During the evolution, the black holes move out of
  the original plane, and the final black hole recoils with velocity
  $v_{\rm kick} \sim 2500 \kms$ in the negative $z$-direction.  }
\label{fig:Jena-kick}
\end{figure}

The Jena group first simulated binaries in this configuration and
demonstrated the phenomenon of superkicks. Using $\ahat_{1,2} \sim
0.8$, they found a resulting kick velocity $v_{\rm kick} \sim 2500
\kms$ \cite{Gonzalez:2007hi}.  Figure~\ref{fig:Jena-kick} shows the
coordinate positions of the black-hole punctures from one of their
simulations.  Notice that the trajectories move out of the initial
orbital plane and that the final black hole is kicked in the $-z$
direction.  The RIT group \cite{Campanelli:2007cg} carried out similar
simulations with $\ahat_{1,2} = 0.5$.  Their results show kicks
perpendicular to the orbital plane with magnitudes up to $v_{\rm kick}
\sim 1800 \kms$; using the expression from \citet{Kidder:1995zr} they
predict a maximum recoil $v_{\rm kick} \sim 4000 \kms$ for the case of
maximally spinning black holes, $\ahat_{1,2} = 1$.

\citet{Schnittman:2007ij} performed a multipolar analysis of recoil
from black-hole mergers, for both unequal masses and non-zero,
non-precessing spins. They found that multipole moments up to and
including $\ell = 4$ are sufficient to accurately reproduce the final
recoil velocity (within $\simeq 2\%$), and that only a few dominant
modes contribute significantly to it (within $\simeq 5\%$).
\citet{Brugmann:2007zj} studied the role of spin in producing
superkicks.  They showed that the recoil velocity is almost entirely
due to the asymmetry between the $(\ell = 2, m=+2)$ and $(\ell = 2, m
= -2)$ modes of $\Psi_4$.  In addition, the major contribution to the
recoil occurs within a period $\sim 30M$ before and after merger,
after the time at which a standard PN treatment of the evolution
breaks down.

\section{Interaction of Numerical Relativity with Post-Newtonian Theory}
\label{sec:nrpn}

Einstein's theory of general relativity implies predictions for the
dynamics and gravitational-wave generation for generic black-hole
binaries at all stages of the merger process.  As we have seen,
numerical relativity now provides an excellent tool for concretely
deriving these predictions beginning in the last orbits and continuing
through the conclusion of the coalescence in the ringdown of the
remnant black hole.  There is no hard limit to how early numerical
relativity can be applied in these simulations, but the considerable
costs of these simulations tends to limit the duration to tens of
orbital periods.  Though the PN expansion provides only an
approximation to general relativity, it succeeds in making accurate
and efficient predictions over very long time-scales, while the black
holes remain well separated. Late in the merger, however, as the
black-hole velocities increase, these errors grow, and PN theory is no
longer reliable

A full understanding of general relativity's predictions for the
merger process must draw on both approaches.  The maturation of
numerical relativity has fueled efforts toward synthesizing a complete
understanding of black-hole mergers built from the results and
techniques of both these fields.  Even a coarse general review of the
PN formalism and results is outside the scope of this presentation.
Decades of work in this area have already been covered
\cite{Blanchet:2009rw,Schaefer:2009dq,Damour:2009ic,Blanchet:2008je,Blanchet:2006LR}.
Here we focus exclusively on key synergistic areas where fruitful
research interactions drawing from both numerical relativity and PN
results and formalisms are yielding a more complete
general-relativistic understanding of black-hole binary systems.  The
broad range of research that interfaces these two approaches can
be grouped into four categories, based on the different ways that PN
theory impacts numerical-relativity research. PN theory: 1) provides
independent results for cross-checking and comparing with NR; 2)
provides models for the initial values needed to begin astrophysically
realistic NR simulations; 3) provides insight for interpreting NR
results; and 4) may provide the basis for empirical models
representing the combined knowledge drawn from PN and NR
investigations.

\subsection{Independent Post-Newtonian Dynamics and Waveforms}
\label{sec:nrpn:independent}

As mentioned, PN theory is based on an approximate expansion
of Einstein's equations in powers of velocity
$\epsilon^n\sim(v/c)^{2n}$ providing general relativistic corrections
to the Newtonian, small-velocity limit.  This approach has a long
history of success as the primary framework for deriving explicit
predictions from general relativity for physical and astrophysical
observations. The relatively small velocities involved in most
observations have made low-order PN theory an ideal tool for testing
general relativity as the standard model for gravitational physics in
many contexts including Earth-orbit experiments, solar-system
dynamical observations, and in even precision pulsar timing
observations of binary neutron star systems \cite{Will:2006LR}.

Gravitational-wave researchers have applied PN theory to represent
general relativity-based signal expectations for the vast majority of
gravitational-wave searches for anticipated observations of black-hole
and neutron-star binary inspirals, and likewise in the development and
planning for current and future gravitational-wave instruments.  The
early epoch of black-hole binary inspiral is also well-described by
low-order PN theory.  For near-circular inspirals,
$\epsilon\sim(v/c)^2\sim M/R$, where the binary separation $R$ is
scaled by the total system mass $M$.  At large $R$, the PN expansion
provides excellent predictions, even at low order.  As the black holes
lose energy and sink closer together, the velocity grows, requiring
higher-order PN terms for sufficiently accurate predictions.
Currently, PN predictions are available up to 3.5~PN order (2.5~PN for
spin-terms).  These higher-order expansions seem to provide PN
predictions sufficiently accurate for the analysis of data from
current instruments for all but the last several orbits.  Once the
separation approaches, say, $R/M\sim 10$, the accuracy of the PN
expansion diminishes.  Even at (hypothetical) arbitrarily high orders,
the expansion may fail, if the expansion parameter $v/c$ exceeds the
series' radius of convergence.  Precisely when the PN approximation
effectively fails depends on the details of the system being studied
and the requisite accuracy, but generally, for the last orbits and
merger PN theory is no longer reliable.

The strongest gravitational radiation is generated in the late stages
of inspiral or merger where the internal consistency of the PN
approach has been, at best, difficult to assess
\cite{Simone:1996db}. Numerical simulation can treat the late portions
of the mergers, with practical resource limitations on the duration of
the simulations, and consequently how far apart the black holes can be
at the start of the simulation.  Would it be practical, however, to
run numerical simulations long enough to ``overlap'' with the part of
the problem treated successfully by PN methods, or would there be an
intermediate region requiring yet another approach
\cite{Brady:1998du}?

The first numerical inspiral results were quickly compared against PN
calculations \cite{Baker:2006yw,Buonanno:2006ui}.  These first
comparisons yielded promising indications that the gap between NR and
PN could be bridged, but also made clear that PN results were not
uniquely determined for the final part of the inspiral.  Numerical
results can be verified by internal consistency studies, examining,
for instance, the convergence of the results toward consistency with
Einstein's equations as the resolution is increased.  External
verification, however, would strengthen the case that these new
results are indeed correct.  Even more importantly, these comparisons
would allow an independent check of the late inspiral PN predictions
which are not easily confirmed by self-consistency studies.

\begin{figure}
\includegraphics*[scale=.28, angle=0]{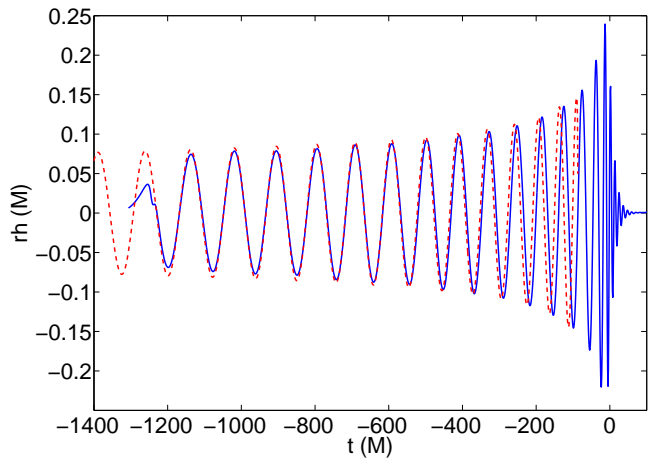}
\caption{Comparison of NR and PN waveforms from \citet{Baker:2006ha}
  provided mutual affirmation of the results from each approach and
  showed that in combination NR and PN results could treat the
  complete signal.  }
\label{fig:PNPRLwf}
\end{figure}

Roughly one year after the first robust numerical-relativity results,
simulations lasting $\sim1000M$ were conducted, and quantitatively
compared with various PN approximations.  Quantitatively
cross-checking waveform comparisons requires some care.  How do the
differences between PN and numerical results compare with intrinsic
numerical error estimate?  How do different variants of PN predictions
compare with the numerical results?  Because each waveform comes with
no meaningful absolute reference in time and phase, how can the
freedom to offset these parameters be controlled for the comparison?
Effective comparisons also require longer waveforms than those
produced in the earliest simulations.  The first comparison addressing
these issues came at the end of 2006 \cite{Baker:2006ha}.  The study
considered the case of equal-mass non-spinning black-hole mergers,
comparing a 3.5~PN waveform with numerical results covering the last
seven orbits.  The compared waveforms are shown in
Fig.~\ref{fig:PNPRLwf}.  In the comparison the waveform phases agreed
within 1 rad of phase drift, for a little over ten gravitational-wave
cycles preceding the last orbit before merger, comparable to
numerical error estimates.  This result gave clear indication that PN
waveforms could be accurate in the last orbits approaching merger, and
that PN and NR, combined, could treat the complete waveform signal.

Not all equally valid variants (approximants) of the approximate PN
waveforms agree this well with numerical results.  To understand this,
it is worthwhile to consider PN results in a little more detail.  The
PN approximation is formally understood as an expansion in powers of
the speed of the merging black holes, $v/c$.  A concrete result of the
theory will typically be a Taylor expansion for a specific dependent
variable, in terms of a chosen independent variable.  The obvious
choice, to express the waveform itself (gravitational-wave strain) as
a function of time, would give a poor result since the sinusoidal
shape of the waveform is difficult to approximate by a polynomial.  PN
waveform results are typically expressed as separate expansions for
the orbital phase, and polarization component amplitudes.  These are
given as expansions in the time to merger $v/c\propto t^{-1/8}$ or
orbital frequency $v/c\propto\Omega^{1/3}$.  The orbital phase
information may also be expressed by an expansion for $\dot\Omega$,
referred to as the chirp-rate; see \citet{Blanchet:2006LR} for a
review of the approach and a particular explicit waveform expansion.
The PN information may also be encoded in an Hamiltonian formulation
description of the dynamics, which may be integrated
numerically. Researchers also choose between Taylor series and other
``resummed'' expressions of the results based, for instance, on Pad\'e
approximants \cite{Damour:1997ub}.

In the comparisons described above, the PN waveform phasing was found
to agree with numerical results derived from a Taylor series expansion
for chirprate $\dot\Omega(\Omega)$ expanded in powers of frequency.
This is known as the TaylorT4 approximant in the language of the
ground-based gravitational-wave community. Later studies with longer
and more accurate numerical simulations provided more precise
tests of phases and amplitudes for a variety of PN approximants
\cite{Hannam:2007ik,Boyle:2007ft}.  \citet{Boyle:2007ft} found even
closer agreement (than found in \cite{Baker:2006ha}) with the TaylorT4
approximant, to within 0.05 rad over nearly 30 cycles before the
orbital frequency reaches $M \Omega = 0.1$ at roughly an orbit before
merger.

\begin{figure}
\includegraphics*[scale=.48, angle=0]{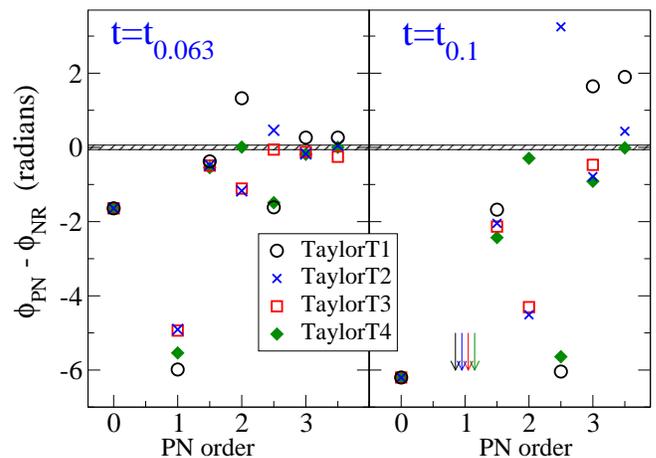}
\caption{Phase differences between a numerical simulation and various
  PN models from \citet{Boyle:2007ft}. The left panel shows phase
  differences from an initial (waveform) angular frequency of
  $M \omega=0.04$ up to $M \omega=0.063$, while the comparison in right
  panel extends this range to $M \omega=0.10$.}
\label{fig:BoylePNComp}
\end{figure}
Figure~\ref{fig:BoylePNComp} shows the results of four PN approximants
with terms kept to various PN orders.  The excellent agreement of the
TaylorT4 approximant with numerical results is not matched by other
equally consistent PN waveform approximants in the late portions of
the waveforms. However, all four approximants agree to within about a
radian if the comparison is cut off at waveform frequency $M \omega =
0.1$, about five orbits before merger [note that this is the frequency
of the dominant $(2,2)$ mode, twice the orbital frequency $M \Omega$].

While most attention has so far been given to the nonspinning cases,
some results exist for other regions of parameter space.  For
equal-mass configurations of black holes with spins aligned with the
orbital angular momentum, the phase agreement of the TaylorT4
approximant is not generally best, and phase differences (over ten
cycles preceding $M \omega = 0.1$) are much larger than those seen in
the nonspinning case, with differences near 1 rad or more for
spins $\ahat_{1,2} \ge 0.5$ \cite{Hannam:2007wf}.  PN spin effects are
not yet derived beyond 2.5~PN order, possibly limiting the accuracy of
the PN waveforms in these comparisons.  Some PN comparisons have also
been examined for other configurations, including a precessing system
\cite{Campanelli:2008nk} and for eccentric configurations of
equal-mass, non-spinning mergers \cite{Hinder:2008kv}.

In comparing Fourier-domain PN approximants with numerical results for
a set of non-spinning mergers with various mass ratios,
\citet{Pan:2007nw} discovered that the abrupt truncation of the
Fourier-domain waveforms, inducing Gibbs oscillations in the
time-domain, can approximately emulate the physical quasinormal
ringdown that terminates the numerical-simulation waveforms.  Even
better agreement is possible if $\eta$, the symmetric mass ratio
\eqref{eq:eta_def}, is allowed to extend beyond its physical range (to
$\eta>0.25$) or if an extra pseudo-4PN order term is added.  More
recent studies have confirmed that these simple models are directly
useful in some gravitational-wave observation applications
\cite{Boyle:2009dg}.  These results have motivated further development
of phenomenological full-waveform models.

\subsection{Analytic Full-Waveform Models}
\label{sec:nrpn:analytic}

For the early part of the waveforms various PN approximants agree at
3.5~PN order, the best currently available, providing excellent
approximations.  At late times, the results of different approximants
diverge, and the numerical-simulation results are the only way to
accurately derive the predictions of general relativity.  Can an
analytic waveform description be developed that simultaneously encodes
the results of both complementary theoretical approaches?  The
generally simple features of the numerically simulated merger
waveforms raise hopes that fairly accurate approximate analytic
waveform descriptions may be produced with simple dependence on the
binary system parameters, providing an efficient means also of
interpolating from a sparse sampling of the full parameter space for
which numerical-simulation studies are completed.  These waveform
models would be computationally efficient (compared to numerical
simulations) and may have applications in a broad class of
observational data analysis applications.

Ajith and collaborators have developed a Fourier-domain full-waveform
model phenomenological approach resembling the treatment of 
\citet{Pan:2007nw}, but with greater emphasis on matching the
numerical waveforms \cite{Ajith:2007kx,Ajith:2007qp,Ajith:2007xh}.
They begin combining information from the PN and NR-based waveforms by
first stitching together time-series data for the dominant
($\ell=2$,$m=2$) component waveforms, including a long PN precursor
joined to a numerical merger waveform.  Fourier transforms of
waveforms are then fit to a parametrized model of the waveform
similar to waveform families previously applied in phenomenological
treatments of purely PN waveforms \cite{Buonanno:2002ft,Arun:2006yw}.
Their waveform model for mergers of non-spinning binaries has the
form
\begin{equation}
h(f) \equiv 
\frac{C(M,\eta)}{d}
{\cal A}_{\rm eff}(f) \, 
	{\mathrm e}^{{\mathrm i}\Psi_{\rm eff}(f)},  
\end{equation}
where $C$ is a constant related to the masses, and $d$ is the distance
to the source.  The effective Fourier amplitude ${\cal A_{\rm eff}}$
is modeled piecewise,
\begin{equation}
{\cal A_{\rm eff}}(f) \equiv 
f_{\rm m}^{-7/6}
\left\{ \begin{array}{ll}
\left(f/f_{\rm m}\right)^{-7/6} &, f < f_{\rm m}\\\\
\left(f/f_{\rm m}\right)^{-2/3} &, f_{\rm m} \leq f < f_{\rm r}\\\\
\frac{\pi \sigma}{2} \left(\frac{f_{\rm r}}{f_{\rm m}}\right)^{-2/3}
 \, {\cal L}(f,f_{\rm r},\sigma) &, f_{\rm r} \leq f
\end{array} \right. ,
\end{equation}
with distinct power-law segments before and after a merger frequency
$f_{\rm m}$, and a Lorentzian $ {\cal L}(f,f_{\rm r},\sigma)$ decay
beyond the transition frequency $f_{\rm r}$ demarking the ringdown of
the final black hole.  The Fourier phase is represented by a single
power-law expansion
\begin{equation}
\Psi_{\rm eff}(f)  =  2 \pi f t_0 + \phi_0 + \sum_{k=0}^7\psi_k \, f^{(k-5)/3}\,.
\end{equation}
For each waveform, the free amplitude and phase coefficients are
determined by a fit to the stitched NR-PN hybrid waveforms. It was
found that the mass-ratio parameter dependence could be modeled by
simple fits to quadratic functions of the symmetric mass ratio
\eqref{eq:eta_def}.  These waveforms have been applied in
gravitational-wave data analysis studies (see
Sec.~\ref{sec:gwda:direct};\cite{Ajith:2009fz}).  Recently
\citet{Ajith:2009bn} have developed a generalization of this model for
non-precessing system of spinning black holes.

Stitching together PN and NR waveforms can be avoided, and closer
contact with the basic physics maintained, by requiring direct PN
consistency in the full waveform model.  Some PN approximants can
resemble the numerical late-merger results.  If such approximants can
be tuned to agree with numerical results right up to merger by careful
choice of adjustable parameters which only affect PN consistency
beyond known PN order, the result would simultaneously encode PN and
numerical-relativity results.  The effective-one-body (EOB) family of
PN Hamiltonian models has promise as a tunable model for encoding both
PN and NR results
\cite{Buonanno:1998gg,Buonanno:2000ef,Damour:2008yg,Damour:2009ic}.
In the time domain, techniques have also been developed for extending
these waveforms into the ringdown of the final black hole. Analysis of
the early numerical results in comparison with an untuned EOB model
gave promising indications that the waveforms could be closely
approximated this way \cite{Buonanno:2006ui}.  With tuning, it appears
that this construction can provide an analytic but potentially very
accurate, approximation to the complete coalescence waveform.

In the EOB model, the binary motion is recast as the motion of a
single {\em effective} body of mass $\mu = M_1 M_2/(M_1+M_2)$ moving
about a central potential, as is familiar from Newtonian mechanics.
In the general relativistic version of this framework, the effective
body's motion follows a geodesic (to 2~PN order) around a modified
version of a Schwarzschild metric.  The motion of the effective body
is described by an effective Hamiltonian, which, for systems of
nonspinning black holes, may take the form
\begin{widetext}
\begin{eqnarray}
\label{eq:genexp}
H_{\rm eff}(\mathbf{r},\mathbf{p}) = \mu\,\sqrt{A (r) \left[ 1 + 
{\mathbf p}^{2} +
\left( \frac{A(r)}{D(r)} - 1 \right) ({\mathbf n} \cdot {\mathbf p})^2
+ Q(r) ({\mathbf n} \cdot {\mathbf p})^4 \right]} \,.
\end{eqnarray}
\end{widetext}
The expressions for $A(r), D(r)$ and $Q(r)$ are chosen so that the
PN-expansion of the Hamiltonian is consistent with the results from PN
theory to known order (3~PN for nonspinning black holes)
\cite{Buonanno:1998gg,Buonanno:2000ef,Damour:2000we}.  The
nonconservative contribution to the motion, arising from the loss of
angular momentum to gravitational radiation, is encoded in an
additional flux term entering as an external force in Hamilton's
equations, which is also constrained to be consistent with PN theory
(typically to 3.5PN order).  Hamilton's equations are integrated to
derive the effective body's motion, and hence the black-hole
trajectories.  Waveforms are constructed using PN extension of the
quadrupole formula relating the motion of the black holes to the
amplitude and phase of the multipolar radiation components. The last
part of the radiation, arising after merger, is completed by
continuously matching a sum of quasinormal ringdown modes to the
waveforms truncated near the point of merger.

\begin{figure}
\includegraphics[width=0.45\textwidth]{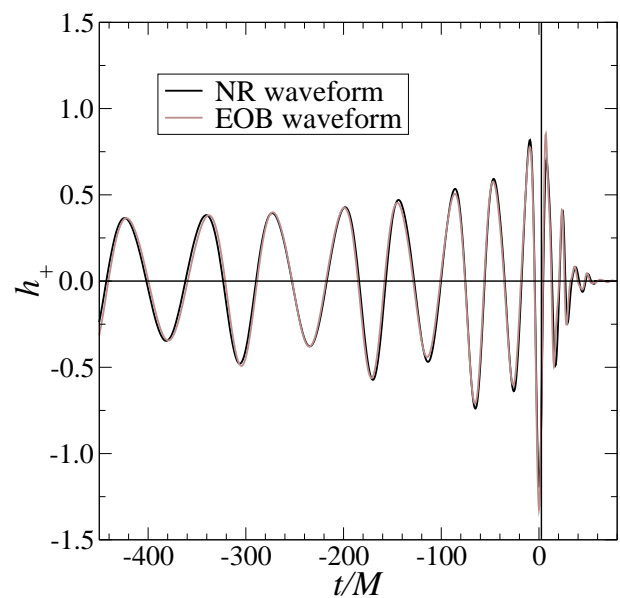}
\caption{Comparison of EOB-model and numerical-relativity waveforms from
\citet{Buonanno:2007pf} for the $h_+$ component of the gravitational-wave
strain from the merger of binary with mass ratio 4:1. The waveform is for an
observer located an at inclination angle $\theta = \pi/3$ from the axis of rotation. 
\label{fig:Faithful}}
\end{figure}

While consistent with PN theory, the formalism can be adjusted to also
match the wave forms derived from numerical-relativity
simulations. The flux function, as well as $A(r)$ and $D(r)$, can be
adjusted by the addition of higher-order PN terms to modify the
strong-field dynamics to agree with NR without violating PN
consistency.  An early implementation of this approach, encoding the
combined results PN and NR for nonspinning black-hole mergers was
presented by \citet{Buonanno:2007pf}. Figure~\ref{fig:Faithful} shows
their result comparing numerical and adjusted EOB-model waveforms for
a 4:1 mass-ratio merger.
Subsequent work involving more accurate numerical simulations, and
more careful tuning of the EOB model has improved EOB model waveforms
\cite{Damour:2007yf,Damour:2007vq,Boyle:2008ge}.  Recent comparisons
of improved EOB models with high-accuracy numerical results from
\citet{Scheel:2008rj} yield differences comparable to numerical errors
showing phase agreement within about $0.01$ rad through $\sim$ 30
gravitational-wave cycles \cite{Damour:2009kr,Buonanno:2009qa}.
\begin{figure*}[t]
\begin{center}
\includegraphics[width=120 mm, height=45mm]{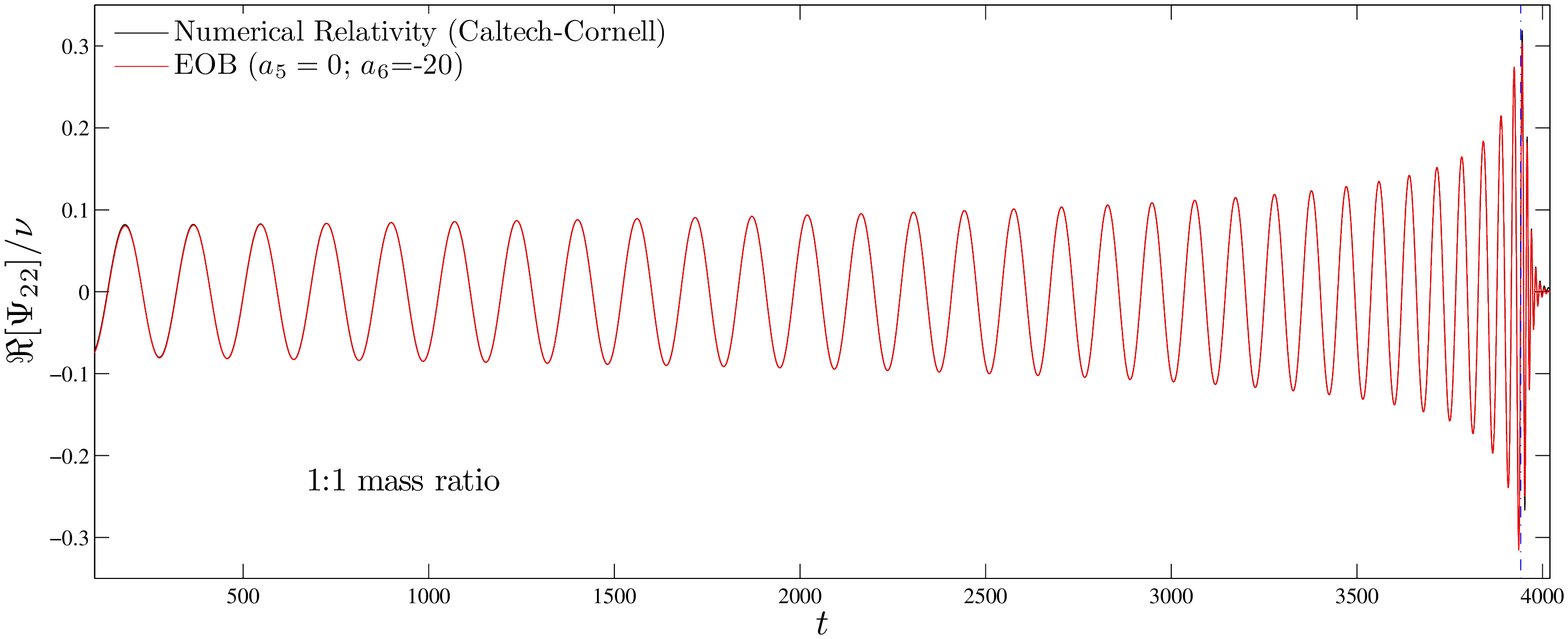}
\hspace{-12 mm}
\includegraphics[width=60 mm, height=45mm]{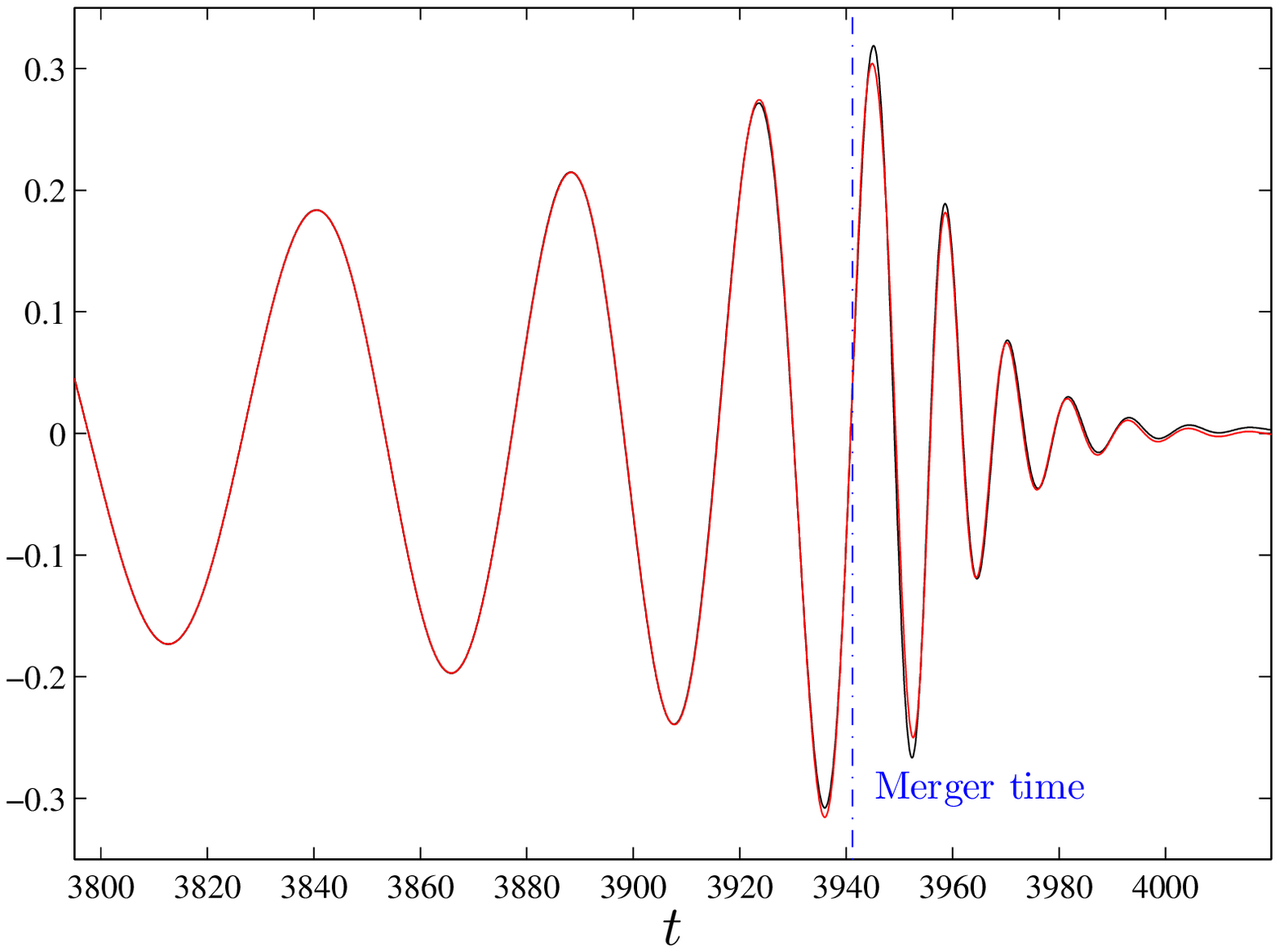}
\caption{Agreement of refined EOB-model $\ell=m=2$ waveform
  \cite{Damour:2009kr} (red) with numerical result from
  Ref.~\cite{Scheel:2008rj} for equal-mass non-spinning merger
  (black).  Slight differences near the merger time are difficult to
  perceive without color.}
\label{fig:DamourNagarEOB}
\end{center}
\hspace{-5mm}
\end{figure*}

Recently the EOB approach has been extended to match cases of
non-precessing spinning black-hole mergers\cite{Pan:2009wj}.  There
remain open questions in extending these waveform models to generic
spin configurations, more extreme mass ratios, and full multipolar
content.

\subsection{Post-Newtonian Models for Numerical Initial Data.}
\label{sec:nrpn:initialdata}

Before a numerical simulation can commence, a numerical relativist
requires a model for the initial configuration of the two black holes
and the geometric fields which represent them.  As discussed in
Sec.~\ref{sec:comp:initialdata} this involves not only solving the
general relativistic constraint equations, but also providing some
ansatz for the free data.  In some cases, some information from PN
theory may be applied to produce an initial data model that more
precisely represents the desired physical configuration.

To understand how PN information can be applied, it will be useful to
revisit the discussion of numerical initial data.  Typical
simulations begin with a modeling ansatz, the Brandt-Br\"ugmann model,
for instance, which produces initial {\em field} data from a given set
of particle-like parameters related to the black holes' masses, spins,
initial positions, and momenta.  Usually, the initial data models make
no attempt to represent the gravitational radiation fields which would
have been previously generated by the motion of the black holes, and
do not utilize PN information in this reduction of the field degrees
of freedom to particle parameters.  Information from PN theory is
frequently applied in choosing the specific particle parameters which
correspond to the sought-after simulation, particularly for circular
inspiral configurations.  Work has also begun toward making richer use
of PN information for improving the field ansatz.

Most of the simulations discussed have been designed to
represent black holes in circularized orbits.  Before long-lasting
numerical-simulation results were available, comparative studies of
numerical initial data sets with PN-derived information provided a
gauge of the results [see for instance
\citet{Baker:2002qf,Cook:2004kt,Caudill:2006hw}]. Without evolutions,
these studies focused on theoretical properties of the black-hole
configuration space such as the ISCO (see
Sec.~\ref{sssec:astro_grav:basic:bh}).

Still subtle imbalances in the initial data, such as excess angular
momentum for the chosen separation, can lead to eccentricity in the
simulation which impacts the simulated radiation waveforms
\cite{Boyle:2008ge,Pfeiffer:2007yz}.  The residual eccentricity can be
reduced via a straightforward iterative procedure
\cite{Pfeiffer:2007yz}, but this requires repeated simulations lasting
several orbits; expensive in both time and computational resources.
To minimize this eccentricity without resorting to iteration, several
groups use trajectory information from PN theory in setting up the
parameters for the numerical initial data.  This approach enables
simulations with eccentricities $e<0.002$ for equal-mass non-spinning
mergers with initial separations $d_{\rm init}>10M$
\cite{Husa:2007rh}.  The technique is helpful in simulations with
unequal masses and nonvanishing spins as well, though the residual
eccentricity is generally larger \cite{Walther:2009ng}.

Work is underway on techniques allowing the richer application of PN
information from numerical initial data modeling.  The common
assumption of a conformally flat spatial metric in numerical initial
data, disagrees with PN results at the 2~PN order
\cite{Damour:2000we}, making this a likely leading source of modeling
error which produces spurious initial transients in the
simulations. Though not yet as well-developed as the widely applied
models, an alternative approach applies metric information from PN
theory for a non-trivial initial conformal metric
\cite{Ohta:1974kp,Schaefer85,Jaranowski:1997ky}.  With these
techniques it is also possible to encode in the initial data
information about the prior radiation generated by the system before
the ``initial'' time of the simulation
\cite{Tichy:2002ec,Kelly:2007uc}.
 
\subsection{Post-Newtonian Theory for Interpretation of Numerical Results.}
\label{sec:nrpn:interp}

We have noted several specific areas where research in numerical simulations
makes contact with PN theory.  These alone do not provide a full picture of
the interplay between the two theoretical approaches.  Most broadly, PN 
theory provides a foundation for interpreting
numerical results.  

Numerical relativists draw widely from PN-based background knowledge
of the black-hole binary.  Many important phenomena, such
as the ISCO, spin-orbit coupling, spin precession, and `gravitational
rocket' kicks were first studied in the PN approximation, providing a
foundation for subsequent numerical studies.  Small examples can be
noted throughout this review. In particular we point to the value of
PN results in interpreting mergers of spinning black holes
(Sec.~\ref{sec:GWs:spinning}) and kicks
(Sec.~\ref{sec:GWs:uneq}). Analytic formulas for approximately
expressing the final state of the black hole, mass, spin, and momentum
have also drawn heavily from insights based on the PN treatment (see
Sec.~\ref{sec:astro:recoil_formulae}.)

\section{Applications to Gravitational Wave Data Analysis}
\label{sec:gwda}

\begin{figure}
\includegraphics*[width=3.0in, angle=0]{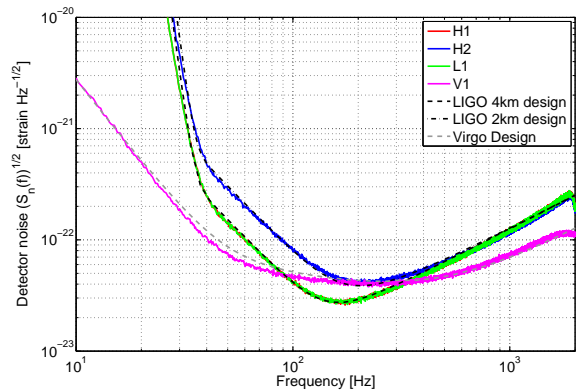}
\caption{Design sensitivity curves for the LIGO \& Virgo detectors
  (dashed lines), as well as the approximate curves used for the NINJA
  project (solid lines). LIGO sensitivity is effectively zero below
  $\sim 30\Hz$, while Virgo does a little better (note that achieved
  sensitivities may not perfectly mirror the design curves).  Taken
  from \citet{Aylott:2009ya}. Reproduced by permission of the Institute of Physics.}
\label{fig:LIGO_Virgo_noise}
\end{figure}

Several detectors are active or in their planning stages to detect the
gravitational-wave signals from astrophysical processes.  Prominent
among these are the ground-based interferometric detectors -- LIGO
(U.S.A.)  \cite{Abbott:2007kv}, GEO (Germany), Virgo (Italy), TAMA
(Japan), and AIGO (Australia) -- sensitive to frequencies in the range
$10^1 - 10^3 \Hz$, as well as next-generation instruments such as the
Einstein Telescope \cite{Freise:2008dk}. Also planned to launch at the
end of the next decade is the Laser Interferometer Space Antenna
(LISA), with complementary frequency sensitivities between $10^{-4} -
10^{-1} \Hz$. All of these instruments are subject to a variety of
noise sources; in the ground-based detectors, these sources will
completely overwhelm the signal, unless filtered
intelligently. Figure~\ref{fig:LIGO_Virgo_noise} shows the design
sensitivity curve resulting from these disparate noise sources, for
the LIGO and Virgo detectors.

Burst analysis can extract many signals, but optimal results in
gravitational-wave data analysis require \emph{matched filtering} of
the noisy input signal
\cite{Wainstein_L:1962,Flanagan:1997sx}. Fundamental to this approach
is some measure of overlap between a physical signal $h_1(t)$ and
filtering template waveform $h_2(t)$. For this it is convenient to use
a frequency-space inner product $\langle \cdot | \cdot \rangle$
between signals, defined as \cite{Cutler:1994ys}
\begin{eqnarray}
\langle h_1 | h_2 \rangle &\equiv& 2 \int_{0}^{\infty} \frac{\tilde{h}_1(f) \tilde{h}_2(f)^* + \tilde{h}_1(f)^* \tilde{h}_2(f)}{S_n(f)} df \nonumber \\
                          &=& 4 {\rm Re} \left[ \int_{0}^{\infty} \frac{\tilde{h}_1(f) \tilde{h}_2(f)^*}{S_n(f)} df\right],
\label{eq:inner_product_DA}
\end{eqnarray}
where $\tilde{h}_1(f)$ and $\tilde{h}_2(f)$ are the Fourier transforms
of the signals, and $S_n(f)$ is the (one-sided) noise power spectral
density of the detector we are interested in.  This can be normalized
to produce the \emph{match} (or \emph{overlap}) \cite{Owen:1995tm}
between two waveforms.

To use \eqref{eq:inner_product_DA} requires the availability of a set
of waveform \emph{templates} -- simple, few-parameter analytic model
waveforms to filter against.  Even for a relatively simple system such as 
a black-hole binary, the combination of possible source and detector
configurations leads to a 17-dimensional parameter space\footnote{It
  is easy to see that there are a total of 17 dimensions. Generally,
  the instantaneous state of each black hole has ten degrees of freedom
  for its mass, position, momentum and spin. Of the 20 degrees of
  freedom for two black holes, the three related to the center-of-mass
  momentum are unmeasurable by distant gravitational-wave
  observations.  The peculiar motion has negligible effect, and the
  proper motion results in a Doppler shift indistinguishable from a
  change in the total mass. Also note that changes in eight more degrees
  of freedom, corresponding to the center-of-mass position in
  spacetime, spatial orientation and total mass have trivial effects
  which can be treated analytically, leaving, in full generality, a
  space of nine parameters that must be covered by simulations.}.  Some
of these parameters are intrinsic to the source: total binary mass $M
= M_1+M_2$, symmetric mass ratio $\eta$ \eqref{eq:eta_def}, spins
$\vec{S}_{1,2}$, time to coalescence $t_c$, eccentricity $e$, and
eccentric phase $\phi_e$.  The remaining parameters have to do with
the relation between source and detector: (luminosity) distance to
source $D_L$, inclination $\iota$, orbital phase $\phi$, waveform
polarization $\psi$, and position of the binary on the detector's sky
$\{\Theta,\Phi\}$.  Restricting consideration to binaries that have
circularized by the time they enter the detector's window
\cite{Peters:1964zz} allows us to neglect the eccentricity parameters
$\{e, \phi_e\}$. Crucially also all observables have a simple
dependence on the total (redshifted) mass, $(1+z)M$. Similarly, the
observed waveforms have a trivial dependence on the time to
coalescence $t_c$. As this means only one theoretical waveform has to
be generated to cover all astrophysical masses and coalescence times,
$(1+z)M$ and $t_c$ are sometimes treated as extrinsic parameters
instead.  Nevertheless, numerical relativists are left with a
seven-dimensional parameter space to cover ($\eta$, $\vec{S}_1$,
$\vec{S}_2$), with no obvious shortcuts to lighten the workload
further.  Moreover, numerical methods will always struggle with finite
accuracy.

Equipped with a parametrized set of template waveforms
$h_m(t;\vec{\lambda})$, we can test an incoming detector data stream
$s(t) = h_e(t) + n(t)$, consisting of a (possible) ``exact''
gravitational wave $h_e(t)$ and detector noise $n(t)$ by calculating
the \emph{signal-to-noise ratio} (SNR) \cite{Cutler:1994ys}:
\begin{equation}
\label{eq:SNR_def}
\rho(\lambda_i) \equiv \frac{\langle h_m(\lambda_i) | s \rangle}{\sqrt{\langle h_m(\lambda_i)|h_m(\lambda_i)\rangle}}.
\end{equation}
In the case where the signal $h_e(t)$ in the data stream corresponds
perfectly (up to overall scaling) with the model waveform
$h_m(t;\lambda_i)$ for some value of the model parameters $\lambda_i =
\lambda_i'$, we obtain the optimal SNR: $\rho_{\rm opt} =
\sqrt{\langle h_e | h_e \rangle}$.  We refer to this simply as
the ``SNR'' $\rho$ for the remainder of this section.

Expected SNRs depend strongly on the detector in question.  Generally
speaking, the ground-based detectors in operation (LIGO, GEO, VIRGO,
etc) are expected to observe with only modest optimal SNRs, while LISA
is expected to observe MBH mergers with SNRs of hundreds \cite{LISA1}.

Of course, real model waveforms will be inaccurate, due to incomplete
knowledge of the underlying physics, or perhaps a desire for
simplicity. In this case, the achieved SNR cannot be
optimal. \citet{Apostolatos:1995pj} defined the \emph{fitting factor}
(FF) as the reduction in signal from such imperfect templates:
\begin{equation}
{\rm FF} \equiv \max_{\lambda_i} \frac{\langle h_m(\lambda_i) | h_e \rangle}{\sqrt{\langle h_m(\lambda_i) | h_m(\lambda_i) \rangle \langle h_e | h_e \rangle  }}
\end{equation}

\begin{figure}
\includegraphics*[width=3in, angle=0]{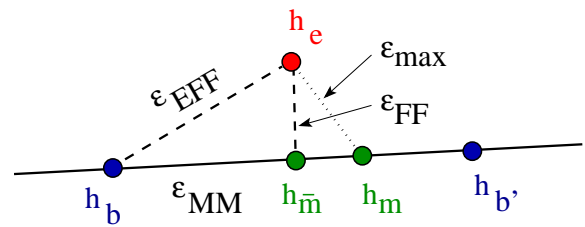}
\caption{Relationship between exact waveform $h_e$, model waveform
  with the same physical parameters $h_m$, actual ``best fit'' model
  waveform $h_{\bar{m}}$, and template bank waveforms $h_b$,
  $h_{b'}$. [Figure taken from \citet{Lindblom:2008cm}]}
\label{fig:Lindblom_Fig3}
\end{figure}

A fitting factor of 1 means that the exact signal lies somewhere in
the model space. However, for a particular detector, it is impossible
to generate templates arbitrarily close together in any one parameter
dimension.  We can develop a sense of how closely spaced a set of
templates must be so that any physical signal will be detected by at
least one of them with high likelihood. \citet{Owen:1995tm} developed
a geometric picture of a template-space metric defining the distance
between neighboring closely-spaced templates.  With these tools, he
was able to link the number of templates $\cal{N}$ to the desired
\emph{minimal match} (MM) -- the worst value of the match between the
signal and any template.  The connection between these concepts is
nicely laid out by \citet{Lindblom:2008cm}.
Figure~\ref{fig:Lindblom_Fig3}, from \citet{Lindblom:2008cm},
demonstrates how $\varepsilon_{\rm MM} \equiv 1 - {\rm MM}$ and
$\varepsilon_{\rm FF} \equiv 1 - {\rm FF}$ combine to produce an effective
mismatch $\varepsilon_{\rm EFF}$.

The percentage of detected signals scales as the cube of the fitting
factor.  For this reason, detector analysts require rather stringent
fitting factors from their template banks, e.g., 97\% (implying
detection of 90\% of all signals present).  For initial LIGO, for
instance, \citet{Baumgarte:2006en} estimate that $\sim$ 100 zero-spin
templates (produced from $\sim$ 10 different numerical simulations)
would suffice to detect all nonspinning merging binaries.  Obviously
we expect that the addition of spins of significant magnitude and
arbitrary direction will greatly increase this number.

Detection is only a first step, although an important one. When signal
strengths are high, as is expected with LISA for massive binaries, we
might expect to be able to read off some of the system's intrinsic and
extrinsic parameters. Achieving this requires templates that
accurately cover the parameter space of the binary, with a one-to-one
relationship with the parameters of the binary -- that is, that the
best-fit model waveform $h_{\bar{m}}$ and closest-parameters model
waveform $h_m$ in Fig.~\ref{fig:Lindblom_Fig3} coincide closely 
(within statistical errors).  It also requires that we have an
understanding of the probabilistic correlations between parameters.

For high-$\rho$ signals, the simplest way to estimate parameter uncertainties is using the \emph{Fisher
information matrix} \cite{Finn:1992wt,Cutler:1994ys}, which uses simple derivatives of the template
waveform shape with respect to the physical parameters:
\begin{equation}
F_{ij} \equiv \left\langle\frac{\partial h}{\partial \theta^i}\left|\frac{\partial h}{\partial \theta^j}\right.\right\rangle.
\end{equation}
In the high-SNR limit, the parameters are assumed to have a
multivariate Gaussian distribution, and the Fisher matrix is the
inverse of the covariance matrix between parameters. More
sophisticated approaches that do not assume high SNR include
Markov-chain Monte Carlo methods \cite{Gilks_MCMC}.

Different requirements can be made on the quality of the developed
templates, as described by \citet{Damour:1997ub}. The first is that
they be \emph{effectual}. Roughly speaking, this means that the
templates must be capable in the bulk of detecting gravitational-wave
signals. In the language of matched filtering above, we demand that
the fitting factor $FF$ be extremely close to unity for any signal.

More stringently, we might demand that the templates be
\emph{faithful}: that is, that the template waveform corresponds to
underlying source parameters that correspond closely to the parameters
of the detected waveform source.  In the language of
\citet{Lindblom:2008cm}, the model waveforms $h_m$ and $h_{\bar{m}}$
coincide closely (see Fig. \ref{fig:Lindblom_Fig3}). This requirement
will be crucial for identifying the astrophysical source of the
radiation.

Traditionally, templates used in the search for comparable-mass
black-hole binaries have depended wholly on PN theory. As detection
strategies are most sensitive to phase discrepancies between template
and observed signal (see below), the determination of waveform phase
to high accuracy has been crucial. Modern PN templates include phase
terms up to 3.5PN order for nonspinning binaries, with 2.5~PN order
spin corrections. Many templates use this high-accuracy phase with the
leading-order quadrupole amplitude to produce \emph{restricted}
templates; however, amplitude corrections are known to 2.5~PN order
(higher for certain low-$\ell$ modes) \cite{Kidder:2007rt}.

While such PN-driven templates offer excellent phase accuracy during
the long, slow inspiral of the binary, their usefulness becomes
questionable as we approach the last pre-merger orbits of the
binary. We may ask, then, what does our new numerical insight into the
final moments of merger bring us?

\subsection{The Direct Impact of Merger Waveforms in Data Analysis}
\label{sec:gwda:direct}

\begin{figure}
\includegraphics*[width=3.5in, angle=0]{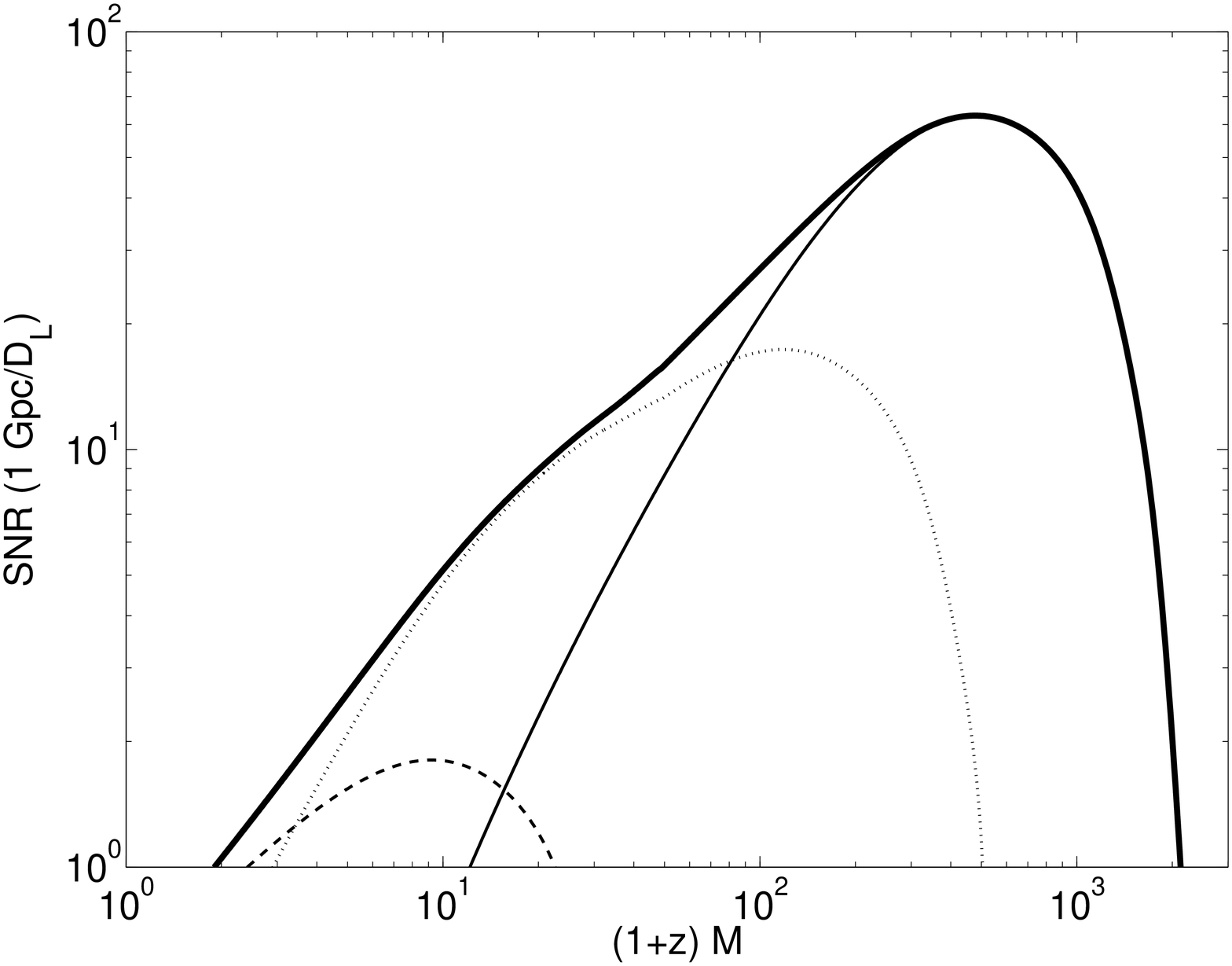}
\includegraphics*[width=3.5in, angle=0]{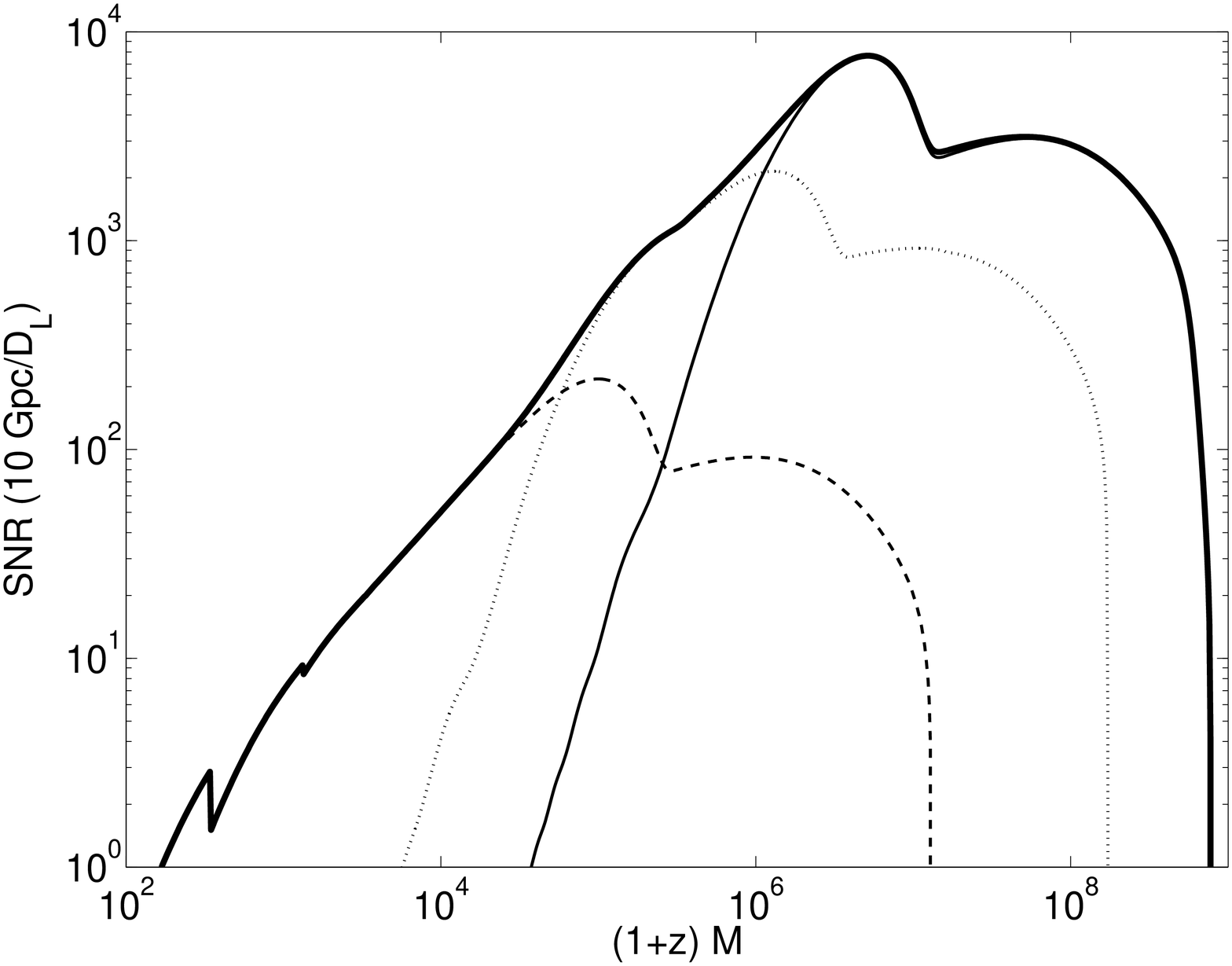}
\caption{The importance of including NR merger waveforms in data
  analysis can be seen here for Advanced LIGO (top) and LISA (bottom),
  taken from \citet{Baker:2006kr}. In both panels, the dashed curve is
  the achieved SNR (for a given redshifted source mass) when only the
  early inspiral signal (up to $\sim 1000 M$ before merger) is used;
  the dotted curve uses the signal between this time and when final
  plunge commences ($1000M$ to $50M$ before merger; the thin solid
  curve uses just the merger waveform (starting $50M$ before
  merger). The thick solid curve is the combined result of using the
  entire waveform.  The SNR, and hence distance reach, of the detector
  is clearly greatly enhanced by the final merger portion.}
\label{fig:LIGOSNR}
\end{figure}

Once full numerical waveforms became available, several groups used
the results to test older predictions of the effect of the merger
segment on observability. For the equal-mass nonspinning case, the
numerical merger gave results consistent with \citet{Flanagan:1997sx}
for initial LIGO, though with a smaller merger SNR
\cite{Baker:2006kr,Buonanno:2006ui}. The boost in SNR from merger is
significantly greater for Advanced LIGO and LISA, as shown in
Fig.~\ref{fig:LIGOSNR}, from \citet{Baker:2006kr}.

\begin{figure}
\includegraphics*[width=3.5in, angle=0]{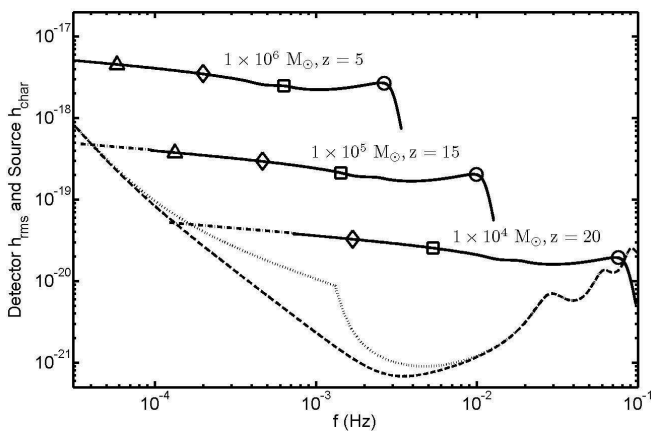}
\caption{The importance of binary mass for the LISA observation
  window, demonstrated by the characteristic amplitudes
  \cite{Baker:2006kr} of three different sources, relative to the rms
  noise amplitude of the LISA detector. On each $h_{\rm char}$ curve,
  we mark times before the peak amplitude (circle) -- one hour
  (square), one day (diamond) and one month (triangle).}
\label{fig:LISAhchar}
\end{figure}

The relevance of the final plunge and merger to the overall SNR of the
binary depends on the total mass of the binary and where it falls in
the window of the detector. Figure~\ref{fig:LISAhchar}, adapted from
\citet{Baker:2006kr}, shows this for LISA, plotting the
``characteristic signal strain'' $h_{\rm char}(f) \equiv 2 f
|\tilde{h}(f)|$ (for the dominant quadrupole radiation).  Lower-mass
binaries have higher-frequency waveforms at all dynamical stages: for
an equal-mass binary with $M \lesssim 10^4 \MSun$, the late-merger
and ringdown signal will fall outside LISA's sensitivity band. For
such low masses, inspiral-only waveforms should be adequate for
data-analysis purposes.

The LIGO and Virgo detectors are sensitive to frequencies above $\sim
30 \Hz$ (see Fig.  \ref{fig:LIGO_Virgo_noise}). The merger of binary
systems is marked by a chirp gravitational-wave signal, whose monotonically increasing
frequency saturates at the dominant quasinormal mode, which depends on
the post-merger hole's mass and spin. For a nonspinning binary of
total mass $M$, this $f_{\rm QNM} \approx 1.75 \times 10^4
(M/\MSun)^{-1} \Hz$.  This means that a merging binary of mass greater
than $\sim 600 \MSun$ will never be seen by LIGO. In contrast, most
post-Newtonian templates stop at $f_{\rm ISCO}$, the frequency at the
innermost stable circular orbit (well defined only for test particles;
see Sec.~\ref{sssec:astro_grav:basic:bh}). This is
\begin{eqnarray}
f_{\rm ISCO} &=& \frac{1}{6\sqrt{6}M} \approx 1.36 \times 10^4 (M/\MSun)^{-1} \Hz.
\end{eqnarray} 
For an incomplete PN waveform to be useful, we want the missing
merger-ringdown section (i.e. the part of the signal with $f > f_{\rm
  ISCO}$) to contribute as little as possible to the total SNR. For
instance, if we take the high end of LIGO's sensitivity window to be
$\sim 800\Hz$, this is $f_{\rm ISCO}$ for a binary of total mass $M
\approx 17 \MSun$. Then, for binaries with $M \gtrsim 17 \MSun$, $f >
f_{\rm ISCO}$ will fall within the visible window; that is, LIGO will
see the final stages of the binary merger, where no acceptable PN
waveform is available.  Recent work by \citet{Buonanno:2009zt}
supplied a more precise answer: inspiral-only PN templates can be
trusted for all LIGO configurations for systems with $M \lesssim
12\MSun$.  Above this critical mass, there is a gap in reliable
information, which must be filled by numerical results.

\begin{figure}
\includegraphics*[width=3.5in, angle=0]{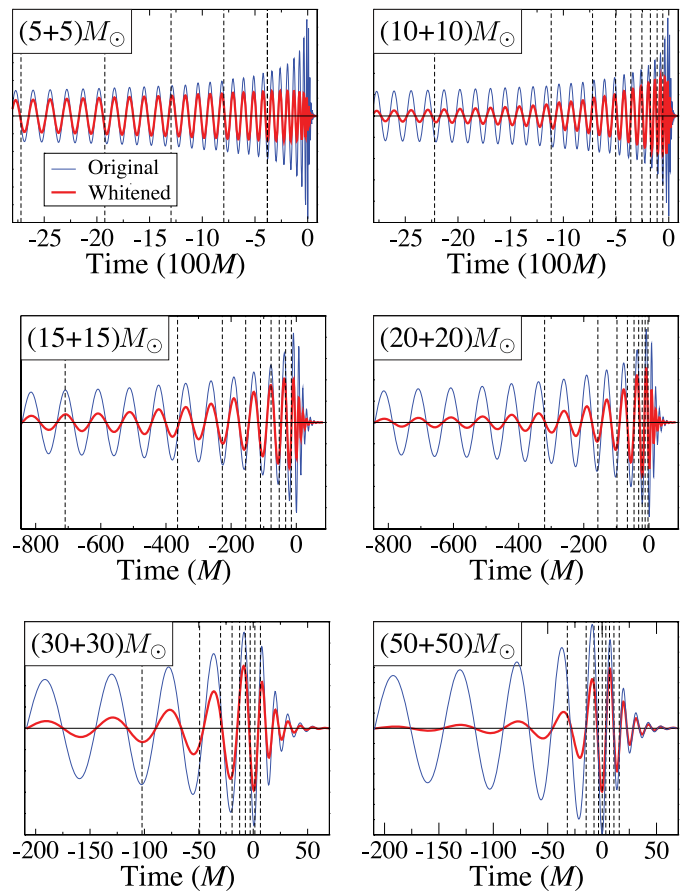}
\caption{The power distribution in 10\% segments of the waveform
  before merger, for a range of LIGO-appropriate equal-mass
  binaries. Note that as the total mass $M$ increases, the final
  merger and ringdown account for a greater fraction of the total
  power.  Adapted from \citet{Pan:2007nw}.}
\label{fig:PanEtAl_SNRdist}
\end{figure}

Figure~\ref{fig:PanEtAl_SNRdist}, adapted from \citet{Pan:2007nw},
shows the importance of the last stages of merger in initial
LIGO as a function of mass range. The thin blue curves are hybrid
NR/EOB waveforms, while the thick red waveforms are generated from
these by ``whitening'': that is, they have been Fourier-transformed to
the frequency domain, then rescaled by $1/\sqrt{S_n(f)}$, and finally
re-transformed to the time domain.  The whitened amplitude in a
segment is proportional to the contribution of that segment to the
total SNR; each marked segment accounts for 10\% of the total. As the
total system mass $M$ increases, the whitened amplitude becomes more
bulked toward the merger time. For the largest total mass shown, $M =
100\MSun$, more than 90\% of the signal power comes from the last
cycle + merger + ringdown.

\begin{figure}
\includegraphics*[width=3.5in, angle=0]{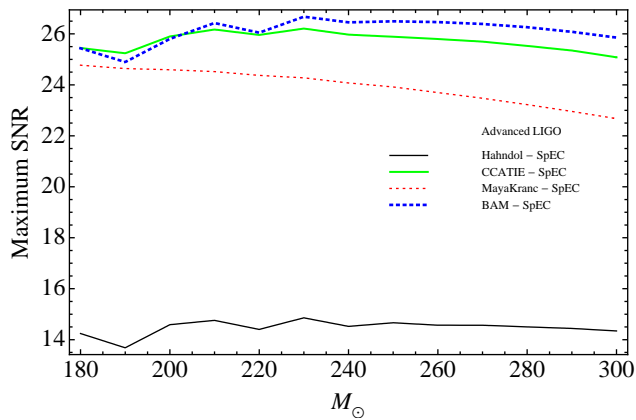}
\caption{Maximum signal-to-noise ratio (SNR) $\rho$ below which
  different ``long'' numerical waveforms are indistinguishable for the
  Advanced LIGO detector. Even the earliest, lowest-accuracy published
  long waveforms are indistinguishable for SNR levels below 14, as
  demonstrated in this figure from the Samurai paper
  \cite{Hannam:2009hh}.}
\label{fig:Samurai_MeasuredALIGO}
\end{figure}

In Sec.~\ref{sec:GWs:longer_wfs}, we referenced the recent ``Samurai''
project \cite{Hannam:2009hh}, which establishes the consistency of all
of the ``long'' (pre-merger duration $\gtrsim 1000 M$) equal-mass
nonspinning waveforms within their stated numerical accuracy. As well
as direct comparisons of phase and amplitude errors over time, they
conducted mismatch tests in the regime of the LIGO and Virgo
ground-based detectors. They demonstrated that quadrupole waveforms
($\ell = 2, m = \pm 2$) from the five numerical codes have tiny
mismatches ($\sim 10^{-3}$) for binary masses above $60 \MSun$ with
``enhanced'' LIGO, and masses above $180 \MSun$ with Advanced LIGO,
Virgo and Advanced Virgo. All waveforms would be indistinguishable for
SNRs below 14, entirely reasonable for the current generation of
ground-based detectors; this is shown for the Advanced LIGO
detector in Fig.~\ref{fig:Samurai_MeasuredALIGO}.

Studies using incomplete (non-merger) PN waveforms have shown that
introducing fuller harmonics of spinning and precessing binary systems
greatly enhances parameter estimation, removing some parameter
correlations and hence reducing certain errors by several orders of
magnitude (though other parameters are significantly less improved)
\cite{Sintes:1999cg,Lang:2006bz,Lang:2007ge,Lang:2008gh,Trias:2008pu}. It
is expected that the introduction of full numerical waveforms for the
merger portion will lead to further improvements, at least in regions
of the detector's window where the binary merger is visible.  A recent
study by \citet{Ajith:2009fz} used phenomenological
merger-inspiral-ringdown templates (discussed in
Sec.~\ref{sec:gwda:templates}) to assess these improvements for
Advanced LIGO and Advanced Virgo. They showed that parameter-estimate
accuracy is improved substantially compared to inspiral-only waveforms
for system masses $\gtrsim 20 \MSun$; in particular, the average
sky-position error for an $M = 100 \MSun$ equal-mass, nonspinning
binary at a luminosity distance of 1 Gpc drops to about one-tenth of a
square degree. Since they only uses the dominant mode for this
investigation, there will be systematic errors as well.

Numerical verification of the extent of parameter estimation
improvements for LISA has led to differing estimates
\cite{Babak:2008bu,Thorpe:2008wh,McWilliams:2009bg}.  In particular,
\citet{Babak:2008bu} find that 50\% of injected signals could be
located on the detector sky within an error box of 3 arc min, while
$t_c$ could be measured to less than a second and $D_L$ within 1.5\%
(ignoring issues of weak lensing). They attributed these impressive
results to a combination of very high SNR (between 900 and 9000) and
the use of higher multipoles of the radiation. In contrast,
\citet{McWilliams:2009bg} found smaller (by about one order of
magnitude) improvements over inspiral-only waveforms. This may in part
be attributable to the use of different mass ratios and
different-length waveforms by the two groups, but has not been fully
resolved.

\subsection{Developing Analytic Inspiral-Merger-Ringdown Gravitational Waveform Templates}
\label{sec:gwda:templates}

One of the eventual aims of numerical-relativity simulations of binary
mergers is the production of a set of gravitational-wave ``templates''
that will cover the many-dimensional space of astrophysical
parameters. Numerical merger simulations are still orders of magnitude
too slow to run on the fly to filter incoming detector data streams;
thus there is a need to develop analytic waveform expressions that
encode the numerics.

It has been demonstrated in the physical regime of nonspinning binary
mergers \cite{Buonanno:2009zt} that inspiral-only PN-based templates
become inconsistent with each other (and with full waveforms) at total
binary masses $M \gtrsim 12 \MSun$ for initial and advanced LIGO
configurations; above this mass, full templates including the
numerical-simulation-based understanding of merger/ringdown become
necessary.

One such set of templates incorporating numerical-relativity data is 
due to \citet{Pan:2007nw}. These were extensions of the
Stationary-Phase Approximation (SPA) templates used in LIGO/Virgo data
analysis, with input from equal-mass numerical waveforms supplied by
Frans Pretorius \cite{Buonanno:2006ui}, as well as equal- and
unequal-mass numerical waveforms from the Goddard group. These
templates are effectual -- they achieved fitting factors of $\gtrsim
0.96$ -- but they did this by extending source parameters into
unphysical space.

To date a handful of useful faithful full-waveform template banks
has been produced, aimed at covering all mass ratios for nonspinning
binaries. The ``phenomenological'' templates of
\citet{Ajith:2007qp,Ajith:2007kx} are simple three-segment curves in
frequency space, with matching parameters tuned by numerical data
generated by the \texttt{BAM} \cite{Bruegmann:2006at} and \texttt{CCATIE}
\cite{Pollney:2007ss} codes.  The templates of
\citet{Buonanno:2007pf,Boyle:2008ge}, coded in the LSC Algorithm
Library as ``EOBNR'', are an extension of effective-one-body
waveforms, with a single ``pseudo-4PN'' parameter tuned by numerical
data (initially from the Goddard Hahndol code, and later using
waveforms from the Caltech-Cornell group. Both the phenomenological
and EOBNR template banks are discussed in
Sec.~\ref{sec:nrpn:analytic}.

\begin{figure}
\includegraphics*[width=3.0in, angle=0]{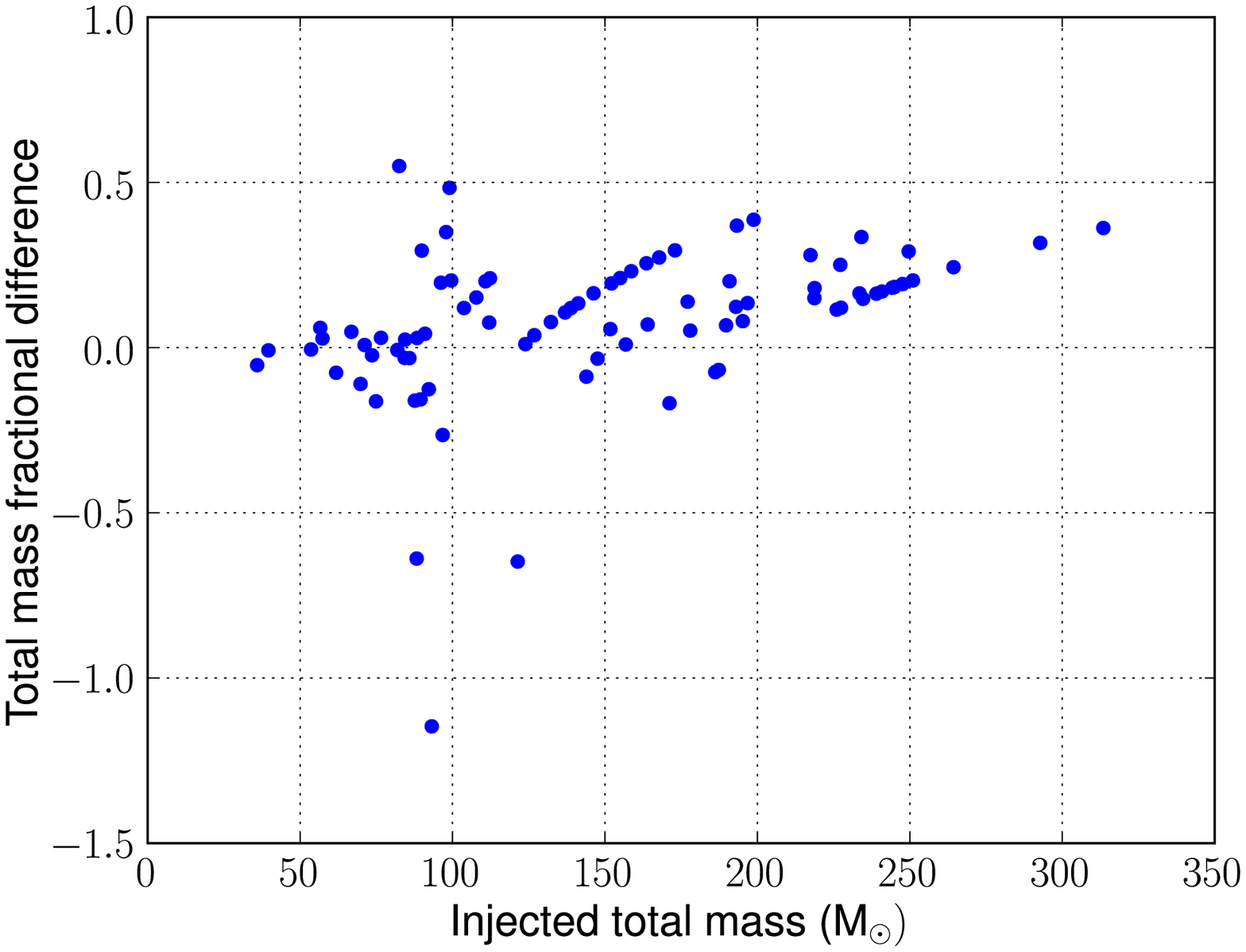}

\includegraphics*[width=3.0in, angle=0]{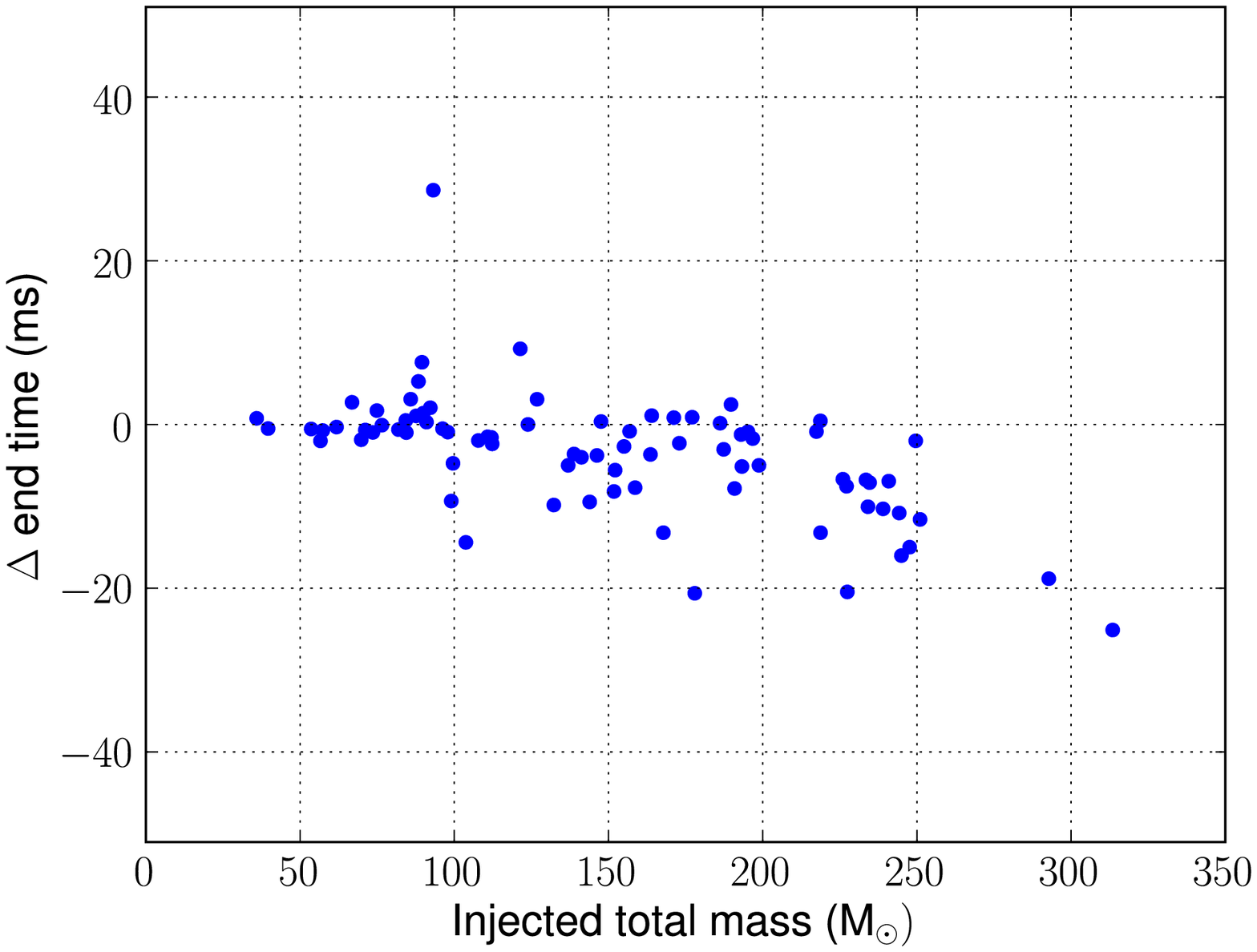}
\caption{Accuracy in parameter estimation for EOBNR templates in the
  NINJA project.  The top panel shows the fractional error in
  estimated total mass, $(M_{\rm injected}-M_{\rm detected})/M_{\rm
    injected}$, for all detected signals, while the lower shows the
  error in merger time. Adapted from \citet{Aylott:2009ya}. Reproduced by permission of the Institute of Physics.}
\label{fig:NINJA_EOBNR_parest}
\end{figure}

Both template banks are faithful in the sense described above, in the
restricted $(M,\eta)$ parameter space, and both have been used in
data-analysis injection tests \cite{Aylott:2009ya,Santamaria:2009tm}.
Currently, EOBNR templates are being used both for injection and
filtering in the high-mass region of LIGO test analysis, and also for
injection for the ringdown band. They were also both used in the
matched-filter analyses of the NINJA project (see Sec.
\ref{sec:gwda:inject}), performing as well as the inspiral-only
templates in detection, and considerably better in the limited
parameter estimation attempted. Figure~\ref{fig:NINJA_EOBNR_parest},
from \citet{Aylott:2009ya}, shows the accuracy with which
the total mass $M$ and time of merger were extracted from all detected
injections in NINJA, when the EOBNR templates were used. Results for
the phenomenological templates were similar for $M$, but
performed slightly less well for time of merger, as it is not an
explicit part of this model (given the relatively small number of
samples in the NINJA tests, such minor differences may not be
significant).

As significant spin is expected in astrophysical black holes at all
scales, it is important to develop templates that encode spin
also. \citet{Vaishnav:2007nm} demonstrated that matched filtering with
nonspinning merger waveforms does a good job in detecting waveforms
from binaries with significant spins. As their spinning waveforms come
from aligned-spin mergers where the total spin is zero (one hole's
spin is aligned with the orbital angular momentum, while the other's
is anti-aligned), their configurations will lack some prominent
spin-related effects, including spin-orbit effects (pulling in or
pushing out the ISCO of the orbit) and precession of the orbital
plane. Their conclusions, then, are likely to be optimistic, and more
generic configurations can be expected to do much worse with
nonspinning templates.  The phenomenological templates of Ajith \etal
have recently been extended to include non-precessing spinning systems
\cite{Ajith:NRDA2009}. These have been shown to be both effectual and
faithful in searches over non-precessing binary hybrid waveforms with
LIGO, where earlier zero-spin templates failed badly.  For maximum
simplicity, these new phenomenological template families are
ultimately parametrized by three numbers: the total mass $M$, the
symmetric mass ratio $\eta$ \eqref{eq:eta_def}, and a single effective
spin
\begin{equation}
\ahat_{\rm eff} \equiv \frac{(1 + \delta)}{2} \ahat_1 + \frac{(1 - \delta)}{2} \ahat_2,
\end{equation}
where $\delta = (M_1 - M_2)/M = \sqrt{1 - 4\eta^2}$.

\subsection{Using Numerical Waveforms in Data Analysis Applications}
\label{sec:gwda:inject}

An obvious application of numerical waveforms for data analysis is in
directly testing analysis algorithms. As current templates are
overwhelmingly based on PN information -- either extended
heuristically to include merger and ringdown, or terminating before
merger -- so current detection and parameter estimation algorithms are
also based on incomplete information.

Recently, a multi-group collaboration (the NINJA project) has
performed the first direct insertion of explicit numerical waveforms
into a simulated LIGO datastream, and investigated detection
efficiency and systematic errors \cite{Aylott:2009ya}. Essentially all
active numerical-relativity groups contributed waveforms; these were
scaled randomly to represent different-mass sources, resulting in 126
different injections.  Nine data-analysis groups analyzed the
post-injection data stream, with a variety of methods. These fell into
three main groups: matched filtering, burst analysis, and Bayesian
techniques.

For the matched filtering, some groups used standard ``TaylorF2''
inspiral-only templates generated by PN theory, some used
ringdown-only templates, and some used full inspiral-merger-ringdown
templates, such as the EOBNR and phenomenological waveforms discussed
below and in Sec.~\ref{sec:nrpn:analytic}.  Both the TaylorF2 and the
full template waveforms yielded quite high detection rates -- around
80 of the 126 injected signals in triple coincidence between the LIGO
detectors -- while the ringdown templates performed more poorly,
detecting only 45 signals in triple coincidence. However for parameter
estimation of the total mass $M$, most of the TaylorF2 estimates have
an error of 40\% or more, significantly more than for the EOBNR and
phenomenological estimates. Figure~\ref{fig:NINJA_EOBNR_parest} shows
errors in detected mass (upper panel) and coalescence time (lower
panel) when using the EOBNR templates. While mass errors are still
quite large, between $\sim -30\%$ and $+50\%$ for the bulk of detected
signals, they are significantly less biased than those from the
TaylorF2 templates (compare with Fig.~8 of \citet{Aylott:2009ya}).
Estimation of the coalescence time $t_c$ was better for the
EOBNR and phenomenological template banks as well, as might be expected
(since TaylorF2 templates must necessarily cut off before
coalescence). However, the EOBNR and phenomenological templates typically
covered a larger mass range as well (up to $200 \MSun$ compared to $90
\MSun$ for TaylorF2), which will bias the results when the injected
signals had a large mass.

Burst analysis is a more abstract approach to the problem, using
partial information about signals, such as dominant frequency and
approximate duration. In NINJA, two burst techniques were used -- the
``Q pipeline'' and the Hilbert-Huang transform (HHT). The Q pipeline fits
signals to a sine-Gaussian model parametrized by a central frequency
and a $Q$ factor ($\sim f \tau$, the number of oscillations under the
Gaussian). At the single-detector level, the detection rate is
comparable to that of the EOBNR and phenomenological matched-filter
searches.  Parameter estimation is limited to estimates of the peak
power time (close to $t_c$) and peak frequency; for these parameters,
performance was comparable to the ringdown matched-filter search. The
Hilbert-Huang transform decomposes input data into ``intrinsic mode
functions'', each characterized by a single frequency scale. Applied
to the signal, the HHT generates a high-resolution time-frequency map
of the data \cite{Camp:2007ee,Stroeer:2009zz}. The HHT applied to the
NINJA data found $\sim 80$ coincident events, competitive with the
best matched-filter results; however, no parameter estimation is
currently possible with this method.

Bayesian analysis attempts to reconstruct the posterior probability
density function of a parametrized signal based on the observed
signal. Two variant Bayesian approaches were taken in NINJA analysis,
involving different Monte Carlo methods -- Markov Chain Monte Carlo
(MCMC) and Nested Sampling. In both cases, a parametrized waveform
model is needed. For the MCMC study, a restricted PN waveform was
used, only 1.5~PN in phase, but including leading spin effects.
Detection of injected signals performed well, but the high mass of the
injections meant that the bulk of the SNR occurred in the
merger-ringdown phase; as a result the inspiral-only model masses
merger time was significantly biased. The Nested Sampling effort
instead uses the TaylorF2 templates (restricted amplitude, 2.0~PN in
phase) as well as phenomenological templates. In detection, the
phenomenological templates significantly out-perform the TaylorF2
model; in parameter estimation, it achieves results consistent with
those of the matched-filter analysis.

Overall, current results are encouraging, with high detection rates of
injected waveforms, and good, if limited, accuracy in estimated
parameters when attempted with faithful template banks.  However, the
project was hampered by the limited length of the contributed
waveforms and unrealistically stationary LIGO and Virgo noise
profiles. Follow-on work in this area will demand longer numerical
waveforms that span the LIGO sensitivity window, or more likely, a
systematic blending of long PN inspiral waveforms to
late-inspiral/merger numerical waveforms, but also may be extended to
other gravitational-wave sources.

\section{Impact on Astrophysics}
\label{sec:astro}

The recent successes in modeling of binary black-hole mergers have
captured the attention of astronomers and astrophysicists.  While
black-hole binaries have been studied both observationally and
theoretically for many years, most efforts have been primarily within
the framework of Newtonian gravity.  The new developments in numerical
relativity now supply the missing piece: the effects of
the final merger in the strong-field general relativistic regime. In
this section we highlight three key areas of current interest:
recoiling black holes, the spin of the merged black hole, and
electromagnetic signatures from the final merger.

\subsection{Recoiling Black Holes}
\label{sec:astro:recoil}

One of the more dramatic implications of gravitational-wave
computations for astrophysical observations is the possibility that
gravitational radiation-induced recoil will eject black holes from
their host galaxies.  The largest recoil velocities predicted by
numerical relativity exceed the escape velocities of many galaxies
($\sim 1000 \kms$).  Calculating the probability of this kind of event
requires a detailed understanding of the recoil on mass ratios, spin
magnitudes, and spin orientations, together with some expectations
about the distributions of these parameters.

\subsubsection{Predicting the Recoil}
\label{sec:astro:recoil_formulae}

It would be useful to have in hand a simple, analytic formula
expressing the relevant dependencies of the recoil velocity.  The
highly nonlinear, strong-field interactions of merger dynamics
determine the final recoil velocity of the merged remnant.  Numerical
simulations are required to accurately compute such results.  However,
in attempting to construct an ansatz for a phenomenological formula,
one might initially assume a form consistent with PN predictions.
This will suggest the functional dependence on mass and spin, and
their leading powers.  For example, PN analysis \cite{Fitchett_1983}
suggests that the recoil due to unequal masses may be proportional to
$v=A\eta^2\sqrt{1-4\eta}(1+B\eta)$ (the ``Fitchett formula"), where
$\eta$ is the symmetric mass ratio \eqref{eq:eta_def}.  Meanwhile, to
leading PN orders, the spins should contribute through components of
$\vec{\Delta} \equiv \vec{S}_2/M_2-\vec{S}_1/M_1$ and
$\vec{S}\equiv\vec{S}_1+\vec{S}_2$
\cite{Kidder:1995zr,Campanelli:2007ew,Racine:2008kj}.

Such an ansatz can be further supported and constrained by symmetry
arguments.  An economical approach is to consider the recoil velocity
as a Taylor expansion in all six spin components (three for each black
hole), where the coefficients are functions of mass and initial
separation.  Rotation, parity, and exchange symmetries of the binary
system then impose relationships between these coefficients
\cite{Boyle:2007sz,Boyle:2007ru}.  By such arguments, for example, the
component of the recoil velocity parallel to the orbital angular
momentum can be shown to be proportional, to leading order in spin, to
the dot product of the coordinate separation vector between the black
holes and some linear combination of the spins
\cite{Boyle:2007sz,Baker:2008md}.

Unknown coefficients in the ansatz can then be fit to numerical
results, and correction terms can be added as needed for numerical
agreement.  With these considerations, a tentative formula has been
taking shape through the combined efforts of the numerical-relativity
community, of the following form
\cite{Gonzalez:2006md,Baker:2007gi,Baker:2008md,Lousto:2009mf,Campanelli:2007cg,Brugmann:2007zj,Lousto:2008dn,vanMeter:2010md}:
\begin{eqnarray}
\vec{V}_{\rm recoil} &=& v_m \, \hat{e}_1 + v_{\perp} (\cos\xi \, \hat{e}_1 + \sin\xi \, \hat{e}_2) + v_{\parallel} \, \hat{e}_2, \label{eq:v_total}\\
      v_m     &=& A \eta^2 \sqrt{1 - 4 \eta} (1 + B \eta), \label{eq:v_mass}\\
v_{\perp}     &=& H \frac{\eta^2}{(1+q)} \left( \alpha_2^{\parallel} - q \alpha_1^{\parallel} \right), \label{eq:v_perp}\\
v_{\parallel} &=& \frac{K_2\eta^2+K_3\eta^3}{q+1} \nonumber\\
              &\times&\left[q \alpha^{\perp}_1 \cos(\phi_1-\Phi_1)-\alpha^{\perp}_2\cos(\phi_2-\Phi_2)\right] \nonumber\\
              &+&\frac{K_S(q-1)\eta^2}{(q+1)^3} \nonumber\\
              &\times&\left[q^2\alpha^{\perp}_1 \cos(\phi_1-\Phi_1)+\alpha^{\perp}_2\cos(\phi_2-\Phi_2)\right] \label{eq:v_parallel}
\end{eqnarray}
where $v_m$ is the contribution due to mass asymmetry, $v_{\perp}$ is
the contribution due to spin that yields a kick perpendicular to the
orbital angular momentum, $v_{\parallel}$ is the contribution due to
spin that yields a kick parallel to the orbital angular momentum,
$\alpha_i^{\parallel}$ is the projection of the dimensionless spin
vector $\vec{\alpha}_i=\vec{S}_i/M_i^2$ of black hole $i$ along the
orbital angular momentum, $\alpha_i^{\perp}$ is the magnitude of its
projection $\vec{\alpha}_i^{\perp}$ into the orbital plane, $\phi_i$
refers to the angle made by $\vec{\alpha}_i^{\perp}$ with respect to
some reference angle in the orbital plane, and $\Phi_1$ and $\Phi_2$
are constants for a given mass ratio and initial separation.  Here the
spins are considered to have been measured at some point before
merger or ideally, arbitrarily close to merger.  $\Phi_1$ and
$\Phi_2$ encode the amount of precession of each spin before merger.

One can then evaluate the formula over ranges of expected mass ratios
and spin magnitudes, and for all spin orientations, to compute the
probability of a given recoil speed.  For example, for mass ratios
between 1 and 10, and spin magnitudes of 0.9, the probability of
exceeding $1000 \kms$ is predicted to be $\sim 10\%$.  Studies of
various model-dependent speed probabilities are given by
\citet{Schnittman:2007nb} and \citet{Baker:2008md}.

\subsubsection{Consequences of Black Hole Recoil}
\label{sec:astro:recoil_conseq}

Among the significant astrophysical consequences of gravitational-radiation
recoil, growth rates of black holes can be affected. For
example, the recoil of massive black holes imparting less than the
escape velocity of their host galaxies tends to have regulatory
effects on the amount of mass accreted \cite{Blecha:2008mg}.  Recoil
velocities that exceed galactic escape velocities can impact massive
black-hole growth in a different way, by reducing the chances of
subsequent mergers \cite{Sesana:2007sh, Volonteri:2007et}.  This in
turn may modestly reduce the rate of coalescence events observable by
LISA \cite{Sesana:2007sh}.

The recoil might also have more directly observable consequences.
Some quasars are thought to originate from the coalescence of massive
black holes during galactic mergers.  Quasars ejected from their host
galaxies might therefore be expected. A study by
\citet{Bonning:2007vt} found no evidence for such events, indicating
that they are rare.  However, recent observations of two rapidly
moving extragalactic quasars \cite{Shields:2009jf} are strong
candidates for such recoiled black holes \cite{Komossa:2008qd}.
 
\subsection{The Spin of the Final Black Hole}
\label{sec:astro:final_spin}

The final spin of the merged remnant of a binary is also of
astrophysical interest.  It would be useful to know the probability of
various spin magnitudes in constructing matched filtering templates
and estimating gravitational signal detectability for LIGO and LISA
\cite{Berti:2007zu}.  The ability to predict a final spin given
initial binary parameters also implies that observation of the spin of
a black hole may help understand its origin.  In particular, an
understanding of the relationship between the final spin of a black
hole and the orbital angular momentum of its binary precursor may help
explain the formation of X-shaped jets \cite{Barausse:2009uz}.

As with the kick, it would be advantageous to construct a simple
analytic formula with which to estimate the spin.  Observations from
numerical simulations suggest this might be possible.  For example, in
the case of nonspinning, equal-mass binaries undergoing circular
inspiral, the final spin is insensitive to the initial separation,
being determined by the terminal dynamics of the merger. Such
evolution results in a universal final spin of $\sim 0.7$
\cite{Pretorius:2005gq,Campanelli:2005dd,Campanelli:2006gf,Baker:2006yw,Campanelli:2006uy}.
Additionally, for unequal-mass, nonspinning binaries, the final spin
scales roughly linearly with symmetric mass ratio
\eqref{eq:eta_def} \cite{Gonzalez:2006md}.

Various approaches to constructing an ansatz for the final spin have
been proposed. For aligned-spin, equal-mass mergers,
\citet{Campanelli:2006uy} produced a simple formula quadratic in the
(common) initial spin; this satisfied the ``cosmic censorship''
hypothesis: $a_f/M_f < 1$. For nonspinning mergers,
\citet{Berti:2007fi} simply assumed $a_f/M_f=a\eta+b\eta^2$, for
fitting parameters $a$ and $b$, which agreed with data from
nonspinning binaries reasonably well. \citet{Lousto:2009mf} arrived at
a more generic ansatz, for initial black holes of arbitrary spins,
using the PN approximation.  \citet{Barausse:2009uz} proposed a
different expression for generic binaries, using a set of assumptions
about the approximate conservation of the magnitudes and relative
angles of certain angular momentum vectors.  Several other formulas
have been suggested
\cite{Buonanno:2007sv,Rezzolla:2007rd,Rezzolla:2007rz,Rezzolla:2008sd,Tichy:2008du},
each in good agreement with some subset of available numerical data.
However, broad agreement on an analytic formula for the final spin from
generically precessing binaries has yet to be achieved.

\subsection{Electromagnetic Counterparts of Black Hole Mergers}
\label{sec:astro:EM} 

We have seen above that black-hole mergers are expected to be
``loud,'' since they produce strong gravitational-wave signals. But
will they also be ``bright?''  That is, will there be an accompanying
display of photons, detectable by telescopes in any frequency range of
the full electromagnetic spectrum?

\subsubsection{Astrophysical Considerations}

The answer to this question depends critically on the amount and
distribution of gas and magnetic fields surrounding the merging
binary. For stellar black-hole binaries, and intermediate-mass black-hole
binaries in stellar clusters, any matter in an accretion disk would be
consumed by one or both black holes and disappear relatively
quickly, making an electromagnetic counterpart highly unlikely.

Massive black-hole binaries formed from galaxy mergers, however, present
a very different situation.  In a gas-rich or ``wet'' merger, there is
likely enough gas available to feed accretion disks around each black
hole that eventually evolve into a circumbinary disk; this provides a
source of gas and magnetic fields that could generate detectable
electromagnetic emission.  Even in the case of gas-poor or ``dry''
mergers, the thin hot gas present is such (elliptical) galaxies might
be sufficient to produce an electromagnetic signature.

Such electromagnetic signatures would be valuable to
astrophysics.  Identification of the source on the sky would help
confirm and characterize the merger, and probe accretion physics.  A
measurement of its redshift using electromagnetic radiation, taken
together with the determination of luminosity distance using
gravitational waves observed by LISA \cite{Lang:2008gh}, would provide
an independent calibration of the distance scale and a precise probe
of cosmology including the nature of the mysterious dark energy
\cite{Kocsis:2007hq,Jonsson:2006vc,Dalal:2006qt,Kocsis:2005vv,
  Holz:2005df,Kocsis:2007yu,Arun:2008xf}. Differences in arrival times
of the electromagnetic and gravitational-wave signatures could also
test fundamental principles such as the relative propagation speed of
photons and gravitational waves.

With the recent successes in numerical-relativity modeling of black-hole
mergers, we find considerable interest in understanding possible
electromagnetic signals from these events. Most work to date has
focused on mechanisms that could produce emission from a surrounding
accretion disk, including signals induced by a recoiling merged black
hole encountering the disk
\cite{Armitage:2002uu,Milosavljevic:2004cg,Dotti:2006zn,Kocsis:2005vv,
  Phinney:2007,Bode:2007,Kocsis:2007yu,Shields:2008va,Lippai:2008fx,
  Schnittman:2008ez,Kocsis:2008va,
  Haiman:2008zy,O'Neill:2008dg,Haiman:2009te,Chang:2009rx,
  Megevand:2009yx}. Moreover, explorations of possible electromagnetic
signals that could arise in the dynamic spacetime near the binary
during its last few orbits and merger are just now beginning.

\subsubsection{Simulations with Magnetic Fields or Gas near the Merging Holes}

\citet{Palenzuela:2009yr} recently studied the effects of a merging
black-hole binary on a surrounding magnetic field.  The equal-mass,
nonspinning holes start out with separation $\approx 6M$ on
quasicircular orbits somewhat outside the ISCO.  The magnetic field is
initially poloidal and assumed to be generated by currents in a
distant circumbinary disk located at $\approx 10^3 M$.  The electric
field around the binary is initially zero.  There is no matter near
the black holes, as predicted when the circumbinary disk is thin
\cite{Milosavljevic:2004cg}.
They solved the coupled Einstein-Maxwell equations for a binary
evolving in presence of externally sourced electromagnetic fields.  As
the system evolves, the inspiralling black holes stir up the fields.
\begin{figure}
\includegraphics*[scale=0.45,angle=0]{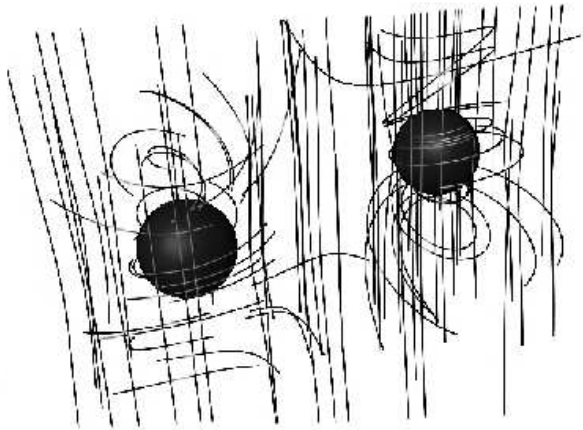}
\caption{Magnetic and electric fields lines around inspiralling
  equal-mass, nonspinning black holes at $\approx 40M$ before the
  merger, from \citet{Palenzuela:2009yr}. The electric field lines are
  twisted around the black holes, while the magnetic field lines are
  mainly aligned with the $z$ axis.}
\label{fig:EM_early}
\end{figure}
\begin{figure}
\includegraphics*[scale=0.45,angle=0]{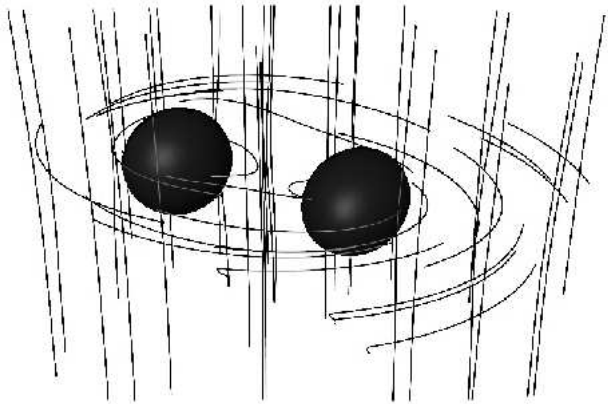}
\caption{Same as Fig.~\ref{fig:EM_early}, except at a later time,
  $\approx 20M$ before merger.}
\label{fig:EM_later}
\end{figure}
Figures~\ref{fig:EM_early} and~\ref{fig:EM_later} show the electric
and magnetic field lines at $\approx 40M$ and $\approx 20M$ before the
merger, respectively.  The magnetic fields are mostly aligned with the
$z$ axis, while the electric fields are twisted around the black
holes.  The binary dynamics induce oscillations in the electromagnetic
energy flux, with a period approximately half of the dominant
(quadrupole) gravitational-wave signal. The energy in the
electromagnetic field is also enhanced gradually.

\citet{Palenzuela:2009yr} have certainly made an interesting start on
this important problem.  Can this scenario generate detectable
electromagnetic emission? The answer awaits more realistic
simulations, including spinning black holes.

Other astrophysical scenarios allow both matter and magnetic fields in
the vicinity of the binary close to merger.  In this case, calculation
of possible electromagnetic signatures requires solving the equations
of general relativistic magnetohydrodynamics in a dynamically evolving
spacetime governed by the Einstein equations.

\citet{vanMeter:2009gu} took a step towards solving this problem
by mapping the flow of pressureless matter, modeled as non-interacting
point particles, in the dynamical spacetime around the merging black
holes.  They started with a distribution of particles around the
black-hole binary, and then evolved the binary using numerical
relativity while tracing the motion of the particles along geodesics
as the binary evolves. To estimate the energetics of the flow, they
detected ``collisions'' by looking for particles within a small distance
(typically, $\lesssim 0.1M$) of each other and then calculating the
Lorentz factors $\gamma_{\rm coll}$ in the center-of-mass frame of
each collision.

They began with equal-mass inspiralling black-hole binaries $\sim 5$
orbits before merger, and considered both black holes to have either
zero spin, or $\ahat_{1,2} = 0.8$ with spin vectors aligned with the
binary orbital angular momentum. A single black hole of mass $M$ is
also evolved as a control case.  Approximately 75,000 geodesic
particles are then initially distributed uniformly throughout a solid
annulus centered on the binary and having inner radius $8M$, outer
radius $25M$, and vertical full thickness $10M$.  Particles within the
inner radius are excluded, to avoid transient signatures from
particles initially near the horizons. Note that these circumbinary
disks are geometrically thick, and thus potentially have high enough
inward radial speeds to keep up with the shrinking separation of the
inspiralling binary, providing a source of gas near the black holes
during their merger \cite{Milosavljevic:2004cg}.

\citet{vanMeter:2009gu} studied two initial velocity configurations
for the particles. In the ``orbital'' configuration, the initial
velocities are randomly distributed around a tangential velocity $V_c$
that would give a circular orbit in a Schwarzschild spacetime of mass
$M$, resulting in a scale height of $5M$. In contrast, the
``isotropic'' configuration is an extreme (that is, very hot) case in
which the particles only have random velocities, with each component
sampled from a Gaussian distribution of standard deviation
$V_c/\sqrt{3}$.

For the orbital initial velocities, the particles remain mostly in
a rotational configuration; some particles do enter the region around
the black holes, but are soon ejected by a gravitational slingshot.
The collisions typically occur at relatively low velocities with
Lorentz factors $\gamma_{\rm coll} \lesssim 1.8$.  By the time of
merger, there are essentially no particles near the black holes, and
thus there is no energetic signature of the merger.  The situation is
quite different with the isotropic initial conditions, which
provide a continual influx of particles throughout the evolution. Moreover, 
there is a clear signature of the merger visible in the maximum
Lorentz factor, which takes on values $\gamma_{\rm coll,max} \sim 2$
during the inspiral and spikes up to $\gamma_{\rm coll,max} \sim 3.5$
just before merger for the nonspinning case.  For rotating
black holes, $\gamma_{\rm coll,max} \sim 3$ during the inspiral,
spiking to $\gamma_{\rm coll,max} \sim 6$ just before merger.  In
addition, gravitational torques from the merging binary effectively
stir the particle distribution, leading to high-velocity outflows with
particles reaching $\gamma_{\rm coll} \sim 4$ or more.

\citet{vanMeter:2009gu} point out that, in realistic astrophysical
disks, viscosity would cause angular momentum transport, bringing
material inward towards the merging black holes.  This could produce a
scenario between the two extremes they studied, with a clear merger
signal.  More recent work by \citet{Bode:2009mt}, using hydrodynamical
simulations of gas around black-hole binaries, provided strong support
in this regard.  Further confirmation of this interesting
possibility, along with other mechanisms involving general
relativistic magnetohydrodynamics, awaits more detailed simulations
with gas and magnetic fields, currently in development.

\section{Frontiers and Future Directions}
\label{sec:frontiers}

Five years ago, we did not know what obstacles might conspire to
prevent the effective application of numerical-relativity simulations
in deriving predictions of general relativity to address questions of
astrophysical black-hole binaries.  Up to that time, the numerical
relativity researchers addressed their research primarily to questions
of theoretical and computational physics.

Then, quite suddenly, the last obstacles vanished.  As
reported, numerical relativity can now be applied to understand
general relativity's predictions of strong-field gravitational physics
and to begin addressing questions of astrophysics and gravitational-wave
data analysis. We expect future work in this field to be guided
not by internal problems in gravitational theory, but by questions of
astrophysics, and other areas where strong-field gravitational theory
applies. These questions will motivate both continued, more detailed,
investigation of some of the phenomena which have already been
revealed, and the development of new capabilities to bring simulation
results to bear on new questions.

\subsection{Gravitational-Wave Astronomy}

Though the work has begun in applying numerical-relativity results to
gravitational-wave observations, much more work is still needed.  It
is now clear that the signal-to-noise (SNR) ratio from the last
moments of the merger will dominate over the inspiral signal in many
potential observations.  Though numerical-relativity simulations have
shed considerable light on the basic features of the burst of
gravitational radiation that completes the predicted signals,
applications in data analysis will require {\em comprehensive}
quantitative knowledge of the merger signals generated over the full
parameter space of mergers.

While we discussed examples of mergers that sample the binary
black-hole parameter space along what may be its principal axes, a
full quantification of the signal space will require a long systematic
program of investigation.  Even with the aid of empirical models for
encoding numerical results, many simulations must be conducted to
qualify possible models.  Initial models suitable for detection and
(perhaps) parameter estimation of low-SNR signals, will require much
further development for application to the high-SNR observations
expected with future instrument upgrades, such as Advanced LIGO, and
future instruments such as LISA. Considerably more simulations of high
quality will be required to achieve these goals.

In addition to understanding the signals generated by mergers through
the bulk of parameter space in greater depth, relativists must also
extend the region of parameter space covered by simulations.  Typical
current simulations are limited in the range of parameters that can be
practically covered. Currently popular initial data models,
particularly those assuming conformal flatness, are limited to
$\ahat_{1,2} \lesssim 0.93$ \cite{Lovelace:2008tw,Dain:2002ee}.
Alternatives more suited to rapidly spinning Kerr black holes are less
well developed in other ways.  While astrophysical limits on spin
magnitudes remain unclear, it is likely that larger spins will have to
be treated with possibly novel initial data models. Other, related
challenges in the simulations may also exist.

Current simulations are also restricted to comparable mass ratios.
Simulations beyond $q \sim 5$ require somewhat heroic investments of
computational resources with present techniques.  Astrophysical
mergers may occur, however, over a broad range of mass ratios.  In the
limit of extremely large mass ratios, approximation techniques are
possible that treat the motion of the smaller object as a perturbed
geodesic in the spacetime of the larger black hole.  Such analytic
methods look promising in this limit, where numerical simulations are
less practical.  However, there is a rather large middle ground, say
$10 < q < 100$, where new techniques may be required.  One problem is
that the small spatial scale of the smaller black hole requires
extremely small timesteps for stable explicit time integrations, even
though there is nothing interesting happening on these timescales.
Alternatives such as implicit schemes \cite{Lau:2008fb} may open the
door to a broader range of applications.

Most of the simulations discussed also focus on circular, or
nearly circular orbits.  While this is clearly an important portion of
the black-hole-binary population, scenarios have been proposed in
which one black hole captures another from nearly parabolic initial
encounters.  Understanding the potential signals from such systems may
require new techniques appropriate for long periods of slow,
effectively Newtonian evolution, punctuated by brief periods of strong
gravitational interaction.

\subsection{Other Astrophysics}

As discussed, the numerical discovery of strong
gravitational-wave recoils in asymmetric mergers of spinning black
holes has had significant impact in areas of astrophysics beyond
direct gravitational-wave observations.  These results have spurred
excitement in the black-hole astronomy community about other possible
applications of numerical relativity. Since other areas of astronomy
are based on (primarily) electromagnetic observations, these
interactions naturally lead to applications involving the interactions
of black holes and electromagnetically visible matter.  There are many
relevant phenomena of interest: accretion disks around black holes,
possibly disturbed by gravitational recoil, black-hole--neutron-star
binaries, neutron-star--neutron-star binaries, jets from active
galactic nuclei, quasars, and the mysterious origins of gamma ray
bursts.

A key direction for future work has been adding physics beyond the
purely gravitational simulations we have reviewed here.  As
discussed, efforts are currently underway to use and develop
magnetohydrodynamics in a dynamical spacetime.  As such simulations
are applied with increasing realism to more astrophysical scenarios,
refinement of the techniques will be required.  Adequately resolving
shocks and turbulence interacting with magnetic fields, achieving
robust accuracy in the presence of very strong magnetic fields, and
enforcing the vanishing divergence of the magnetic fields, all in
curved spacetime, are among the challenges that need to be met.

\subsection{Other Physics}

Astrophysical mergers typically involve nearly circular orbits, or in
extreme scenarios, initially parabolic encounters.  Numerical
simulations can be applied to other sorts of black-hole interactions
as well.

\begin{figure}
\includegraphics*[scale=0.45,angle=0]{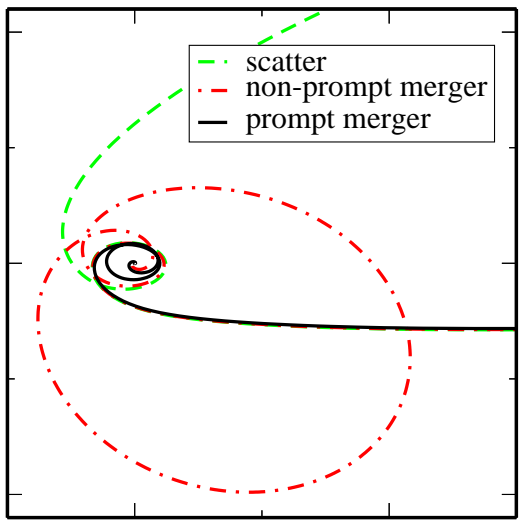}
\caption{Numerical simulations are applied to study the critical
  transition separating mergers from scattering events in
  high-velocity black-hole encounters \cite{Sperhake:2009jz}. The
  curves show the path of one black hole in each of three simulations
  begun with different initial conditions near the critical impact
  parameter.  The trajectories track closely together coming in from
  the right until the black holes encounter each other near the
  origin. Figure kindly provided by U. Sperhake.}
\label{fig:crit_traj}
\end{figure}

For high-velocity black-hole encounters, one expects a phase
transition dividing the spacetimes in which the black holes merge from
those in which they pass each other hyperbolically, never to approach
each other again: see Fig.~\ref{fig:crit_traj}. Numerical simulation
studies are beginning to elucidate critical behavior near this phase
transition \cite{Pretorius:2006tp,Pretorius:2007jn,Sperhake:2009jz}.

These fascinating phenomena at the extreme limit of strong
gravitational dynamics have been discussed in the context of upcoming
experimental measurements.  In the trans-Planckian energy limit, some
particle collisions may be describable by classical black-hole
dynamics \cite{Banks:1999gd,Eardley:2002re}.  In highly speculative
theories involving large extra spatial dimensions, it is conceivable
that TeV scale physics may be sufficient to produce such
black-hole-like collisions
\cite{Giddings:2001ih,Giddings:2001bu}. These possibilities motivate
the first numerical-simulation studies of ultra-relativistic
black-hole collisions
\cite{Sperhake:2008ga,Shibata:2008rq,Sperhake:2009jz}.

\subsection{Strong Gravity as Observational Science}

Einstein gave us general relativity nearly 100 years ago.  Over the
past century this theory, our standard model of gravitational physics,
has passed experimental and observational tests over ranging from
laboratory-scale physics experiments and solar-system tests to
observations of compact astronomical objects and cosmology
\cite{Will:2006LR}.  These measurements have, so far, provided no need
for a refinement of Einstein's theory of gravity [though some models
to explain cosmic acceleration do involve alternative gravity
\cite{Silvestri:2009hh}].

Numerical simulations are now revealing the detailed predictions of
Einstein's theory for the final merger dynamics of a black-hole
binary, and its record in the emitted gravitational radiation.
Gravitational-wave observations of these events will expose such
gravitational phenomena to measurements in a strong-gravity regime far
beyond anything which has previously been tested.  These include the
first tests of higher-order terms in the PN expansion and, indeed, of
the physical predictions which numerical relativity is just now
revealing.

The possibility that general relativity will find its limits in these
observations motivates a better understanding of what waveforms might
be predicted by alternative theories of gravity \cite{Yunes:2009ke}.
Alternative-gravity models of black-hole mergers would also likely
require numerical simulation to derive waveform predictions, but
little work has been done in this direction so far [see one related
example from \citet{Salgado:2008xh}].

If Einstein's theory proves correct in predicting the detailed physics
revealed by numerical relativity, it would stand as a truly incredible
achievement of scientific induction, divining the details of phenomena
vastly removed from physical observations on which the theory was
founded.  If not, then these observations, together with a confident
understanding of Einstein's predictions founded in numerical
relativity simulations, may indeed lay the foundation for the next
theory of gravity.

\acknowledgments

This review draws on the work of a broad research community and would
not have been possible without the many contributions and support of
our colleagues.  We especially want to thank a few individuals who
have made particularly valuable contributions.  William D. Boggs,
Bernd Br\"{u}gmann, Alessandra Buonanno,
Mark Hannam, Richard Matzner, Cole Miller, and
Harald Pfeiffer gave insightful and helpful comments on our
article.  Manuela Campanelli and Harald Pfeiffer supplied us with
figures from their simulations that were not in the published
literature.  We also benefited from many useful discussions with Sean
McWilliams.  We acknowledge support from NASA Grant No. 06-BEFS06-19. B.J.K.
was supported in part by an appointment to the NASA Postdoctoral
Program at the Goddard Space Flight Center, administered by Oak Ridge
Associated Universities through a contract with NASA.

%

\end{document}